\documentclass[11pt]{article}
\usepackage{amsmath}
\usepackage{tipa}
\usepackage{cite}
\usepackage{amsfonts}
\usepackage{color}
\usepackage{epsfig,colordvi,graphics,textcomp}
\usepackage{hyperref}
\usepackage{graphicx}
\graphicspath{{images/}{./img/}}
\usepackage{wrapfig}
\usepackage{figbib}
\usepackage{subcaption}
\usepackage[a4paper, left=2.3cm, right=2.0cm, top=2.3cm,bottom=2.3cm]{geometry}
\usepackage{cleveref}
\usepackage{cases}

\DeclareMathOperator{\tr}{tr}

\newcommand{\Om}{\Omega }
\newcommand{\om}{\omega }

\newtheorem{lemma}{Lemma}[section]

\newtheorem{rmk}[lemma]{Remark}

\newtheorem{notation}{Notation}

\newcommand{\R}{\mathbb{R}}

\newcommand{\dd}{\mathrm{d}}

\newcommand{\Bmb}{\mathbb{B}}

\newcommand{\ds}{\displaystyle}
\newcommand{\ud}{\, {\mathrm{d}}}

\newcommand{\eps}{\varepsilon}

\renewcommand{\div}{\mbox{div}}

\def\esc{\!\cdot\!}

\newcommand{\DW}[1][]{\mathbb{D}_{W#1}}

\newcommand{\ints}[1]{\ensuremath{\left\langle #1 \right\rangle}}

\newcommand{\bs}[1]{\pmb{#1}}

\newcommand{\trans}[1][]{^{#1\top}}

\newcommand{\basisFull}{\bs{a}}
\newcommand{\basis}{\bs{b}}
\newcommand{\basiscomp}[1][]{b_{#1}}

\newcommand{\momFull}{\bs{u}}
\newcommand{\mom}{\bs{w}}
\newcommand{\momcomp}[1][]{w_{#1}}

\newcommand{\mplFull}{\bs{\alpha}}
\newcommand{\mpl}{\bs{\beta}}
\newcommand{\mplcomp}[1][]{\beta_{#1}}

\def\relincludepath{data/postprocessing/}
\usepackage{tikz}
\usepackage{pgfplots}
\usepackage{ifthen}

\pgfplotsset{compat=1.13}
\usetikzlibrary{calc}
\usetikzlibrary{shapes}
\usepgfplotslibrary{colorbrewer}
\usepgfplotslibrary{groupplots}

\newlength{\figureonecolhorizontalsep}
\newlength{\figuretwocolhorizontalsep}
\newlength{\figurethreecolhorizontalsep}
\newlength{\figurethreecol}
\newlength{\figuretwocol}
\newlength{\figureonecol}
\newlength{\figuretwocollegend}
\newlength{\figurelegendhorizontalsep}
\newlength{\figureheight}
\newlength{\figurewidth}
\newlength{\figurehorizontalsep}
\newcommand{\overwritefiguresize}{false}

\newcommand{\overwritefiguresizefalse}{false}

\newcommand{\setdefaultfiguresize}[3][\figuretwocolhorizontalsep]{
	\ifx\overwritefiguresize\overwritefiguresizefalse%
		\setlength{\figureheight}{#2}%
		\setlength{\figurewidth}{#2}%
		\setlength{\figurehorizontalsep}{#1}
	\fi
	#3
}

\newcommand{\setdefaultfigurelayout}[2]{%
	\ifx\overwritefiguresize\overwritefiguresizefalse%
		\ifthenelse{\equal{#1}{onecol}}{%
			\setlength{\figureheight}{\figureonecol}%
			\setlength{\figurewidth}{\figureonecol}%
			\setlength{\figurehorizontalsep}{\figureonecolhorizontalsep}%
		}{}%
		\ifthenelse{\equal{#1}{twocol}}{%
			\setlength{\figureheight}{\figuretwocol}%
			\setlength{\figurewidth}{\figuretwocol}%
			\setlength{\figurehorizontalsep}{\figuretwocolhorizontalsep}%
		}{}%
		\ifthenelse{\equal{#1}{threecol}}{%
			\setlength{\figureheight}{\figurethreecol}%
			\setlength{\figurewidth}{\figurethreecol}%
			\setlength{\figurehorizontalsep}{\figurethreecolhorizontalsep}%
		}{}%
		\ifthenelse{\equal{#1}{twocollegend}}{%
			\setlength{\figureheight}{\figuretwocollegend}%
			\setlength{\figurewidth}{\figuretwocollegend}%
			\setlength{\figurehorizontalsep}{\figurelegendhorizontalsep}%
		}{}%
	\fi%
	#2
}

\newcommand{\figureticklabelfont}[1]{%
	\ifdim #1 < 3cm%
		\tiny
	\else%
		\ifdim #1 < 5cm%
			\scriptsize
		\else%
			\footnotesize
		\fi%
	\fi%
}

\newcommand{\figurelabelfont}[1]{%
	\ifdim #1 < 3cm%
		\footnotesize
	\else%
		\ifdim #1 < 5cm%
			\small
		\else%
			\normalsize
		\fi%
	\fi%
}

\newcounter{tikzsubfigcounter}[figure]
\renewcommand{\thetikzsubfigcounter}{\thechapter.\the\numexpr\value{figure}+1\relax\alph{tikzsubfigcounter}}

\providecommand{\tikztitle}[1]{ %
	\refstepcounter{tikzsubfigcounter}
	\textbf{(\alph{tikzsubfigcounter})}\space\space #1 
}

\newcounter{tikzsubfigcounterinvisible}[figure]
\renewcommand{\thetikzsubfigcounterinvisible}{\thechapter.\the\numexpr\value{figure}+1\relax\alph{tikzsubfigcounterinvisible}}

\pgfplotsset{
	colormap={cmviridis}{
rgb(0pt)=(0.267004,0.004874,0.329415);
rgb(1pt)=(0.268510,0.009605,0.335427);
rgb(2pt)=(0.269944,0.014625,0.341379);
rgb(3pt)=(0.271305,0.019942,0.347269);
rgb(4pt)=(0.272594,0.025563,0.353093);
rgb(5pt)=(0.273809,0.031497,0.358853);
rgb(6pt)=(0.274952,0.037752,0.364543);
rgb(7pt)=(0.276022,0.044167,0.370164);
rgb(8pt)=(0.277018,0.050344,0.375715);
rgb(9pt)=(0.277941,0.056324,0.381191);
rgb(10pt)=(0.278791,0.062145,0.386592);
rgb(11pt)=(0.279566,0.067836,0.391917);
rgb(12pt)=(0.280267,0.073417,0.397163);
rgb(13pt)=(0.280894,0.078907,0.402329);
rgb(14pt)=(0.281446,0.084320,0.407414);
rgb(15pt)=(0.281924,0.089666,0.412415);
rgb(16pt)=(0.282327,0.094955,0.417331);
rgb(17pt)=(0.282656,0.100196,0.422160);
rgb(18pt)=(0.282910,0.105393,0.426902);
rgb(19pt)=(0.283091,0.110553,0.431554);
rgb(20pt)=(0.283197,0.115680,0.436115);
rgb(21pt)=(0.283229,0.120777,0.440584);
rgb(22pt)=(0.283187,0.125848,0.444960);
rgb(23pt)=(0.283072,0.130895,0.449241);
rgb(24pt)=(0.282884,0.135920,0.453427);
rgb(25pt)=(0.282623,0.140926,0.457517);
rgb(26pt)=(0.282290,0.145912,0.461510);
rgb(27pt)=(0.281887,0.150881,0.465405);
rgb(28pt)=(0.281412,0.155834,0.469201);
rgb(29pt)=(0.280868,0.160771,0.472899);
rgb(30pt)=(0.280255,0.165693,0.476498);
rgb(31pt)=(0.279574,0.170599,0.479997);
rgb(32pt)=(0.278826,0.175490,0.483397);
rgb(33pt)=(0.278012,0.180367,0.486697);
rgb(34pt)=(0.277134,0.185228,0.489898);
rgb(35pt)=(0.276194,0.190074,0.493001);
rgb(36pt)=(0.275191,0.194905,0.496005);
rgb(37pt)=(0.274128,0.199721,0.498911);
rgb(38pt)=(0.273006,0.204520,0.501721);
rgb(39pt)=(0.271828,0.209303,0.504434);
rgb(40pt)=(0.270595,0.214069,0.507052);
rgb(41pt)=(0.269308,0.218818,0.509577);
rgb(42pt)=(0.267968,0.223549,0.512008);
rgb(43pt)=(0.266580,0.228262,0.514349);
rgb(44pt)=(0.265145,0.232956,0.516599);
rgb(45pt)=(0.263663,0.237631,0.518762);
rgb(46pt)=(0.262138,0.242286,0.520837);
rgb(47pt)=(0.260571,0.246922,0.522828);
rgb(48pt)=(0.258965,0.251537,0.524736);
rgb(49pt)=(0.257322,0.256130,0.526563);
rgb(50pt)=(0.255645,0.260703,0.528312);
rgb(51pt)=(0.253935,0.265254,0.529983);
rgb(52pt)=(0.252194,0.269783,0.531579);
rgb(53pt)=(0.250425,0.274290,0.533103);
rgb(54pt)=(0.248629,0.278775,0.534556);
rgb(55pt)=(0.246811,0.283237,0.535941);
rgb(56pt)=(0.244972,0.287675,0.537260);
rgb(57pt)=(0.243113,0.292092,0.538516);
rgb(58pt)=(0.241237,0.296485,0.539709);
rgb(59pt)=(0.239346,0.300855,0.540844);
rgb(60pt)=(0.237441,0.305202,0.541921);
rgb(61pt)=(0.235526,0.309527,0.542944);
rgb(62pt)=(0.233603,0.313828,0.543914);
rgb(63pt)=(0.231674,0.318106,0.544834);
rgb(64pt)=(0.229739,0.322361,0.545706);
rgb(65pt)=(0.227802,0.326594,0.546532);
rgb(66pt)=(0.225863,0.330805,0.547314);
rgb(67pt)=(0.223925,0.334994,0.548053);
rgb(68pt)=(0.221989,0.339161,0.548752);
rgb(69pt)=(0.220057,0.343307,0.549413);
rgb(70pt)=(0.218130,0.347432,0.550038);
rgb(71pt)=(0.216210,0.351535,0.550627);
rgb(72pt)=(0.214298,0.355619,0.551184);
rgb(73pt)=(0.212395,0.359683,0.551710);
rgb(74pt)=(0.210503,0.363727,0.552206);
rgb(75pt)=(0.208623,0.367752,0.552675);
rgb(76pt)=(0.206756,0.371758,0.553117);
rgb(77pt)=(0.204903,0.375746,0.553533);
rgb(78pt)=(0.203063,0.379716,0.553925);
rgb(79pt)=(0.201239,0.383670,0.554294);
rgb(80pt)=(0.199430,0.387607,0.554642);
rgb(81pt)=(0.197636,0.391528,0.554969);
rgb(82pt)=(0.195860,0.395433,0.555276);
rgb(83pt)=(0.194100,0.399323,0.555565);
rgb(84pt)=(0.192357,0.403199,0.555836);
rgb(85pt)=(0.190631,0.407061,0.556089);
rgb(86pt)=(0.188923,0.410910,0.556326);
rgb(87pt)=(0.187231,0.414746,0.556547);
rgb(88pt)=(0.185556,0.418570,0.556753);
rgb(89pt)=(0.183898,0.422383,0.556944);
rgb(90pt)=(0.182256,0.426184,0.557120);
rgb(91pt)=(0.180629,0.429975,0.557282);
rgb(92pt)=(0.179019,0.433756,0.557430);
rgb(93pt)=(0.177423,0.437527,0.557565);
rgb(94pt)=(0.175841,0.441290,0.557685);
rgb(95pt)=(0.174274,0.445044,0.557792);
rgb(96pt)=(0.172719,0.448791,0.557885);
rgb(97pt)=(0.171176,0.452530,0.557965);
rgb(98pt)=(0.169646,0.456262,0.558030);
rgb(99pt)=(0.168126,0.459988,0.558082);
rgb(100pt)=(0.166617,0.463708,0.558119);
rgb(101pt)=(0.165117,0.467423,0.558141);
rgb(102pt)=(0.163625,0.471133,0.558148);
rgb(103pt)=(0.162142,0.474838,0.558140);
rgb(104pt)=(0.160665,0.478540,0.558115);
rgb(105pt)=(0.159194,0.482237,0.558073);
rgb(106pt)=(0.157729,0.485932,0.558013);
rgb(107pt)=(0.156270,0.489624,0.557936);
rgb(108pt)=(0.154815,0.493313,0.557840);
rgb(109pt)=(0.153364,0.497000,0.557724);
rgb(110pt)=(0.151918,0.500685,0.557587);
rgb(111pt)=(0.150476,0.504369,0.557430);
rgb(112pt)=(0.149039,0.508051,0.557250);
rgb(113pt)=(0.147607,0.511733,0.557049);
rgb(114pt)=(0.146180,0.515413,0.556823);
rgb(115pt)=(0.144759,0.519093,0.556572);
rgb(116pt)=(0.143343,0.522773,0.556295);
rgb(117pt)=(0.141935,0.526453,0.555991);
rgb(118pt)=(0.140536,0.530132,0.555659);
rgb(119pt)=(0.139147,0.533812,0.555298);
rgb(120pt)=(0.137770,0.537492,0.554906);
rgb(121pt)=(0.136408,0.541173,0.554483);
rgb(122pt)=(0.135066,0.544853,0.554029);
rgb(123pt)=(0.133743,0.548535,0.553541);
rgb(124pt)=(0.132444,0.552216,0.553018);
rgb(125pt)=(0.131172,0.555899,0.552459);
rgb(126pt)=(0.129933,0.559582,0.551864);
rgb(127pt)=(0.128729,0.563265,0.551229);
rgb(128pt)=(0.127568,0.566949,0.550556);
rgb(129pt)=(0.126453,0.570633,0.549841);
rgb(130pt)=(0.125394,0.574318,0.549086);
rgb(131pt)=(0.124395,0.578002,0.548287);
rgb(132pt)=(0.123463,0.581687,0.547445);
rgb(133pt)=(0.122606,0.585371,0.546557);
rgb(134pt)=(0.121831,0.589055,0.545623);
rgb(135pt)=(0.121148,0.592739,0.544641);
rgb(136pt)=(0.120565,0.596422,0.543611);
rgb(137pt)=(0.120092,0.600104,0.542530);
rgb(138pt)=(0.119738,0.603785,0.541400);
rgb(139pt)=(0.119512,0.607464,0.540218);
rgb(140pt)=(0.119423,0.611141,0.538982);
rgb(141pt)=(0.119483,0.614817,0.537692);
rgb(142pt)=(0.119699,0.618490,0.536347);
rgb(143pt)=(0.120081,0.622161,0.534946);
rgb(144pt)=(0.120638,0.625828,0.533488);
rgb(145pt)=(0.121380,0.629492,0.531973);
rgb(146pt)=(0.122312,0.633153,0.530398);
rgb(147pt)=(0.123444,0.636809,0.528763);
rgb(148pt)=(0.124780,0.640461,0.527068);
rgb(149pt)=(0.126326,0.644107,0.525311);
rgb(150pt)=(0.128087,0.647749,0.523491);
rgb(151pt)=(0.130067,0.651384,0.521608);
rgb(152pt)=(0.132268,0.655014,0.519661);
rgb(153pt)=(0.134692,0.658636,0.517649);
rgb(154pt)=(0.137339,0.662252,0.515571);
rgb(155pt)=(0.140210,0.665859,0.513427);
rgb(156pt)=(0.143303,0.669459,0.511215);
rgb(157pt)=(0.146616,0.673050,0.508936);
rgb(158pt)=(0.150148,0.676631,0.506589);
rgb(159pt)=(0.153894,0.680203,0.504172);
rgb(160pt)=(0.157851,0.683765,0.501686);
rgb(161pt)=(0.162016,0.687316,0.499129);
rgb(162pt)=(0.166383,0.690856,0.496502);
rgb(163pt)=(0.170948,0.694384,0.493803);
rgb(164pt)=(0.175707,0.697900,0.491033);
rgb(165pt)=(0.180653,0.701402,0.488189);
rgb(166pt)=(0.185783,0.704891,0.485273);
rgb(167pt)=(0.191090,0.708366,0.482284);
rgb(168pt)=(0.196571,0.711827,0.479221);
rgb(169pt)=(0.202219,0.715272,0.476084);
rgb(170pt)=(0.208030,0.718701,0.472873);
rgb(171pt)=(0.214000,0.722114,0.469588);
rgb(172pt)=(0.220124,0.725509,0.466226);
rgb(173pt)=(0.226397,0.728888,0.462789);
rgb(174pt)=(0.232815,0.732247,0.459277);
rgb(175pt)=(0.239374,0.735588,0.455688);
rgb(176pt)=(0.246070,0.738910,0.452024);
rgb(177pt)=(0.252899,0.742211,0.448284);
rgb(178pt)=(0.259857,0.745492,0.444467);
rgb(179pt)=(0.266941,0.748751,0.440573);
rgb(180pt)=(0.274149,0.751988,0.436601);
rgb(181pt)=(0.281477,0.755203,0.432552);
rgb(182pt)=(0.288921,0.758394,0.428426);
rgb(183pt)=(0.296479,0.761561,0.424223);
rgb(184pt)=(0.304148,0.764704,0.419943);
rgb(185pt)=(0.311925,0.767822,0.415586);
rgb(186pt)=(0.319809,0.770914,0.411152);
rgb(187pt)=(0.327796,0.773980,0.406640);
rgb(188pt)=(0.335885,0.777018,0.402049);
rgb(189pt)=(0.344074,0.780029,0.397381);
rgb(190pt)=(0.352360,0.783011,0.392636);
rgb(191pt)=(0.360741,0.785964,0.387814);
rgb(192pt)=(0.369214,0.788888,0.382914);
rgb(193pt)=(0.377779,0.791781,0.377939);
rgb(194pt)=(0.386433,0.794644,0.372886);
rgb(195pt)=(0.395174,0.797475,0.367757);
rgb(196pt)=(0.404001,0.800275,0.362552);
rgb(197pt)=(0.412913,0.803041,0.357269);
rgb(198pt)=(0.421908,0.805774,0.351910);
rgb(199pt)=(0.430983,0.808473,0.346476);
rgb(200pt)=(0.440137,0.811138,0.340967);
rgb(201pt)=(0.449368,0.813768,0.335384);
rgb(202pt)=(0.458674,0.816363,0.329727);
rgb(203pt)=(0.468053,0.818921,0.323998);
rgb(204pt)=(0.477504,0.821444,0.318195);
rgb(205pt)=(0.487026,0.823929,0.312321);
rgb(206pt)=(0.496615,0.826376,0.306377);
rgb(207pt)=(0.506271,0.828786,0.300362);
rgb(208pt)=(0.515992,0.831158,0.294279);
rgb(209pt)=(0.525776,0.833491,0.288127);
rgb(210pt)=(0.535621,0.835785,0.281908);
rgb(211pt)=(0.545524,0.838039,0.275626);
rgb(212pt)=(0.555484,0.840254,0.269281);
rgb(213pt)=(0.565498,0.842430,0.262877);
rgb(214pt)=(0.575563,0.844566,0.256415);
rgb(215pt)=(0.585678,0.846661,0.249897);
rgb(216pt)=(0.595839,0.848717,0.243329);
rgb(217pt)=(0.606045,0.850733,0.236712);
rgb(218pt)=(0.616293,0.852709,0.230052);
rgb(219pt)=(0.626579,0.854645,0.223353);
rgb(220pt)=(0.636902,0.856542,0.216620);
rgb(221pt)=(0.647257,0.858400,0.209861);
rgb(222pt)=(0.657642,0.860219,0.203082);
rgb(223pt)=(0.668054,0.861999,0.196293);
rgb(224pt)=(0.678489,0.863742,0.189503);
rgb(225pt)=(0.688944,0.865448,0.182725);
rgb(226pt)=(0.699415,0.867117,0.175971);
rgb(227pt)=(0.709898,0.868751,0.169257);
rgb(228pt)=(0.720391,0.870350,0.162603);
rgb(229pt)=(0.730889,0.871916,0.156029);
rgb(230pt)=(0.741388,0.873449,0.149561);
rgb(231pt)=(0.751884,0.874951,0.143228);
rgb(232pt)=(0.762373,0.876424,0.137064);
rgb(233pt)=(0.772852,0.877868,0.131109);
rgb(234pt)=(0.783315,0.879285,0.125405);
rgb(235pt)=(0.793760,0.880678,0.120005);
rgb(236pt)=(0.804182,0.882046,0.114965);
rgb(237pt)=(0.814576,0.883393,0.110347);
rgb(238pt)=(0.824940,0.884720,0.106217);
rgb(239pt)=(0.835270,0.886029,0.102646);
rgb(240pt)=(0.845561,0.887322,0.099702);
rgb(241pt)=(0.855810,0.888601,0.097452);
rgb(242pt)=(0.866013,0.889868,0.095953);
rgb(243pt)=(0.876168,0.891125,0.095250);
rgb(244pt)=(0.886271,0.892374,0.095374);
rgb(245pt)=(0.896320,0.893616,0.096335);
rgb(246pt)=(0.906311,0.894855,0.098125);
rgb(247pt)=(0.916242,0.896091,0.100717);
rgb(248pt)=(0.926106,0.897330,0.104071);
rgb(249pt)=(0.935904,0.898570,0.108131);
rgb(250pt)=(0.945636,0.899815,0.112838);
rgb(251pt)=(0.955300,0.901065,0.118128);
rgb(252pt)=(0.964894,0.902323,0.123941);
rgb(253pt)=(0.974417,0.903590,0.130215);
rgb(254pt)=(0.983868,0.904867,0.136897);
rgb(255pt)=(0.993248,0.906157,0.143936);
}
}
\pgfplotsset{
colormap={cmgray}{
rgb(000pt)=(0.000000,0.000000,0.000000);
rgb(001pt)=(0.003922,0.003922,0.003922);
rgb(002pt)=(0.007843,0.007843,0.007843);
rgb(003pt)=(0.011765,0.011765,0.011765);
rgb(004pt)=(0.015686,0.015686,0.015686);
rgb(005pt)=(0.019608,0.019608,0.019608);
rgb(006pt)=(0.023529,0.023529,0.023529);
rgb(007pt)=(0.027451,0.027451,0.027451);
rgb(008pt)=(0.031373,0.031373,0.031373);
rgb(009pt)=(0.035294,0.035294,0.035294);
rgb(010pt)=(0.039216,0.039216,0.039216);
rgb(011pt)=(0.043137,0.043137,0.043137);
rgb(012pt)=(0.047059,0.047059,0.047059);
rgb(013pt)=(0.050980,0.050980,0.050980);
rgb(014pt)=(0.054902,0.054902,0.054902);
rgb(015pt)=(0.058824,0.058824,0.058824);
rgb(016pt)=(0.062745,0.062745,0.062745);
rgb(017pt)=(0.066667,0.066667,0.066667);
rgb(018pt)=(0.070588,0.070588,0.070588);
rgb(019pt)=(0.074510,0.074510,0.074510);
rgb(020pt)=(0.078431,0.078431,0.078431);
rgb(021pt)=(0.082353,0.082353,0.082353);
rgb(022pt)=(0.086275,0.086275,0.086275);
rgb(023pt)=(0.090196,0.090196,0.090196);
rgb(024pt)=(0.094118,0.094118,0.094118);
rgb(025pt)=(0.098039,0.098039,0.098039);
rgb(026pt)=(0.101961,0.101961,0.101961);
rgb(027pt)=(0.105882,0.105882,0.105882);
rgb(028pt)=(0.109804,0.109804,0.109804);
rgb(029pt)=(0.113725,0.113725,0.113725);
rgb(030pt)=(0.117647,0.117647,0.117647);
rgb(031pt)=(0.121569,0.121569,0.121569);
rgb(032pt)=(0.125490,0.125490,0.125490);
rgb(033pt)=(0.129412,0.129412,0.129412);
rgb(034pt)=(0.133333,0.133333,0.133333);
rgb(035pt)=(0.137255,0.137255,0.137255);
rgb(036pt)=(0.141176,0.141176,0.141176);
rgb(037pt)=(0.145098,0.145098,0.145098);
rgb(038pt)=(0.149020,0.149020,0.149020);
rgb(039pt)=(0.152941,0.152941,0.152941);
rgb(040pt)=(0.156863,0.156863,0.156863);
rgb(041pt)=(0.160784,0.160784,0.160784);
rgb(042pt)=(0.164706,0.164706,0.164706);
rgb(043pt)=(0.168627,0.168627,0.168627);
rgb(044pt)=(0.172549,0.172549,0.172549);
rgb(045pt)=(0.176471,0.176471,0.176471);
rgb(046pt)=(0.180392,0.180392,0.180392);
rgb(047pt)=(0.184314,0.184314,0.184314);
rgb(048pt)=(0.188235,0.188235,0.188235);
rgb(049pt)=(0.192157,0.192157,0.192157);
rgb(050pt)=(0.196078,0.196078,0.196078);
rgb(051pt)=(0.200000,0.200000,0.200000);
rgb(052pt)=(0.203922,0.203922,0.203922);
rgb(053pt)=(0.207843,0.207843,0.207843);
rgb(054pt)=(0.211765,0.211765,0.211765);
rgb(055pt)=(0.215686,0.215686,0.215686);
rgb(056pt)=(0.219608,0.219608,0.219608);
rgb(057pt)=(0.223529,0.223529,0.223529);
rgb(058pt)=(0.227451,0.227451,0.227451);
rgb(059pt)=(0.231373,0.231373,0.231373);
rgb(060pt)=(0.235294,0.235294,0.235294);
rgb(061pt)=(0.239216,0.239216,0.239216);
rgb(062pt)=(0.243137,0.243137,0.243137);
rgb(063pt)=(0.247059,0.247059,0.247059);
rgb(064pt)=(0.250980,0.250980,0.250980);
rgb(065pt)=(0.254902,0.254902,0.254902);
rgb(066pt)=(0.258824,0.258824,0.258824);
rgb(067pt)=(0.262745,0.262745,0.262745);
rgb(068pt)=(0.266667,0.266667,0.266667);
rgb(069pt)=(0.270588,0.270588,0.270588);
rgb(070pt)=(0.274510,0.274510,0.274510);
rgb(071pt)=(0.278431,0.278431,0.278431);
rgb(072pt)=(0.282353,0.282353,0.282353);
rgb(073pt)=(0.286275,0.286275,0.286275);
rgb(074pt)=(0.290196,0.290196,0.290196);
rgb(075pt)=(0.294118,0.294118,0.294118);
rgb(076pt)=(0.298039,0.298039,0.298039);
rgb(077pt)=(0.301961,0.301961,0.301961);
rgb(078pt)=(0.305882,0.305882,0.305882);
rgb(079pt)=(0.309804,0.309804,0.309804);
rgb(080pt)=(0.313725,0.313725,0.313725);
rgb(081pt)=(0.317647,0.317647,0.317647);
rgb(082pt)=(0.321569,0.321569,0.321569);
rgb(083pt)=(0.325490,0.325490,0.325490);
rgb(084pt)=(0.329412,0.329412,0.329412);
rgb(085pt)=(0.333333,0.333333,0.333333);
rgb(086pt)=(0.337255,0.337255,0.337255);
rgb(087pt)=(0.341176,0.341176,0.341176);
rgb(088pt)=(0.345098,0.345098,0.345098);
rgb(089pt)=(0.349020,0.349020,0.349020);
rgb(090pt)=(0.352941,0.352941,0.352941);
rgb(091pt)=(0.356863,0.356863,0.356863);
rgb(092pt)=(0.360784,0.360784,0.360784);
rgb(093pt)=(0.364706,0.364706,0.364706);
rgb(094pt)=(0.368627,0.368627,0.368627);
rgb(095pt)=(0.372549,0.372549,0.372549);
rgb(096pt)=(0.376471,0.376471,0.376471);
rgb(097pt)=(0.380392,0.380392,0.380392);
rgb(098pt)=(0.384314,0.384314,0.384314);
rgb(099pt)=(0.388235,0.388235,0.388235);
rgb(100pt)=(0.392157,0.392157,0.392157);
rgb(101pt)=(0.396078,0.396078,0.396078);
rgb(102pt)=(0.400000,0.400000,0.400000);
rgb(103pt)=(0.403922,0.403922,0.403922);
rgb(104pt)=(0.407843,0.407843,0.407843);
rgb(105pt)=(0.411765,0.411765,0.411765);
rgb(106pt)=(0.415686,0.415686,0.415686);
rgb(107pt)=(0.419608,0.419608,0.419608);
rgb(108pt)=(0.423529,0.423529,0.423529);
rgb(109pt)=(0.427451,0.427451,0.427451);
rgb(110pt)=(0.431373,0.431373,0.431373);
rgb(111pt)=(0.435294,0.435294,0.435294);
rgb(112pt)=(0.439216,0.439216,0.439216);
rgb(113pt)=(0.443137,0.443137,0.443137);
rgb(114pt)=(0.447059,0.447059,0.447059);
rgb(115pt)=(0.450980,0.450980,0.450980);
rgb(116pt)=(0.454902,0.454902,0.454902);
rgb(117pt)=(0.458824,0.458824,0.458824);
rgb(118pt)=(0.462745,0.462745,0.462745);
rgb(119pt)=(0.466667,0.466667,0.466667);
rgb(120pt)=(0.470588,0.470588,0.470588);
rgb(121pt)=(0.474510,0.474510,0.474510);
rgb(122pt)=(0.478431,0.478431,0.478431);
rgb(123pt)=(0.482353,0.482353,0.482353);
rgb(124pt)=(0.486275,0.486275,0.486275);
rgb(125pt)=(0.490196,0.490196,0.490196);
rgb(126pt)=(0.494118,0.494118,0.494118);
rgb(127pt)=(0.498039,0.498039,0.498039);
rgb(128pt)=(0.501961,0.501961,0.501961);
rgb(129pt)=(0.505882,0.505882,0.505882);
rgb(130pt)=(0.509804,0.509804,0.509804);
rgb(131pt)=(0.513725,0.513725,0.513725);
rgb(132pt)=(0.517647,0.517647,0.517647);
rgb(133pt)=(0.521569,0.521569,0.521569);
rgb(134pt)=(0.525490,0.525490,0.525490);
rgb(135pt)=(0.529412,0.529412,0.529412);
rgb(136pt)=(0.533333,0.533333,0.533333);
rgb(137pt)=(0.537255,0.537255,0.537255);
rgb(138pt)=(0.541176,0.541176,0.541176);
rgb(139pt)=(0.545098,0.545098,0.545098);
rgb(140pt)=(0.549020,0.549020,0.549020);
rgb(141pt)=(0.552941,0.552941,0.552941);
rgb(142pt)=(0.556863,0.556863,0.556863);
rgb(143pt)=(0.560784,0.560784,0.560784);
rgb(144pt)=(0.564706,0.564706,0.564706);
rgb(145pt)=(0.568627,0.568627,0.568627);
rgb(146pt)=(0.572549,0.572549,0.572549);
rgb(147pt)=(0.576471,0.576471,0.576471);
rgb(148pt)=(0.580392,0.580392,0.580392);
rgb(149pt)=(0.584314,0.584314,0.584314);
rgb(150pt)=(0.588235,0.588235,0.588235);
rgb(151pt)=(0.592157,0.592157,0.592157);
rgb(152pt)=(0.596078,0.596078,0.596078);
rgb(153pt)=(0.600000,0.600000,0.600000);
rgb(154pt)=(0.603922,0.603922,0.603922);
rgb(155pt)=(0.607843,0.607843,0.607843);
rgb(156pt)=(0.611765,0.611765,0.611765);
rgb(157pt)=(0.615686,0.615686,0.615686);
rgb(158pt)=(0.619608,0.619608,0.619608);
rgb(159pt)=(0.623529,0.623529,0.623529);
rgb(160pt)=(0.627451,0.627451,0.627451);
rgb(161pt)=(0.631373,0.631373,0.631373);
rgb(162pt)=(0.635294,0.635294,0.635294);
rgb(163pt)=(0.639216,0.639216,0.639216);
rgb(164pt)=(0.643137,0.643137,0.643137);
rgb(165pt)=(0.647059,0.647059,0.647059);
rgb(166pt)=(0.650980,0.650980,0.650980);
rgb(167pt)=(0.654902,0.654902,0.654902);
rgb(168pt)=(0.658824,0.658824,0.658824);
rgb(169pt)=(0.662745,0.662745,0.662745);
rgb(170pt)=(0.666667,0.666667,0.666667);
rgb(171pt)=(0.670588,0.670588,0.670588);
rgb(172pt)=(0.674510,0.674510,0.674510);
rgb(173pt)=(0.678431,0.678431,0.678431);
rgb(174pt)=(0.682353,0.682353,0.682353);
rgb(175pt)=(0.686275,0.686275,0.686275);
rgb(176pt)=(0.690196,0.690196,0.690196);
rgb(177pt)=(0.694118,0.694118,0.694118);
rgb(178pt)=(0.698039,0.698039,0.698039);
rgb(179pt)=(0.701961,0.701961,0.701961);
rgb(180pt)=(0.705882,0.705882,0.705882);
rgb(181pt)=(0.709804,0.709804,0.709804);
rgb(182pt)=(0.713725,0.713725,0.713725);
rgb(183pt)=(0.717647,0.717647,0.717647);
rgb(184pt)=(0.721569,0.721569,0.721569);
rgb(185pt)=(0.725490,0.725490,0.725490);
rgb(186pt)=(0.729412,0.729412,0.729412);
rgb(187pt)=(0.733333,0.733333,0.733333);
rgb(188pt)=(0.737255,0.737255,0.737255);
rgb(189pt)=(0.741176,0.741176,0.741176);
rgb(190pt)=(0.745098,0.745098,0.745098);
rgb(191pt)=(0.749020,0.749020,0.749020);
rgb(192pt)=(0.752941,0.752941,0.752941);
rgb(193pt)=(0.756863,0.756863,0.756863);
rgb(194pt)=(0.760784,0.760784,0.760784);
rgb(195pt)=(0.764706,0.764706,0.764706);
rgb(196pt)=(0.768627,0.768627,0.768627);
rgb(197pt)=(0.772549,0.772549,0.772549);
rgb(198pt)=(0.776471,0.776471,0.776471);
rgb(199pt)=(0.780392,0.780392,0.780392);
rgb(200pt)=(0.784314,0.784314,0.784314);
rgb(201pt)=(0.788235,0.788235,0.788235);
rgb(202pt)=(0.792157,0.792157,0.792157);
rgb(203pt)=(0.796078,0.796078,0.796078);
rgb(204pt)=(0.800000,0.800000,0.800000);
rgb(205pt)=(0.803922,0.803922,0.803922);
rgb(206pt)=(0.807843,0.807843,0.807843);
rgb(207pt)=(0.811765,0.811765,0.811765);
rgb(208pt)=(0.815686,0.815686,0.815686);
rgb(209pt)=(0.819608,0.819608,0.819608);
rgb(210pt)=(0.823529,0.823529,0.823529);
rgb(211pt)=(0.827451,0.827451,0.827451);
rgb(212pt)=(0.831373,0.831373,0.831373);
rgb(213pt)=(0.835294,0.835294,0.835294);
rgb(214pt)=(0.839216,0.839216,0.839216);
rgb(215pt)=(0.843137,0.843137,0.843137);
rgb(216pt)=(0.847059,0.847059,0.847059);
rgb(217pt)=(0.850980,0.850980,0.850980);
rgb(218pt)=(0.854902,0.854902,0.854902);
rgb(219pt)=(0.858824,0.858824,0.858824);
rgb(220pt)=(0.862745,0.862745,0.862745);
rgb(221pt)=(0.866667,0.866667,0.866667);
rgb(222pt)=(0.870588,0.870588,0.870588);
rgb(223pt)=(0.874510,0.874510,0.874510);
rgb(224pt)=(0.878431,0.878431,0.878431);
rgb(225pt)=(0.882353,0.882353,0.882353);
rgb(226pt)=(0.886275,0.886275,0.886275);
rgb(227pt)=(0.890196,0.890196,0.890196);
rgb(228pt)=(0.894118,0.894118,0.894118);
rgb(229pt)=(0.898039,0.898039,0.898039);
rgb(230pt)=(0.901961,0.901961,0.901961);
rgb(231pt)=(0.905882,0.905882,0.905882);
rgb(232pt)=(0.909804,0.909804,0.909804);
rgb(233pt)=(0.913725,0.913725,0.913725);
rgb(234pt)=(0.917647,0.917647,0.917647);
rgb(235pt)=(0.921569,0.921569,0.921569);
rgb(236pt)=(0.925490,0.925490,0.925490);
rgb(237pt)=(0.929412,0.929412,0.929412);
rgb(238pt)=(0.933333,0.933333,0.933333);
rgb(239pt)=(0.937255,0.937255,0.937255);
rgb(240pt)=(0.941176,0.941176,0.941176);
rgb(241pt)=(0.945098,0.945098,0.945098);
rgb(242pt)=(0.949020,0.949020,0.949020);
rgb(243pt)=(0.952941,0.952941,0.952941);
rgb(244pt)=(0.956863,0.956863,0.956863);
rgb(245pt)=(0.960784,0.960784,0.960784);
rgb(246pt)=(0.964706,0.964706,0.964706);
rgb(247pt)=(0.968627,0.968627,0.968627);
rgb(248pt)=(0.972549,0.972549,0.972549);
rgb(249pt)=(0.976471,0.976471,0.976471);
rgb(250pt)=(0.980392,0.980392,0.980392);
rgb(251pt)=(0.984314,0.984314,0.984314);
rgb(252pt)=(0.988235,0.988235,0.988235);
rgb(253pt)=(0.992157,0.992157,0.992157);
rgb(254pt)=(0.996078,0.996078,0.996078);
rgb(255pt)=(1.000000,1.000000,1.000000);
}
}

\pgfplotsset{
colormap={cmRdBu_r}{
rgb(000pt)=(0.019608,0.188235,0.380392);
rgb(001pt)=(0.023914,0.196540,0.391926);
rgb(002pt)=(0.028220,0.204844,0.403460);
rgb(003pt)=(0.032526,0.213149,0.414994);
rgb(004pt)=(0.036832,0.221453,0.426528);
rgb(005pt)=(0.041138,0.229758,0.438062);
rgb(006pt)=(0.045444,0.238062,0.449596);
rgb(007pt)=(0.049750,0.246367,0.461130);
rgb(008pt)=(0.054056,0.254671,0.472664);
rgb(009pt)=(0.058362,0.262976,0.484198);
rgb(010pt)=(0.062668,0.271280,0.495732);
rgb(011pt)=(0.066974,0.279585,0.507266);
rgb(012pt)=(0.071280,0.287889,0.518800);
rgb(013pt)=(0.075586,0.296194,0.530335);
rgb(014pt)=(0.079892,0.304498,0.541869);
rgb(015pt)=(0.084198,0.312803,0.553403);
rgb(016pt)=(0.088504,0.321107,0.564937);
rgb(017pt)=(0.092810,0.329412,0.576471);
rgb(018pt)=(0.097116,0.337716,0.588005);
rgb(019pt)=(0.101423,0.346021,0.599539);
rgb(020pt)=(0.105729,0.354325,0.611073);
rgb(021pt)=(0.110035,0.362630,0.622607);
rgb(022pt)=(0.114341,0.370934,0.634141);
rgb(023pt)=(0.118647,0.379239,0.645675);
rgb(024pt)=(0.122953,0.387543,0.657209);
rgb(025pt)=(0.127259,0.395848,0.668743);
rgb(026pt)=(0.132026,0.403460,0.676278);
rgb(027pt)=(0.137255,0.410381,0.679815);
rgb(028pt)=(0.142484,0.417301,0.683353);
rgb(029pt)=(0.147712,0.424221,0.686890);
rgb(030pt)=(0.152941,0.431142,0.690427);
rgb(031pt)=(0.158170,0.438062,0.693964);
rgb(032pt)=(0.163399,0.444983,0.697501);
rgb(033pt)=(0.168627,0.451903,0.701038);
rgb(034pt)=(0.173856,0.458824,0.704575);
rgb(035pt)=(0.179085,0.465744,0.708112);
rgb(036pt)=(0.184314,0.472664,0.711649);
rgb(037pt)=(0.189542,0.479585,0.715186);
rgb(038pt)=(0.194771,0.486505,0.718724);
rgb(039pt)=(0.200000,0.493426,0.722261);
rgb(040pt)=(0.205229,0.500346,0.725798);
rgb(041pt)=(0.210458,0.507266,0.729335);
rgb(042pt)=(0.215686,0.514187,0.732872);
rgb(043pt)=(0.220915,0.521107,0.736409);
rgb(044pt)=(0.226144,0.528028,0.739946);
rgb(045pt)=(0.231373,0.534948,0.743483);
rgb(046pt)=(0.236601,0.541869,0.747020);
rgb(047pt)=(0.241830,0.548789,0.750557);
rgb(048pt)=(0.247059,0.555709,0.754095);
rgb(049pt)=(0.252288,0.562630,0.757632);
rgb(050pt)=(0.257516,0.569550,0.761169);
rgb(051pt)=(0.262745,0.576471,0.764706);
rgb(052pt)=(0.274894,0.584160,0.768858);
rgb(053pt)=(0.287043,0.591849,0.773010);
rgb(054pt)=(0.299193,0.599539,0.777163);
rgb(055pt)=(0.311342,0.607228,0.781315);
rgb(056pt)=(0.323491,0.614917,0.785467);
rgb(057pt)=(0.335640,0.622607,0.789619);
rgb(058pt)=(0.347789,0.630296,0.793772);
rgb(059pt)=(0.359939,0.637985,0.797924);
rgb(060pt)=(0.372088,0.645675,0.802076);
rgb(061pt)=(0.384237,0.653364,0.806228);
rgb(062pt)=(0.396386,0.661053,0.810381);
rgb(063pt)=(0.408535,0.668743,0.814533);
rgb(064pt)=(0.420684,0.676432,0.818685);
rgb(065pt)=(0.432834,0.684122,0.822837);
rgb(066pt)=(0.444983,0.691811,0.826990);
rgb(067pt)=(0.457132,0.699500,0.831142);
rgb(068pt)=(0.469281,0.707190,0.835294);
rgb(069pt)=(0.481430,0.714879,0.839446);
rgb(070pt)=(0.493579,0.722568,0.843599);
rgb(071pt)=(0.505729,0.730258,0.847751);
rgb(072pt)=(0.517878,0.737947,0.851903);
rgb(073pt)=(0.530027,0.745636,0.856055);
rgb(074pt)=(0.542176,0.753326,0.860208);
rgb(075pt)=(0.554325,0.761015,0.864360);
rgb(076pt)=(0.566474,0.768704,0.868512);
rgb(077pt)=(0.577393,0.775010,0.871972);
rgb(078pt)=(0.587082,0.779931,0.874740);
rgb(079pt)=(0.596770,0.784852,0.877509);
rgb(080pt)=(0.606459,0.789773,0.880277);
rgb(081pt)=(0.616148,0.794694,0.883045);
rgb(082pt)=(0.625836,0.799616,0.885813);
rgb(083pt)=(0.635525,0.804537,0.888581);
rgb(084pt)=(0.645213,0.809458,0.891349);
rgb(085pt)=(0.654902,0.814379,0.894118);
rgb(086pt)=(0.664591,0.819300,0.896886);
rgb(087pt)=(0.674279,0.824221,0.899654);
rgb(088pt)=(0.683968,0.829143,0.902422);
rgb(089pt)=(0.693656,0.834064,0.905190);
rgb(090pt)=(0.703345,0.838985,0.907958);
rgb(091pt)=(0.713033,0.843906,0.910727);
rgb(092pt)=(0.722722,0.848827,0.913495);
rgb(093pt)=(0.732411,0.853749,0.916263);
rgb(094pt)=(0.742099,0.858670,0.919031);
rgb(095pt)=(0.751788,0.863591,0.921799);
rgb(096pt)=(0.761476,0.868512,0.924567);
rgb(097pt)=(0.771165,0.873433,0.927336);
rgb(098pt)=(0.780854,0.878354,0.930104);
rgb(099pt)=(0.790542,0.883276,0.932872);
rgb(100pt)=(0.800231,0.888197,0.935640);
rgb(101pt)=(0.809919,0.893118,0.938408);
rgb(102pt)=(0.819608,0.898039,0.941176);
rgb(103pt)=(0.825452,0.900807,0.942253);
rgb(104pt)=(0.831296,0.903576,0.943329);
rgb(105pt)=(0.837140,0.906344,0.944406);
rgb(106pt)=(0.842983,0.909112,0.945483);
rgb(107pt)=(0.848827,0.911880,0.946559);
rgb(108pt)=(0.854671,0.914648,0.947636);
rgb(109pt)=(0.860515,0.917416,0.948712);
rgb(110pt)=(0.866359,0.920185,0.949789);
rgb(111pt)=(0.872203,0.922953,0.950865);
rgb(112pt)=(0.878047,0.925721,0.951942);
rgb(113pt)=(0.883891,0.928489,0.953018);
rgb(114pt)=(0.889735,0.931257,0.954095);
rgb(115pt)=(0.895579,0.934025,0.955171);
rgb(116pt)=(0.901423,0.936794,0.956248);
rgb(117pt)=(0.907266,0.939562,0.957324);
rgb(118pt)=(0.913110,0.942330,0.958401);
rgb(119pt)=(0.918954,0.945098,0.959477);
rgb(120pt)=(0.924798,0.947866,0.960554);
rgb(121pt)=(0.930642,0.950634,0.961630);
rgb(122pt)=(0.936486,0.953403,0.962707);
rgb(123pt)=(0.942330,0.956171,0.963783);
rgb(124pt)=(0.948174,0.958939,0.964860);
rgb(125pt)=(0.954018,0.961707,0.965936);
rgb(126pt)=(0.959862,0.964475,0.967013);
rgb(127pt)=(0.965705,0.967243,0.968089);
rgb(128pt)=(0.969089,0.966474,0.964937);
rgb(129pt)=(0.970012,0.962168,0.957555);
rgb(130pt)=(0.970934,0.957862,0.950173);
rgb(131pt)=(0.971857,0.953556,0.942791);
rgb(132pt)=(0.972780,0.949250,0.935409);
rgb(133pt)=(0.973702,0.944944,0.928028);
rgb(134pt)=(0.974625,0.940638,0.920646);
rgb(135pt)=(0.975548,0.936332,0.913264);
rgb(136pt)=(0.976471,0.932026,0.905882);
rgb(137pt)=(0.977393,0.927720,0.898501);
rgb(138pt)=(0.978316,0.923414,0.891119);
rgb(139pt)=(0.979239,0.919108,0.883737);
rgb(140pt)=(0.980161,0.914802,0.876355);
rgb(141pt)=(0.981084,0.910496,0.868973);
rgb(142pt)=(0.982007,0.906190,0.861592);
rgb(143pt)=(0.982930,0.901884,0.854210);
rgb(144pt)=(0.983852,0.897578,0.846828);
rgb(145pt)=(0.984775,0.893272,0.839446);
rgb(146pt)=(0.985698,0.888966,0.832065);
rgb(147pt)=(0.986621,0.884660,0.824683);
rgb(148pt)=(0.987543,0.880354,0.817301);
rgb(149pt)=(0.988466,0.876048,0.809919);
rgb(150pt)=(0.989389,0.871742,0.802537);
rgb(151pt)=(0.990311,0.867436,0.795156);
rgb(152pt)=(0.991234,0.863130,0.787774);
rgb(153pt)=(0.992157,0.858824,0.780392);
rgb(154pt)=(0.990773,0.850519,0.769781);
rgb(155pt)=(0.989389,0.842215,0.759170);
rgb(156pt)=(0.988005,0.833910,0.748558);
rgb(157pt)=(0.986621,0.825606,0.737947);
rgb(158pt)=(0.985236,0.817301,0.727336);
rgb(159pt)=(0.983852,0.808997,0.716724);
rgb(160pt)=(0.982468,0.800692,0.706113);
rgb(161pt)=(0.981084,0.792388,0.695502);
rgb(162pt)=(0.979700,0.784083,0.684890);
rgb(163pt)=(0.978316,0.775779,0.674279);
rgb(164pt)=(0.976932,0.767474,0.663668);
rgb(165pt)=(0.975548,0.759170,0.653057);
rgb(166pt)=(0.974164,0.750865,0.642445);
rgb(167pt)=(0.972780,0.742561,0.631834);
rgb(168pt)=(0.971396,0.734256,0.621223);
rgb(169pt)=(0.970012,0.725952,0.610611);
rgb(170pt)=(0.968627,0.717647,0.600000);
rgb(171pt)=(0.967243,0.709343,0.589389);
rgb(172pt)=(0.965859,0.701038,0.578777);
rgb(173pt)=(0.964475,0.692734,0.568166);
rgb(174pt)=(0.963091,0.684429,0.557555);
rgb(175pt)=(0.961707,0.676125,0.546944);
rgb(176pt)=(0.960323,0.667820,0.536332);
rgb(177pt)=(0.958939,0.659516,0.525721);
rgb(178pt)=(0.957555,0.651211,0.515110);
rgb(179pt)=(0.954556,0.641753,0.505729);
rgb(180pt)=(0.949942,0.631142,0.497578);
rgb(181pt)=(0.945329,0.620531,0.489427);
rgb(182pt)=(0.940715,0.609919,0.481276);
rgb(183pt)=(0.936102,0.599308,0.473126);
rgb(184pt)=(0.931488,0.588697,0.464975);
rgb(185pt)=(0.926874,0.578085,0.456824);
rgb(186pt)=(0.922261,0.567474,0.448674);
rgb(187pt)=(0.917647,0.556863,0.440523);
rgb(188pt)=(0.913033,0.546251,0.432372);
rgb(189pt)=(0.908420,0.535640,0.424221);
rgb(190pt)=(0.903806,0.525029,0.416071);
rgb(191pt)=(0.899193,0.514418,0.407920);
rgb(192pt)=(0.894579,0.503806,0.399769);
rgb(193pt)=(0.889965,0.493195,0.391619);
rgb(194pt)=(0.885352,0.482584,0.383468);
rgb(195pt)=(0.880738,0.471972,0.375317);
rgb(196pt)=(0.876125,0.461361,0.367166);
rgb(197pt)=(0.871511,0.450750,0.359016);
rgb(198pt)=(0.866897,0.440138,0.350865);
rgb(199pt)=(0.862284,0.429527,0.342714);
rgb(200pt)=(0.857670,0.418916,0.334564);
rgb(201pt)=(0.853057,0.408305,0.326413);
rgb(202pt)=(0.848443,0.397693,0.318262);
rgb(203pt)=(0.843829,0.387082,0.310112);
rgb(204pt)=(0.839216,0.376471,0.301961);
rgb(205pt)=(0.833679,0.365398,0.296732);
rgb(206pt)=(0.828143,0.354325,0.291503);
rgb(207pt)=(0.822607,0.343253,0.286275);
rgb(208pt)=(0.817070,0.332180,0.281046);
rgb(209pt)=(0.811534,0.321107,0.275817);
rgb(210pt)=(0.805998,0.310035,0.270588);
rgb(211pt)=(0.800461,0.298962,0.265359);
rgb(212pt)=(0.794925,0.287889,0.260131);
rgb(213pt)=(0.789389,0.276817,0.254902);
rgb(214pt)=(0.783852,0.265744,0.249673);
rgb(215pt)=(0.778316,0.254671,0.244444);
rgb(216pt)=(0.772780,0.243599,0.239216);
rgb(217pt)=(0.767243,0.232526,0.233987);
rgb(218pt)=(0.761707,0.221453,0.228758);
rgb(219pt)=(0.756171,0.210381,0.223529);
rgb(220pt)=(0.750634,0.199308,0.218301);
rgb(221pt)=(0.745098,0.188235,0.213072);
rgb(222pt)=(0.739562,0.177163,0.207843);
rgb(223pt)=(0.734025,0.166090,0.202614);
rgb(224pt)=(0.728489,0.155017,0.197386);
rgb(225pt)=(0.722953,0.143945,0.192157);
rgb(226pt)=(0.717416,0.132872,0.186928);
rgb(227pt)=(0.711880,0.121799,0.181699);
rgb(228pt)=(0.706344,0.110727,0.176471);
rgb(229pt)=(0.700807,0.099654,0.171242);
rgb(230pt)=(0.692272,0.092272,0.167705);
rgb(231pt)=(0.680738,0.088581,0.165859);
rgb(232pt)=(0.669204,0.084890,0.164014);
rgb(233pt)=(0.657670,0.081200,0.162168);
rgb(234pt)=(0.646136,0.077509,0.160323);
rgb(235pt)=(0.634602,0.073818,0.158478);
rgb(236pt)=(0.623068,0.070127,0.156632);
rgb(237pt)=(0.611534,0.066436,0.154787);
rgb(238pt)=(0.600000,0.062745,0.152941);
rgb(239pt)=(0.588466,0.059054,0.151096);
rgb(240pt)=(0.576932,0.055363,0.149250);
rgb(241pt)=(0.565398,0.051672,0.147405);
rgb(242pt)=(0.553864,0.047982,0.145559);
rgb(243pt)=(0.542330,0.044291,0.143714);
rgb(244pt)=(0.530796,0.040600,0.141869);
rgb(245pt)=(0.519262,0.036909,0.140023);
rgb(246pt)=(0.507728,0.033218,0.138178);
rgb(247pt)=(0.496194,0.029527,0.136332);
rgb(248pt)=(0.484660,0.025836,0.134487);
rgb(249pt)=(0.473126,0.022145,0.132641);
rgb(250pt)=(0.461592,0.018454,0.130796);
rgb(251pt)=(0.450058,0.014764,0.128950);
rgb(252pt)=(0.438524,0.011073,0.127105);
rgb(253pt)=(0.426990,0.007382,0.125260);
rgb(254pt)=(0.415456,0.003691,0.123414);
rgb(255pt)=(0.403922,0.000000,0.121569);
}
}

\pgfplotsset{
colormap={cmRdBu_r_plus}{
rgb(000pt)=(0.969089,0.966474,0.964937);
rgb(001pt)=(0.969089,0.966474,0.964937);
rgb(002pt)=(0.970012,0.962168,0.957555);
rgb(003pt)=(0.970012,0.962168,0.957555);
rgb(004pt)=(0.970934,0.957862,0.950173);
rgb(005pt)=(0.970934,0.957862,0.950173);
rgb(006pt)=(0.971857,0.953556,0.942791);
rgb(007pt)=(0.971857,0.953556,0.942791);
rgb(008pt)=(0.972780,0.949250,0.935409);
rgb(009pt)=(0.972780,0.949250,0.935409);
rgb(010pt)=(0.973702,0.944944,0.928028);
rgb(011pt)=(0.973702,0.944944,0.928028);
rgb(012pt)=(0.974625,0.940638,0.920646);
rgb(013pt)=(0.974625,0.940638,0.920646);
rgb(014pt)=(0.975548,0.936332,0.913264);
rgb(015pt)=(0.975548,0.936332,0.913264);
rgb(016pt)=(0.976471,0.932026,0.905882);
rgb(017pt)=(0.976471,0.932026,0.905882);
rgb(018pt)=(0.977393,0.927720,0.898501);
rgb(019pt)=(0.977393,0.927720,0.898501);
rgb(020pt)=(0.978316,0.923414,0.891119);
rgb(021pt)=(0.978316,0.923414,0.891119);
rgb(022pt)=(0.979239,0.919108,0.883737);
rgb(023pt)=(0.979239,0.919108,0.883737);
rgb(024pt)=(0.980161,0.914802,0.876355);
rgb(025pt)=(0.980161,0.914802,0.876355);
rgb(026pt)=(0.981084,0.910496,0.868973);
rgb(027pt)=(0.981084,0.910496,0.868973);
rgb(028pt)=(0.982007,0.906190,0.861592);
rgb(029pt)=(0.982007,0.906190,0.861592);
rgb(030pt)=(0.982930,0.901884,0.854210);
rgb(031pt)=(0.982930,0.901884,0.854210);
rgb(032pt)=(0.983852,0.897578,0.846828);
rgb(033pt)=(0.983852,0.897578,0.846828);
rgb(034pt)=(0.984775,0.893272,0.839446);
rgb(035pt)=(0.984775,0.893272,0.839446);
rgb(036pt)=(0.985698,0.888966,0.832065);
rgb(037pt)=(0.985698,0.888966,0.832065);
rgb(038pt)=(0.986621,0.884660,0.824683);
rgb(039pt)=(0.986621,0.884660,0.824683);
rgb(040pt)=(0.987543,0.880354,0.817301);
rgb(041pt)=(0.987543,0.880354,0.817301);
rgb(042pt)=(0.988466,0.876048,0.809919);
rgb(043pt)=(0.988466,0.876048,0.809919);
rgb(044pt)=(0.989389,0.871742,0.802537);
rgb(045pt)=(0.989389,0.871742,0.802537);
rgb(046pt)=(0.990311,0.867436,0.795156);
rgb(047pt)=(0.990311,0.867436,0.795156);
rgb(048pt)=(0.991234,0.863130,0.787774);
rgb(049pt)=(0.991234,0.863130,0.787774);
rgb(050pt)=(0.992157,0.858824,0.780392);
rgb(051pt)=(0.992157,0.858824,0.780392);
rgb(052pt)=(0.990773,0.850519,0.769781);
rgb(053pt)=(0.990773,0.850519,0.769781);
rgb(054pt)=(0.989389,0.842215,0.759170);
rgb(055pt)=(0.989389,0.842215,0.759170);
rgb(056pt)=(0.988005,0.833910,0.748558);
rgb(057pt)=(0.988005,0.833910,0.748558);
rgb(058pt)=(0.986621,0.825606,0.737947);
rgb(059pt)=(0.986621,0.825606,0.737947);
rgb(060pt)=(0.985236,0.817301,0.727336);
rgb(061pt)=(0.985236,0.817301,0.727336);
rgb(062pt)=(0.983852,0.808997,0.716724);
rgb(063pt)=(0.983852,0.808997,0.716724);
rgb(064pt)=(0.982468,0.800692,0.706113);
rgb(065pt)=(0.982468,0.800692,0.706113);
rgb(066pt)=(0.981084,0.792388,0.695502);
rgb(067pt)=(0.981084,0.792388,0.695502);
rgb(068pt)=(0.979700,0.784083,0.684890);
rgb(069pt)=(0.979700,0.784083,0.684890);
rgb(070pt)=(0.978316,0.775779,0.674279);
rgb(071pt)=(0.978316,0.775779,0.674279);
rgb(072pt)=(0.976932,0.767474,0.663668);
rgb(073pt)=(0.976932,0.767474,0.663668);
rgb(074pt)=(0.975548,0.759170,0.653057);
rgb(075pt)=(0.975548,0.759170,0.653057);
rgb(076pt)=(0.974164,0.750865,0.642445);
rgb(077pt)=(0.974164,0.750865,0.642445);
rgb(078pt)=(0.972780,0.742561,0.631834);
rgb(079pt)=(0.972780,0.742561,0.631834);
rgb(080pt)=(0.971396,0.734256,0.621223);
rgb(081pt)=(0.971396,0.734256,0.621223);
rgb(082pt)=(0.970012,0.725952,0.610611);
rgb(083pt)=(0.970012,0.725952,0.610611);
rgb(084pt)=(0.968627,0.717647,0.600000);
rgb(085pt)=(0.968627,0.717647,0.600000);
rgb(086pt)=(0.967243,0.709343,0.589389);
rgb(087pt)=(0.967243,0.709343,0.589389);
rgb(088pt)=(0.965859,0.701038,0.578777);
rgb(089pt)=(0.965859,0.701038,0.578777);
rgb(090pt)=(0.964475,0.692734,0.568166);
rgb(091pt)=(0.964475,0.692734,0.568166);
rgb(092pt)=(0.963091,0.684429,0.557555);
rgb(093pt)=(0.963091,0.684429,0.557555);
rgb(094pt)=(0.961707,0.676125,0.546944);
rgb(095pt)=(0.961707,0.676125,0.546944);
rgb(096pt)=(0.960323,0.667820,0.536332);
rgb(097pt)=(0.960323,0.667820,0.536332);
rgb(098pt)=(0.958939,0.659516,0.525721);
rgb(099pt)=(0.958939,0.659516,0.525721);
rgb(100pt)=(0.957555,0.651211,0.515110);
rgb(101pt)=(0.957555,0.651211,0.515110);
rgb(102pt)=(0.954556,0.641753,0.505729);
rgb(103pt)=(0.954556,0.641753,0.505729);
rgb(104pt)=(0.949942,0.631142,0.497578);
rgb(105pt)=(0.949942,0.631142,0.497578);
rgb(106pt)=(0.945329,0.620531,0.489427);
rgb(107pt)=(0.945329,0.620531,0.489427);
rgb(108pt)=(0.940715,0.609919,0.481276);
rgb(109pt)=(0.940715,0.609919,0.481276);
rgb(110pt)=(0.936102,0.599308,0.473126);
rgb(111pt)=(0.936102,0.599308,0.473126);
rgb(112pt)=(0.931488,0.588697,0.464975);
rgb(113pt)=(0.931488,0.588697,0.464975);
rgb(114pt)=(0.926874,0.578085,0.456824);
rgb(115pt)=(0.926874,0.578085,0.456824);
rgb(116pt)=(0.922261,0.567474,0.448674);
rgb(117pt)=(0.922261,0.567474,0.448674);
rgb(118pt)=(0.917647,0.556863,0.440523);
rgb(119pt)=(0.917647,0.556863,0.440523);
rgb(120pt)=(0.913033,0.546251,0.432372);
rgb(121pt)=(0.913033,0.546251,0.432372);
rgb(122pt)=(0.908420,0.535640,0.424221);
rgb(123pt)=(0.908420,0.535640,0.424221);
rgb(124pt)=(0.903806,0.525029,0.416071);
rgb(125pt)=(0.903806,0.525029,0.416071);
rgb(126pt)=(0.899193,0.514418,0.407920);
rgb(127pt)=(0.899193,0.514418,0.407920);
rgb(128pt)=(0.894579,0.503806,0.399769);
rgb(129pt)=(0.894579,0.503806,0.399769);
rgb(130pt)=(0.889965,0.493195,0.391619);
rgb(131pt)=(0.889965,0.493195,0.391619);
rgb(132pt)=(0.885352,0.482584,0.383468);
rgb(133pt)=(0.885352,0.482584,0.383468);
rgb(134pt)=(0.880738,0.471972,0.375317);
rgb(135pt)=(0.880738,0.471972,0.375317);
rgb(136pt)=(0.876125,0.461361,0.367166);
rgb(137pt)=(0.876125,0.461361,0.367166);
rgb(138pt)=(0.871511,0.450750,0.359016);
rgb(139pt)=(0.871511,0.450750,0.359016);
rgb(140pt)=(0.866897,0.440138,0.350865);
rgb(141pt)=(0.866897,0.440138,0.350865);
rgb(142pt)=(0.862284,0.429527,0.342714);
rgb(143pt)=(0.862284,0.429527,0.342714);
rgb(144pt)=(0.857670,0.418916,0.334564);
rgb(145pt)=(0.857670,0.418916,0.334564);
rgb(146pt)=(0.853057,0.408305,0.326413);
rgb(147pt)=(0.853057,0.408305,0.326413);
rgb(148pt)=(0.848443,0.397693,0.318262);
rgb(149pt)=(0.848443,0.397693,0.318262);
rgb(150pt)=(0.843829,0.387082,0.310112);
rgb(151pt)=(0.843829,0.387082,0.310112);
rgb(152pt)=(0.839216,0.376471,0.301961);
rgb(153pt)=(0.839216,0.376471,0.301961);
rgb(154pt)=(0.833679,0.365398,0.296732);
rgb(155pt)=(0.833679,0.365398,0.296732);
rgb(156pt)=(0.828143,0.354325,0.291503);
rgb(157pt)=(0.828143,0.354325,0.291503);
rgb(158pt)=(0.822607,0.343253,0.286275);
rgb(159pt)=(0.822607,0.343253,0.286275);
rgb(160pt)=(0.817070,0.332180,0.281046);
rgb(161pt)=(0.817070,0.332180,0.281046);
rgb(162pt)=(0.811534,0.321107,0.275817);
rgb(163pt)=(0.811534,0.321107,0.275817);
rgb(164pt)=(0.805998,0.310035,0.270588);
rgb(165pt)=(0.805998,0.310035,0.270588);
rgb(166pt)=(0.800461,0.298962,0.265359);
rgb(167pt)=(0.800461,0.298962,0.265359);
rgb(168pt)=(0.794925,0.287889,0.260131);
rgb(169pt)=(0.794925,0.287889,0.260131);
rgb(170pt)=(0.789389,0.276817,0.254902);
rgb(171pt)=(0.789389,0.276817,0.254902);
rgb(172pt)=(0.783852,0.265744,0.249673);
rgb(173pt)=(0.783852,0.265744,0.249673);
rgb(174pt)=(0.778316,0.254671,0.244444);
rgb(175pt)=(0.778316,0.254671,0.244444);
rgb(176pt)=(0.772780,0.243599,0.239216);
rgb(177pt)=(0.772780,0.243599,0.239216);
rgb(178pt)=(0.767243,0.232526,0.233987);
rgb(179pt)=(0.767243,0.232526,0.233987);
rgb(180pt)=(0.761707,0.221453,0.228758);
rgb(181pt)=(0.761707,0.221453,0.228758);
rgb(182pt)=(0.756171,0.210381,0.223529);
rgb(183pt)=(0.756171,0.210381,0.223529);
rgb(184pt)=(0.750634,0.199308,0.218301);
rgb(185pt)=(0.750634,0.199308,0.218301);
rgb(186pt)=(0.745098,0.188235,0.213072);
rgb(187pt)=(0.745098,0.188235,0.213072);
rgb(188pt)=(0.739562,0.177163,0.207843);
rgb(189pt)=(0.739562,0.177163,0.207843);
rgb(190pt)=(0.734025,0.166090,0.202614);
rgb(191pt)=(0.734025,0.166090,0.202614);
rgb(192pt)=(0.728489,0.155017,0.197386);
rgb(193pt)=(0.728489,0.155017,0.197386);
rgb(194pt)=(0.722953,0.143945,0.192157);
rgb(195pt)=(0.722953,0.143945,0.192157);
rgb(196pt)=(0.717416,0.132872,0.186928);
rgb(197pt)=(0.717416,0.132872,0.186928);
rgb(198pt)=(0.711880,0.121799,0.181699);
rgb(199pt)=(0.711880,0.121799,0.181699);
rgb(200pt)=(0.706344,0.110727,0.176471);
rgb(201pt)=(0.706344,0.110727,0.176471);
rgb(202pt)=(0.700807,0.099654,0.171242);
rgb(203pt)=(0.700807,0.099654,0.171242);
rgb(204pt)=(0.692272,0.092272,0.167705);
rgb(205pt)=(0.692272,0.092272,0.167705);
rgb(206pt)=(0.680738,0.088581,0.165859);
rgb(207pt)=(0.680738,0.088581,0.165859);
rgb(208pt)=(0.669204,0.084890,0.164014);
rgb(209pt)=(0.669204,0.084890,0.164014);
rgb(210pt)=(0.657670,0.081200,0.162168);
rgb(211pt)=(0.657670,0.081200,0.162168);
rgb(212pt)=(0.646136,0.077509,0.160323);
rgb(213pt)=(0.646136,0.077509,0.160323);
rgb(214pt)=(0.634602,0.073818,0.158478);
rgb(215pt)=(0.634602,0.073818,0.158478);
rgb(216pt)=(0.623068,0.070127,0.156632);
rgb(217pt)=(0.623068,0.070127,0.156632);
rgb(218pt)=(0.611534,0.066436,0.154787);
rgb(219pt)=(0.611534,0.066436,0.154787);
rgb(220pt)=(0.600000,0.062745,0.152941);
rgb(221pt)=(0.600000,0.062745,0.152941);
rgb(222pt)=(0.588466,0.059054,0.151096);
rgb(223pt)=(0.588466,0.059054,0.151096);
rgb(224pt)=(0.576932,0.055363,0.149250);
rgb(225pt)=(0.576932,0.055363,0.149250);
rgb(226pt)=(0.565398,0.051672,0.147405);
rgb(227pt)=(0.565398,0.051672,0.147405);
rgb(228pt)=(0.553864,0.047982,0.145559);
rgb(229pt)=(0.553864,0.047982,0.145559);
rgb(230pt)=(0.542330,0.044291,0.143714);
rgb(231pt)=(0.542330,0.044291,0.143714);
rgb(232pt)=(0.530796,0.040600,0.141869);
rgb(233pt)=(0.530796,0.040600,0.141869);
rgb(234pt)=(0.519262,0.036909,0.140023);
rgb(235pt)=(0.519262,0.036909,0.140023);
rgb(236pt)=(0.507728,0.033218,0.138178);
rgb(237pt)=(0.507728,0.033218,0.138178);
rgb(238pt)=(0.496194,0.029527,0.136332);
rgb(239pt)=(0.496194,0.029527,0.136332);
rgb(240pt)=(0.484660,0.025836,0.134487);
rgb(241pt)=(0.484660,0.025836,0.134487);
rgb(242pt)=(0.473126,0.022145,0.132641);
rgb(243pt)=(0.473126,0.022145,0.132641);
rgb(244pt)=(0.461592,0.018454,0.130796);
rgb(245pt)=(0.461592,0.018454,0.130796);
rgb(246pt)=(0.450058,0.014764,0.128950);
rgb(247pt)=(0.450058,0.014764,0.128950);
rgb(248pt)=(0.438524,0.011073,0.127105);
rgb(249pt)=(0.438524,0.011073,0.127105);
rgb(250pt)=(0.426990,0.007382,0.125260);
rgb(251pt)=(0.426990,0.007382,0.125260);
rgb(252pt)=(0.415456,0.003691,0.123414);
rgb(253pt)=(0.415456,0.003691,0.123414);
rgb(254pt)=(0.403922,0.000000,0.121569);
rgb(255pt)=(0.403922,0.000000,0.121569);
}
}

\pgfplotsset{
colormap={cmRdBu_r_minus}{
rgb(000pt)=(0.019608,0.188235,0.380392);
rgb(001pt)=(0.019608,0.188235,0.380392);
rgb(002pt)=(0.023914,0.196540,0.391926);
rgb(003pt)=(0.023914,0.196540,0.391926);
rgb(004pt)=(0.028220,0.204844,0.403460);
rgb(005pt)=(0.028220,0.204844,0.403460);
rgb(006pt)=(0.032526,0.213149,0.414994);
rgb(007pt)=(0.032526,0.213149,0.414994);
rgb(008pt)=(0.036832,0.221453,0.426528);
rgb(009pt)=(0.036832,0.221453,0.426528);
rgb(010pt)=(0.041138,0.229758,0.438062);
rgb(011pt)=(0.041138,0.229758,0.438062);
rgb(012pt)=(0.045444,0.238062,0.449596);
rgb(013pt)=(0.045444,0.238062,0.449596);
rgb(014pt)=(0.049750,0.246367,0.461130);
rgb(015pt)=(0.049750,0.246367,0.461130);
rgb(016pt)=(0.054056,0.254671,0.472664);
rgb(017pt)=(0.054056,0.254671,0.472664);
rgb(018pt)=(0.058362,0.262976,0.484198);
rgb(019pt)=(0.058362,0.262976,0.484198);
rgb(020pt)=(0.062668,0.271280,0.495732);
rgb(021pt)=(0.062668,0.271280,0.495732);
rgb(022pt)=(0.066974,0.279585,0.507266);
rgb(023pt)=(0.066974,0.279585,0.507266);
rgb(024pt)=(0.071280,0.287889,0.518800);
rgb(025pt)=(0.071280,0.287889,0.518800);
rgb(026pt)=(0.075586,0.296194,0.530335);
rgb(027pt)=(0.075586,0.296194,0.530335);
rgb(028pt)=(0.079892,0.304498,0.541869);
rgb(029pt)=(0.079892,0.304498,0.541869);
rgb(030pt)=(0.084198,0.312803,0.553403);
rgb(031pt)=(0.084198,0.312803,0.553403);
rgb(032pt)=(0.088504,0.321107,0.564937);
rgb(033pt)=(0.088504,0.321107,0.564937);
rgb(034pt)=(0.092810,0.329412,0.576471);
rgb(035pt)=(0.092810,0.329412,0.576471);
rgb(036pt)=(0.097116,0.337716,0.588005);
rgb(037pt)=(0.097116,0.337716,0.588005);
rgb(038pt)=(0.101423,0.346021,0.599539);
rgb(039pt)=(0.101423,0.346021,0.599539);
rgb(040pt)=(0.105729,0.354325,0.611073);
rgb(041pt)=(0.105729,0.354325,0.611073);
rgb(042pt)=(0.110035,0.362630,0.622607);
rgb(043pt)=(0.110035,0.362630,0.622607);
rgb(044pt)=(0.114341,0.370934,0.634141);
rgb(045pt)=(0.114341,0.370934,0.634141);
rgb(046pt)=(0.118647,0.379239,0.645675);
rgb(047pt)=(0.118647,0.379239,0.645675);
rgb(048pt)=(0.122953,0.387543,0.657209);
rgb(049pt)=(0.122953,0.387543,0.657209);
rgb(050pt)=(0.127259,0.395848,0.668743);
rgb(051pt)=(0.127259,0.395848,0.668743);
rgb(052pt)=(0.132026,0.403460,0.676278);
rgb(053pt)=(0.132026,0.403460,0.676278);
rgb(054pt)=(0.137255,0.410381,0.679815);
rgb(055pt)=(0.137255,0.410381,0.679815);
rgb(056pt)=(0.142484,0.417301,0.683353);
rgb(057pt)=(0.142484,0.417301,0.683353);
rgb(058pt)=(0.147712,0.424221,0.686890);
rgb(059pt)=(0.147712,0.424221,0.686890);
rgb(060pt)=(0.152941,0.431142,0.690427);
rgb(061pt)=(0.152941,0.431142,0.690427);
rgb(062pt)=(0.158170,0.438062,0.693964);
rgb(063pt)=(0.158170,0.438062,0.693964);
rgb(064pt)=(0.163399,0.444983,0.697501);
rgb(065pt)=(0.163399,0.444983,0.697501);
rgb(066pt)=(0.168627,0.451903,0.701038);
rgb(067pt)=(0.168627,0.451903,0.701038);
rgb(068pt)=(0.173856,0.458824,0.704575);
rgb(069pt)=(0.173856,0.458824,0.704575);
rgb(070pt)=(0.179085,0.465744,0.708112);
rgb(071pt)=(0.179085,0.465744,0.708112);
rgb(072pt)=(0.184314,0.472664,0.711649);
rgb(073pt)=(0.184314,0.472664,0.711649);
rgb(074pt)=(0.189542,0.479585,0.715186);
rgb(075pt)=(0.189542,0.479585,0.715186);
rgb(076pt)=(0.194771,0.486505,0.718724);
rgb(077pt)=(0.194771,0.486505,0.718724);
rgb(078pt)=(0.200000,0.493426,0.722261);
rgb(079pt)=(0.200000,0.493426,0.722261);
rgb(080pt)=(0.205229,0.500346,0.725798);
rgb(081pt)=(0.205229,0.500346,0.725798);
rgb(082pt)=(0.210458,0.507266,0.729335);
rgb(083pt)=(0.210458,0.507266,0.729335);
rgb(084pt)=(0.215686,0.514187,0.732872);
rgb(085pt)=(0.215686,0.514187,0.732872);
rgb(086pt)=(0.220915,0.521107,0.736409);
rgb(087pt)=(0.220915,0.521107,0.736409);
rgb(088pt)=(0.226144,0.528028,0.739946);
rgb(089pt)=(0.226144,0.528028,0.739946);
rgb(090pt)=(0.231373,0.534948,0.743483);
rgb(091pt)=(0.231373,0.534948,0.743483);
rgb(092pt)=(0.236601,0.541869,0.747020);
rgb(093pt)=(0.236601,0.541869,0.747020);
rgb(094pt)=(0.241830,0.548789,0.750557);
rgb(095pt)=(0.241830,0.548789,0.750557);
rgb(096pt)=(0.247059,0.555709,0.754095);
rgb(097pt)=(0.247059,0.555709,0.754095);
rgb(098pt)=(0.252288,0.562630,0.757632);
rgb(099pt)=(0.252288,0.562630,0.757632);
rgb(100pt)=(0.257516,0.569550,0.761169);
rgb(101pt)=(0.257516,0.569550,0.761169);
rgb(102pt)=(0.262745,0.576471,0.764706);
rgb(103pt)=(0.262745,0.576471,0.764706);
rgb(104pt)=(0.274894,0.584160,0.768858);
rgb(105pt)=(0.274894,0.584160,0.768858);
rgb(106pt)=(0.287043,0.591849,0.773010);
rgb(107pt)=(0.287043,0.591849,0.773010);
rgb(108pt)=(0.299193,0.599539,0.777163);
rgb(109pt)=(0.299193,0.599539,0.777163);
rgb(110pt)=(0.311342,0.607228,0.781315);
rgb(111pt)=(0.311342,0.607228,0.781315);
rgb(112pt)=(0.323491,0.614917,0.785467);
rgb(113pt)=(0.323491,0.614917,0.785467);
rgb(114pt)=(0.335640,0.622607,0.789619);
rgb(115pt)=(0.335640,0.622607,0.789619);
rgb(116pt)=(0.347789,0.630296,0.793772);
rgb(117pt)=(0.347789,0.630296,0.793772);
rgb(118pt)=(0.359939,0.637985,0.797924);
rgb(119pt)=(0.359939,0.637985,0.797924);
rgb(120pt)=(0.372088,0.645675,0.802076);
rgb(121pt)=(0.372088,0.645675,0.802076);
rgb(122pt)=(0.384237,0.653364,0.806228);
rgb(123pt)=(0.384237,0.653364,0.806228);
rgb(124pt)=(0.396386,0.661053,0.810381);
rgb(125pt)=(0.396386,0.661053,0.810381);
rgb(126pt)=(0.408535,0.668743,0.814533);
rgb(127pt)=(0.408535,0.668743,0.814533);
rgb(128pt)=(0.420684,0.676432,0.818685);
rgb(129pt)=(0.420684,0.676432,0.818685);
rgb(130pt)=(0.432834,0.684122,0.822837);
rgb(131pt)=(0.432834,0.684122,0.822837);
rgb(132pt)=(0.444983,0.691811,0.826990);
rgb(133pt)=(0.444983,0.691811,0.826990);
rgb(134pt)=(0.457132,0.699500,0.831142);
rgb(135pt)=(0.457132,0.699500,0.831142);
rgb(136pt)=(0.469281,0.707190,0.835294);
rgb(137pt)=(0.469281,0.707190,0.835294);
rgb(138pt)=(0.481430,0.714879,0.839446);
rgb(139pt)=(0.481430,0.714879,0.839446);
rgb(140pt)=(0.493579,0.722568,0.843599);
rgb(141pt)=(0.493579,0.722568,0.843599);
rgb(142pt)=(0.505729,0.730258,0.847751);
rgb(143pt)=(0.505729,0.730258,0.847751);
rgb(144pt)=(0.517878,0.737947,0.851903);
rgb(145pt)=(0.517878,0.737947,0.851903);
rgb(146pt)=(0.530027,0.745636,0.856055);
rgb(147pt)=(0.530027,0.745636,0.856055);
rgb(148pt)=(0.542176,0.753326,0.860208);
rgb(149pt)=(0.542176,0.753326,0.860208);
rgb(150pt)=(0.554325,0.761015,0.864360);
rgb(151pt)=(0.554325,0.761015,0.864360);
rgb(152pt)=(0.566474,0.768704,0.868512);
rgb(153pt)=(0.566474,0.768704,0.868512);
rgb(154pt)=(0.577393,0.775010,0.871972);
rgb(155pt)=(0.577393,0.775010,0.871972);
rgb(156pt)=(0.587082,0.779931,0.874740);
rgb(157pt)=(0.587082,0.779931,0.874740);
rgb(158pt)=(0.596770,0.784852,0.877509);
rgb(159pt)=(0.596770,0.784852,0.877509);
rgb(160pt)=(0.606459,0.789773,0.880277);
rgb(161pt)=(0.606459,0.789773,0.880277);
rgb(162pt)=(0.616148,0.794694,0.883045);
rgb(163pt)=(0.616148,0.794694,0.883045);
rgb(164pt)=(0.625836,0.799616,0.885813);
rgb(165pt)=(0.625836,0.799616,0.885813);
rgb(166pt)=(0.635525,0.804537,0.888581);
rgb(167pt)=(0.635525,0.804537,0.888581);
rgb(168pt)=(0.645213,0.809458,0.891349);
rgb(169pt)=(0.645213,0.809458,0.891349);
rgb(170pt)=(0.654902,0.814379,0.894118);
rgb(171pt)=(0.654902,0.814379,0.894118);
rgb(172pt)=(0.664591,0.819300,0.896886);
rgb(173pt)=(0.664591,0.819300,0.896886);
rgb(174pt)=(0.674279,0.824221,0.899654);
rgb(175pt)=(0.674279,0.824221,0.899654);
rgb(176pt)=(0.683968,0.829143,0.902422);
rgb(177pt)=(0.683968,0.829143,0.902422);
rgb(178pt)=(0.693656,0.834064,0.905190);
rgb(179pt)=(0.693656,0.834064,0.905190);
rgb(180pt)=(0.703345,0.838985,0.907958);
rgb(181pt)=(0.703345,0.838985,0.907958);
rgb(182pt)=(0.713033,0.843906,0.910727);
rgb(183pt)=(0.713033,0.843906,0.910727);
rgb(184pt)=(0.722722,0.848827,0.913495);
rgb(185pt)=(0.722722,0.848827,0.913495);
rgb(186pt)=(0.732411,0.853749,0.916263);
rgb(187pt)=(0.732411,0.853749,0.916263);
rgb(188pt)=(0.742099,0.858670,0.919031);
rgb(189pt)=(0.742099,0.858670,0.919031);
rgb(190pt)=(0.751788,0.863591,0.921799);
rgb(191pt)=(0.751788,0.863591,0.921799);
rgb(192pt)=(0.761476,0.868512,0.924567);
rgb(193pt)=(0.761476,0.868512,0.924567);
rgb(194pt)=(0.771165,0.873433,0.927336);
rgb(195pt)=(0.771165,0.873433,0.927336);
rgb(196pt)=(0.780854,0.878354,0.930104);
rgb(197pt)=(0.780854,0.878354,0.930104);
rgb(198pt)=(0.790542,0.883276,0.932872);
rgb(199pt)=(0.790542,0.883276,0.932872);
rgb(200pt)=(0.800231,0.888197,0.935640);
rgb(201pt)=(0.800231,0.888197,0.935640);
rgb(202pt)=(0.809919,0.893118,0.938408);
rgb(203pt)=(0.809919,0.893118,0.938408);
rgb(204pt)=(0.819608,0.898039,0.941176);
rgb(205pt)=(0.819608,0.898039,0.941176);
rgb(206pt)=(0.825452,0.900807,0.942253);
rgb(207pt)=(0.825452,0.900807,0.942253);
rgb(208pt)=(0.831296,0.903576,0.943329);
rgb(209pt)=(0.831296,0.903576,0.943329);
rgb(210pt)=(0.837140,0.906344,0.944406);
rgb(211pt)=(0.837140,0.906344,0.944406);
rgb(212pt)=(0.842983,0.909112,0.945483);
rgb(213pt)=(0.842983,0.909112,0.945483);
rgb(214pt)=(0.848827,0.911880,0.946559);
rgb(215pt)=(0.848827,0.911880,0.946559);
rgb(216pt)=(0.854671,0.914648,0.947636);
rgb(217pt)=(0.854671,0.914648,0.947636);
rgb(218pt)=(0.860515,0.917416,0.948712);
rgb(219pt)=(0.860515,0.917416,0.948712);
rgb(220pt)=(0.866359,0.920185,0.949789);
rgb(221pt)=(0.866359,0.920185,0.949789);
rgb(222pt)=(0.872203,0.922953,0.950865);
rgb(223pt)=(0.872203,0.922953,0.950865);
rgb(224pt)=(0.878047,0.925721,0.951942);
rgb(225pt)=(0.878047,0.925721,0.951942);
rgb(226pt)=(0.883891,0.928489,0.953018);
rgb(227pt)=(0.883891,0.928489,0.953018);
rgb(228pt)=(0.889735,0.931257,0.954095);
rgb(229pt)=(0.889735,0.931257,0.954095);
rgb(230pt)=(0.895579,0.934025,0.955171);
rgb(231pt)=(0.895579,0.934025,0.955171);
rgb(232pt)=(0.901423,0.936794,0.956248);
rgb(233pt)=(0.901423,0.936794,0.956248);
rgb(234pt)=(0.907266,0.939562,0.957324);
rgb(235pt)=(0.907266,0.939562,0.957324);
rgb(236pt)=(0.913110,0.942330,0.958401);
rgb(237pt)=(0.913110,0.942330,0.958401);
rgb(238pt)=(0.918954,0.945098,0.959477);
rgb(239pt)=(0.918954,0.945098,0.959477);
rgb(240pt)=(0.924798,0.947866,0.960554);
rgb(241pt)=(0.924798,0.947866,0.960554);
rgb(242pt)=(0.930642,0.950634,0.961630);
rgb(243pt)=(0.930642,0.950634,0.961630);
rgb(244pt)=(0.936486,0.953403,0.962707);
rgb(245pt)=(0.936486,0.953403,0.962707);
rgb(246pt)=(0.942330,0.956171,0.963783);
rgb(247pt)=(0.942330,0.956171,0.963783);
rgb(248pt)=(0.948174,0.958939,0.964860);
rgb(249pt)=(0.948174,0.958939,0.964860);
rgb(250pt)=(0.954018,0.961707,0.965936);
rgb(251pt)=(0.954018,0.961707,0.965936);
rgb(252pt)=(0.959862,0.964475,0.967013);
rgb(253pt)=(0.959862,0.964475,0.967013);
rgb(254pt)=(0.965705,0.967243,0.968089);
rgb(255pt)=(0.969089,0.966474,0.964937);
}
}

\pgfplotscreateplotcyclelist{marklist}{% 
	mark=x\\
	mark=o\\
	mark=triangle\\
	mark=diamond\\
	mark=square\\
}
\pgfplotsset{
	cycle list/Set1-5,
}

\pgfplotsset{groupplotdefault/.style={
		scale only axis,
		enlargelimits=false,
		width = \figurewidth,
		height = \figureheight,
		title style = {yshift = 0.6cm, font=\figurelabelfont{\figurewidth}},
		xlabel style = {font=\figurelabelfont{\figurewidth}}, 
		ylabel style = {font=\figurelabelfont{\figurewidth}}, 
		xticklabel style={font=\figureticklabelfont{\figurewidth}},
		yticklabel style={font=\figureticklabelfont{\figurewidth}},
		axis on top,
		every x tick/.style={white!50!black},
		every y tick/.style={white!50!black},
}}

\def\lincolorplot#1#2#3#4#5#6{
	\nextgroupplot[
	xmin=-1.5,
	xmax=1.5,
	ymin=-1.5,
	ymax=1.5,
	xlabel={$x / cm$},
	ylabel={$y / cm$},
	title={#1},
	point meta min={#3},
	point meta max={#4},
	colorbar horizontal,
	colorbar style={,
		colormap name={cmviridis},  
		at={(0,1.02)},
		anchor=south west,
		height=0.02\textwidth,
		width=\figurewidth,
		xticklabel style={font=\figureticklabelfont{\figurewidth},anchor=south},
		ticklabel pos=right,
		max space between ticks=1000pt,
		try min ticks=3,
	},
	#5
	]
	\addplot graphics[xmin=-1.5,xmax=1.5,ymin=-1.5,ymax=1.5, #6] {\relincludepath #2};
}

\def\logerrplot#1#2#3#4#5#6{
	\nextgroupplot[
	xmin=-1.5,
	xmax=1.5,
	ymin=-1.5,
	ymax=1.5,
	xlabel={$x / cm$},
	ylabel={$y / cm$},
	title={\tikztitle{#1}},
	colorbar horizontal,
	colorbar style={
		colormap name ={cmRdBu_r_minus},
		point meta min ={-#4}, 
		point meta max ={0},
		at={(0,1.02)},
		anchor=south west,
		height=0.02\textwidth,
		width=0.49\figurewidth,
		ticklabel pos=right,
		xticklabel style={font=\figureticklabelfont{\figurewidth},anchor=south, /pgf/number format/.cd,	/tikz/.cd},
		xticklabel={
			\pgfmathsetmacro\upBound{0.7*\pgfkeysvalueof{/pgfplots/xmax}+0.3*\pgfkeysvalueof{/pgfplots/xmin}}
			\ifdim \tick pt < \upBound pt % suppress labels that are too close to the right
			$-10^{\pgfmathparse{-\tick+#3}\pgfmathprintnumber\pgfmathresult}$ % display -10^{-x+offset}
			\fi
		},
		try min ticks=2,
	},
	colorbar/draw/.append code={
		\begin{axis}[
			every colorbar, % make this axis a colorbar
			at={(1., 1.02)},
			anchor=south east,
			colormap name={cmRdBu_r_plus},
			colorbar horizontal,
			point meta min={0},
			point meta max={#4},
			colorbar=false,
			height=0.02\textwidth,
			width=0.49\figurewidth,
			ticklabel pos=right,
			xticklabel style={font=\figureticklabelfont{\figurewidth},anchor=south, /pgf/number format/.cd, /tikz/.cd},
			try min ticks=2, 
			xticklabel={
				\pgfmathsetmacro\loBound{0.3*\pgfkeysvalueof{/pgfplots/xmax}+0.7*\pgfkeysvalueof{/pgfplots/xmin}}
				\ifdim \tick pt > \loBound pt 
				$10^{\pgfmathparse{\tick+#3}\pgfmathprintnumber\pgfmathresult}$ % display 10^{x+offset}
				\fi
			},
			extra x ticks=0., % make the central tick
			extra x tick labels=$\pm10^{#3}$,
			] % \begin{axis}        
			\pgfkeysvalueof{/pgfplots/colorbar addplot}
		\end{axis}
	},		
	#5
	] %\nextgroupplot
	\addplot graphics[xmin=-1,xmax=1,ymin=-1,ymax=1, #6] {\relincludepath #2};	
}

\title{Modeling glioma invasion with anisotropy- and hypoxia-triggered motility enhancement: from subcellular dynamics
	to macroscopic PDEs with multiple taxis}
\author{G. Corbin, C. Engwer, A. Klar, J. Nieto, J. Soler, C. Surulescu, and M. Wenske}

\date{\today}

\begin{document}

\newcommand{\cmg}[1]{\textcolor{magenta}{#1}} %surulescu@mathematik.uni-kl.de
\newcommand{\cbl}[1]{\textcolor{violet}{#1}} 	  %m_wens01@wwu.de

\maketitle

\abstract
We deduce a model for glioma invasion making use of DTI data and accounting for the dynamics of brain tissue being actively degraded by tumor cells via excessive acidity production, but also according to the local orientation of tissue fibers. Our approach has a multiscale character: we start with a microscopic description of single cell dynamics including  biochemical and/or biophysical effects of the tumor microenvironment, translated on the one hand into cell stress and corresponding forces and on the other hand into receptor binding dynamics; these lead on the mesoscopic level to kinetic  equations involving transport terms w.r.t. all kinetic variables and eventually, by appropriate upscaling, to a macroscopic reaction-diffusion equation for glioma density with multiple taxis, coupled to (integro-)differential equations characterizing the evolution of acidity and macro- and mesoscopic tissue. Our approach also allows 
for a switch between fast and slower moving regimes, %diffusion- and drift-dominated regimes, 
according to the local tissue anisotropy. We perform numerical simulations to investigate the behavior of solutions w.r.t. various scenarios of tissue dynamics and the dominance of each of the tactic terms, also suggesting how the model can be used to perform a numerical necrosis-based tumor grading or support radiotherapy planning by dose painting. We also provide a discussion about alternative ways of including cell level environmental influences in such multiscale modeling approach, ultimately leading in the macroscopic limit to (multiple) taxis. 

\vspace{0.5cm}
\section{Introduction}\label{intro}

Glioma is a common type of primary, fast growing brain tumor with a poor prognosis, the median survival time 
with the most frequent (and most aggressive) form called glioblastome multiforme 
amounting at 60 weeks, in spite of modern treatment involving resection, radio-, and chemotherapy 
(see e.g., \cite{wrensch} and references therein). An exhaustive removal of the tumor is in general impossible, 
due to the rapid advancement of glioma cells through their cycle and to the diffuse 
tumor infiltration. This leads to serious clinical challenges and to rather modest treatment outcomes. The correct 
assessment of tumor margins and of the related GTV, CTV, and PTV\footnote{GTV=gross tumor volume; CTV=
clinical target volume; PTV=planning target volume} is therefore of utmost importance. Noninvasive medical 
imaging techniques like MRI and CT are valuable diagnostic tools, however they only provide a macroscopic classification 
of neoplastic regions and the surounding oedema, without being able to evaluate the actual tumor extent which is 
heavily influenced by the biological, mechanical, and chemical processes on single cell and subcellular levels. Moreover, every patient has a 
different brain structure and there is abundant evidence that glioma follow the white matter tracts made 
up of myelinated axon bundles \cite{giese-k,giese-w}. Therefore, the personalized prediction of the tumor burden 
is necessary for an enhanced therapy planning and mathematical models are called 
 upon to provide such information via numerical simulations, thereby relying on medical data, of which 
 diffusion tensor imaging (DTI) is quite common \cite{sundgren}. It measures the spatial diffusion of water
molecules by MRI per volume element (voxel). The information obtained in this way can be expressed in the form of a diffusion tensor containing the full (apparent) diffusion information along six
directions. One way of visualizing diffusion tensors is by the so-called \textit{fractional anisotropy index} FA which is a measure for the local tissue alignment and can directly be obtained from DTI data; it is 
a scalar value between zero and one, calculated by using the eigenvalues of the apparent diffusion tensor.
A value close to one means high anisotropy, i.e., a strong preference for a specific
direction, whereas a very small value corresponds to the nearly isotropic case. For more details we refer e.g., to \cite{beppu}, see also \cite{EHKS}. 
%The eigenvalues and eigenvectors of the diffusion tensor are also used to calculate vector orientation maps, which are generated by mapping the major eigenvector directional components into 
%colour channels (RGB or other colour scheme) and weighting the colour brightness by FA.
For more about DTI and its visualization approaches along with advantages, drawbacks and extensions we refer e.g., to \cite{basser,descoteaux,jellison}.
 
 Several approaches to modeling glioma invasion 
 have been considered so far; they range from discrete formulations \cite{hatzik-deutsch,kansal,kansal-b} over 
 hybrid settings \cite{frieboes,tanaka,zhang} combining 
 decriptions of individual cells moving on a lattice with PDEs describing  the evolution of some 
 stimulus (e.g., a chemoattractant) and up to pure continuum models coupling several types of differential 
 equations. The latter category comprises in turn several classes of models, according to the scales taken 
 into account and to the type of PDEs employed. Pure macroscopic models using reaction-diffusion equations (possibly with space- and time-dependent
diffusion coefficients, or accounting for the interaction with the surrounding tissue by letting the diffusion coefficient be proportional to the water diffusion tensor assessed by DTI) have been 
introduced e.g., in \cite{bondiau,jbabdi,konukoglu,wang} to characterize the evolution of glioma density; in \cite{clatz} some biomechanical effects have been included as well. While these models impose the 
precise form of the equations and their coefficients, other approaches \cite{hillen-painter,Painter-Hillen} deduce them by using a scaling argument from a mesoscopic setting relying on kinetic transport equations for the glioma density function depending -supplementary to 
time and position- also on the cell velocities. These models have been further extended in \cite{EHKS,EHS,EKS,Hunt} to take into account subcellular level effects on the mesoscopic and ultimately macroscopic 
behavior of the tumor. Concretely, these multiscale settings included on the microscopic level receptor binding to tissue fibers, which was related to the mesoscale by way of an additional transport term and by 
the turning rate. Appropriate scalings then led to the macroscale dynamics, whose deduced coefficients are still carrying subcellular level information. In \cite{NieUrr} different types of scaling have been 
considered for 
this modeling approach, and a well-posedness result was provided for a model also involving the time-space evolution of tissue. A more general analytical result in less regular function spaces 
for a complex micro-meso-macroscale model involving both chemo- and haptotaxis introduced on the mesolevel has been proved in \cite{LorSur}. 

Here we propose a multiscale model starting from a kinetic transport equation formulation and accounting, too, for receptor binding dynamics. Unlike in \cite{EHKS,EHS,EKS,Hunt} where the space-dependent 
tissue was given by DTI data at a fixed time point and the glioma cells were migrating on this brain structure, in the current model the tissue is allowed to evolve in time as well and the glioma 
cells degrade it according to their respective orientation to the tissue fibers, similarly to the way introduced in \cite{Hillen} and readdressed in \cite{LorSur,NieUrr}. Moreover, in this work we also 
include a characterization of velocity changes via a transport term involving derivatives with respect to the velocity variable, hence allowing for external forces to act on the cells. Specifically, these external 
actions are decisively influenced by the spatial gradients of tissue fibers in the tumor surroundings and by the bias induced through the acidity gradient.  
Reference \cite{CHP} considered a similar approach accounting for a given chemotactic force acting on the cells and being proportional to the concentration gradient of some non-evolving chemoattractant. Here we pay more detailed attention to the influence of time- and space-dependent acidity  and tissue on the cell migration, proliferation, and depletion, thereby also looking into the effect of tissue fiber orientation, hence dealing with both macroscopic and mesoscopic tissue evolution. By formal upscaling we then obtain a system of ordinary and partial (integro-)differential equations with (myopic) diffusion, multiple taxis, and nonlinear source terms, which couples macroscopic dynamics of tumor cells, necrotic matter, acidity, and normal tissue with that of mesoscopic tissue.  The rest of the paper is organized as follows: in Section \ref{sec:setup} we set up the micro- and mesoscopic descriptions of (sub)cellular glioma dynamics in the kinetic theory of active particles (KTAP) framework developed e.g., in \cite{bellom1,bellom2} and also provide descriptions of the other involved model variables: tissue, necrotic matter, and acidity. Section \ref{sec:upscaling} is dedicated to deducing effective equations for the macroscopic glioma density influenced by tissue and acidity dynamics. Formal parabolic and hyporbolic upscalings are performed and a low-order moment approximation is deduced for the characterization of mesoscopic tissue, in order to reduce the complexity of the system. Numerical simulations are conducted in Section \ref{sec:numerics} for three different scenarios of increasing complexity, and the obtained results are compared and commented. Finally, Section \ref{sec:discussion} provides a discussion of this work's outcome, model extensions, and perspectives, along with a review of various ways of multiscale modeling of tactic cell behavior in the KTAP framework.  %Some of diffusion and taxis coefficients feature nonlocality w.r.t. the cell velocity, which is a kinetic variable. 

\section{Set up of the micro-meso model}\label{sec:setup} 

Our modeling approach aims at obtaining a macroscopic description of the tumor cell evolution in interaction 
with the underlying tissue. The resulting PDE should carry in its coefficients information about relevant 
processes occuring on the lower levels (subcellular/microscopic, individual cells/mesoscopic) and 
should be deduced from mathematical descriptions of the dynamics on the latter. On the macroscopic scale the two main components of cell motion in a tissue should be retrieved, 
namely diffusion and haptotaxis. 

% follows the evolution of a tumoral population moving along the extracellular matrix, 
% which is remodelled by the interactions with the cells. 
% 
% Then, our system consists of two different equations: one for the cell population, and other for the 
% microscopic density of the ECM fibers, responsible for the haptotactic response of cells.
We start by introducing some notations for the various variables and functions involved in this work.
\begin{notation}\label{notation}\emph{
\begin{itemize}
\item Mechanical variables: time $t\in \R^+$, space $x\in \R^N$, velocity $v\in V:=s\mathbb S ^{N-1}$, where $\mathbb S ^{N-1}$ is the unit sphere in $\R^N$, and, as in \cite{EHKS,EHS,EKS,Hillen,HS}, we assume the cell speed $s>0$ to be constant.
\item $\hat v := \frac{v}{|v|}$ a unitary vector denoting the direction of 
a vector $v\in V$.
\item $Q(t,x)$: macroscopic tissue density.
\item Biological (activity) variable $y$: volume fraction of bound receptors on the cell membrane. Its dynamics 
is given by %the function $G$ in the right hand side of the 
an ODE obtained for $y$ by mass action kinetics 
of receptor bindings and detachments (with rates $k_+$ and $k_-$, respectively): % (what to do with $Q$?):
\begin{equation}\label{y-dynamics0}
\dot y= k_+ (R_0-y) Q(t,x) - k_-y,\qquad y\in (0,R_0). \end{equation}
This represents in our model the subcellular dynamics.
Thereby, as in previous works \cite{KelSur, NieUrr}, we assume for simplicity that the 
total amount of cell membrane receptors $R_0$ is constant and 
%denote the latter by $R_0$. 
we rescale $y\to y/R_0$, thus obtaining\footnote{Observe that the activity variable $y$ belongs to an open set; indeed, $y=0$ would mean that no receptors are bound (which is not possible for cells living in tissue), while $y=1$ would correspond to all receptors being occupied (which actually never happens, as it would impede migration, proliferation, and even cell survival, due to anoikis). For an actual mathematical justification we refer to \cite{KelSur,LorSur}.} 
\begin{equation}\label{y-dynamics1}
\dot y= G(y,Q):=k_+ (1-y) Q(t,x) - k_-y,\qquad y\in Y:=(0,1) \end{equation}
\item $p(t,x,v,y)$: density function of tumor cells (mesoscopic tumor cell density). We will also use  $\bar p(t,x,y):=\int _Vp(t,x,v,y)dv$.
\item $\rho (t,x):=\int _Y\bar p(t,x,y)dy=\int _Y\int _Vp(t,x,v,y)dvdy$: macroscopic density of tumor cells.
\item $q(t,x,\theta)$: directional distribution function of tissue fibers with orientation 
$\theta \in \mathbb{S}^{N-1}$, where $\mathbb{S}^{N-1}$ denotes the unit sphere in $\R^N$. Hence $\int _{\mathbb S^{N-1}}q(t,x,\theta )d\theta =1$.
\item $\mathbb E_q(t,x):=\int _{\mathbb S^{N-1}}\theta q(t,x,\theta )d\theta$: mean fiber orientation.
\item $\mathbb V_q(t,x):=\int _{\mathbb S^{N-1}}(\theta -\mathbb E_q(t,x))\otimes (\theta -\mathbb E_q(t,x))q(t,x,\theta )d\theta $: variance-covariance matrix for orientation distribution of tissue fibers.
%\item $Q(t,x)$: macroscopic tissue density. %Not directly related with $q$. Does not appear in the (macro) model!!! Yes it does now!
\item $h(t,x)$: concentration of acidity (protons), a genuinely macroscopic quantity.
\item $n(t,x)$: density of necrotic tissue (also a macroscopic quantity).
\item $\lambda(x,y)$: cell turning rate. 
%\item $\hat v := \frac{v}{|v|}$ for $v\in \R^N$ a non--null vector, hence $\hat v$ denotes the direction of 
%a vector $v\in V:=s\mathbb S^{N-1}$. As in \cite{EHKS,EHS,EKS,Hillen,HS} we assume the cell speed $s$ to be constant.
\end{itemize}
}
\end{notation}
\smallskip 

Next we characterize innovations of cell density function due to changes in velocity and activity variable under the influence of the tissue structure. Depending on their nature, these 
are incorporated into the model in different ways: %\cb {(this needs to be readressed)}
\begin{enumerate}\label{enu-i}
\item[(i)] As transport parts of the mesoscopic equation for the cell density function $p$: a $\div_y$ term accounts for the binding of cell receptors to tissue fibers, while a $\div_v$ term describes biomechanical effects, such as cell stress and directionality dictated by the spatial variations in (mesoscopic) tissue density and macroscopic acidity concentration.
%%In the transport part of the mesoscopic equation for the cell density function $p$, in form of a $\div_y$ term. This accounts for the biochemical interaction aspect: cell receptors are binding to unsoluble matrix components (tissue fibers).
%%\item[(ii)] Still in the transport part of the equation for $p$, in form of a $\div_v$ term. This refers rather to biomechanical effects: cell stress, together with  directionality dictated by the spatial variations in (mesoscopic) tissue density and macroscopic acidity concentration.  %This approach is easier to justify when deducing the model from the microscopic level.
\item[(ii)] As a jump process, being modeled as an integral operator $L_0$ on the right-hand side of the equation. The kernel of that operator will depend as in previous works \cite{EHKS,EHS,EKS,Hillen,HS} 
on the tissue structure, by way of the fiber directional distribution $q(t,x,\hat v)$, as will be explained below. 
%In Section \ref{sec:hyp-scale} we also present, however, a different modeling approach considering a conservative form of the operator $L_0$ which does not involve the tissue effects. 
% \item[(iii)] As a jump process, including on the right hand side the operator  $L_0$ from (ii) and a supplementary integral operator $L_1$. 
% This provides a new degree of freedom (the kernel of the said operator), allowing to include   
% different types of drift caused by the tissue structure (see Subsection \ref{subseq:withL1}).
\item[(iii)] As a source term describing proliferation/decay and being modeled as another integral operator $\mathcal P$. This can be done for instance similarly to the approach in \cite{EHS}, where the interaction of cells with the surrounding tissue was considered to be at the onset of cell survival and mitosis. Here we modify that approach in order to account for the unfavorable effect of acidity on these processes. The concrete form of the corresponding term $\mathcal P(p,Q,h,\rho)$ will be specified later.
\end{enumerate}

% We choose the latter for our model, being the terms given by the former 
% added in {\color{red} red color} just for the sake of completness.
% 
% (justification, modeling, definitions...)
Thus, the equation for $p$ takes the following form:
\begin{align}\label{KinEq1}
\partial_t p +  v \esc \nabla_x p + \partial_y (G(y,Q)p) &+ \alpha\div_v \Big (\mathbb B(v)%(|v|^2\mathbb I_N-v\otimes v)
(\nabla_x q(t,x,\hat v)-\nabla _xh(t,x))p\Big ) \notag \\
&= \lambda (x,y) L_0(p)+\mathcal P(p,Q,h,\rho),%+a \lambda(x,y) L_1[\nabla_x q(t,x,\hat v)](p),
\end{align}
where we used the notation $\mathbb B(v):=|v|^2\mathbb I_N-v\otimes v$, and where %$a\in \{0,1\}$ allows to choose between taking into account only situations (i) and (ii) or all situations (i)-(iii) enumerated above. 
the transport coefficients are obtained from the subcellular (microlevel) dynamics \footnote{Strictly speaking we understand the ODEs for $v$ and $y$ as being written for tissue $q$, $Q$ and acidity $h$, scaled with their reference quantities (e.g., tissue carrying capacity $K_Q$ and acidity threshold $h_0$ introduced below), but for the sake of simplifying the notation we do not explicitly write those denominators. Same applies to the complementary solution components involved in the source terms of the forthcoming equations for cell density, acidity, and necrotic matter.}:
\begin{align}\label{microlevel-dyn}
\frac{dx}{dt}=v;\quad \frac{dv}{dt}=\alpha \mathbb B(v)(\nabla _xq(t,x,\hat v)-\nabla_xh(t,x)); \quad \frac{dy}{dt}=G(y,Q), 
\end{align}
the latter equation in \eqref{microlevel-dyn} being just \eqref{y-dynamics0} and the middle equation characterizing the dynamics of $v$ influenced  
by the space gradients of the directional distribution of tissue fibers and of the acidity. Thereby, the tensor $v\otimes v$ represents the active cell stress, while $-|v|^2\mathbb I_N$ is the isotropic part.  Indeed, concerning cell velocities we 
are foremost paying attention to the direction of the vector $v$ and consider that it is mainly influenced by 
the fiber distribution (and in particular by its spatial variations), as well as by the acidity gradient. The latter is known to drive the tumor cells and seems to have a bidirectional effect: Low extracellular pH favorizes both their migration and proliferation \cite{estrella,stock-schwab,webb}, but acidosis can also inhibit cell proliferation, induce stress response, and apoptosis, see e.g. \cite{ohtsubo}. Specifically in glioblastoma multiforme (GBM) it can trigger motion in the opposite direction of the acidity gradient. Indeed, GBM cells form typical, garland-like structures called  pseudopalisades, which are centered around the highly hypoxic occlusion site of a capillary \cite{brat,wippold}. The presence of such histological patterns is an indication for poor prognosis of patient survival \cite{kleih}. In order not to complicate the exposition too much we will only consider here the latter situation, with the repellent effect of acidity, hence the motion of glioma cells being biased towards $-\nabla _xh$.

The dimensionless parameter $\alpha$ will play  an important role, since it incorporates information from the different scales on which the tumor density, fiber directional distribution or acidity are correlated. While the tumor has volume dimensions, the path of travel made up of targeting fibers has a thinner structure close to the flat one. We can assume that the relationship with acidity is similar in scale.  Therefore, the gradients of $q$ and $h$ w.r.t. $x$ contribute with this new scale parameter $\alpha$ among the agents of this process, which must be, as a consequence,  a large quantity.

Observe that the dynamics of $v$ in \eqref{microlevel-dyn} ensures that $\frac{d|v|^2}{dt}$, hence the cell speed $|v|=s$ is constant (in line with the above assumption $V=s\mathbb S^{N-1}$), hence only the direction matters. %Observe that the multiplication with the tensor $v\otimes v$ ensures the correct physical units.

For the evolution of acidity concentration $h$ we also need to provide an equation; it will take the form of a reaction-diffusion PDE describing the facts that protons diffuse through the entire available space with diffusion parameter $D_H$, acidity is produced by cancer cells, and may infer decay, e.g. through uptake by vasculature or normal cells. We denote by $b>0$ the uptake rate and consider a limited production by cancer cells irrespective of their motility and/or receptor binding state, with a constant rate $a>0$. In \eqref{eq:acid} the constant $h_0$ denotes an acidity threshold which is critical for non-cancerous matter; the proton buffering is enhanced when that threshold is exceeded, and it is limited by the availability of  normal tissue and cells, which decreases under hypoxia (as explained later on):
\begin{equation}\label{eq:acid}
h_t=D_H\Delta h+\frac{a\rho }{1+\rho}-bQ(h-h_0).
\end{equation}
The turning operator $L_0$ on the right hand side in \eqref{KinEq1} is 
taken as a relaxation-type operator describing the velocity-jump discontinuities in the cell density function $p$ due to contact guidance, that is, these velocity alterations
are assumed to be caused by the cells interacting with tissue fibers and adapting their motion to
%
%consists of an %one (for $a=0$) or two (for $a=1$) 
%integral operator describing the velocity-induced changes in the cell density function $p$. %We will present two different ways of modeling the respective 
%integral operator. 
%characterizes the velocity innovations and uses an integral operator. 
%These changes (above all in velocity orientations) are assumed to be caused by the cells interacting with tissue fibers and adapting their motion to 
biochemical (biomechanical, etc.) cues available at sites in their direct proximity. More precisely, it is given by
\begin{equation}
\label{L0}
L_0(p) := \bar p(t,x,y) M(t,x,v) - p(t,x,v,y), 
\end{equation}
%We model this behavior with the aid of a relaxation-type operator 
%$$ L_0(p) := \bar p(t,x,y) M(t,x,v) - p(t,x,v,y), $$
%with $M(t,x, v)$ a positive function verifying
%\begin{equation}\label{PropM}
%$\int _VM(t,x,v) \ud v =1$. %, \qquad \int _VvM(t,x,v) \ud v = 0.
%\end{equation}
where $M(t,x, v)$ depends on the fiber directional distribution $q$:
\[
M(t,x,v) := \frac{q(t,x, \hat v)}{\omega}, \qquad \omega := \int _Vq(t,x, \hat v) \ud v, 
\]
thus $\int _VM(t,x,v) \ud v =1$. Notice that this corresponds to considering, 
as in \cite{Hillen,EHKS,EHS,EKS,Painter-Hillen}, a turning operator of type
$$
L_0(p):=-p(t,x,v,y)+\int _V\tilde K(v,v')p(t,x,v',y)dv'
$$
with the turning kernel defined by $\tilde K(v,v')=\frac{q(t,x,\hat v)}{\omega }$, that is, the random migration from any velocity $v'$ to a new one $v$ is given by the (normalized) orientational distribution of tissue fibers on this last velocity $v$, so through contact guidance. 
We recall that this turning operator is multiplied by the cell turning rate $\lambda(x,y)$. It will play an essential role in our attempt  
to characterize the alternation between %different cell migration regimes, %drift- and diffusion-dominated motion of tumor cells, 
slow and enhanced migration of tumor cells, which is mainly due to the tissue anisotropy. To capture this dependence we use the \textit{fractional anisotropy} index FA mentioned in Section \ref{intro}. 
Henceforth we deal with a turning rate of the form
$$\lambda(x,y)= \frac{\kappa y}{FA(x)+y},$$ which is in agreement with the assumption of reduced 
turning in highly aligned regions and with its dependence on the receptor binding state $y$, the latter with a 
certain saturation modeled by the Monod type factor. Thereby, $\kappa $ is a constant that refers to the maximum turning frequency.

Finally, we introduce the proliferation/decay operator. Similarly to the approach in \cite{EHS}, both proliferation and decay of tumor cells originate (at least partially) in the interaction of cells with the surrounding tissue, but here we modify that approach in order to account for the unfavorable effects of acidity on these processes. Concretely, $\mathcal P(p,Q,h,\rho)$ is given by
\begin{equation}
\mathcal P(p,Q,h,\rho):=\mu (\rho,h)\int _Z\chi(x,z,z')p(t,x,v,z')Q(t,x)dz',
\end{equation}
where $\chi (x,z,z')$ is a kernel with respect to $z$ and represents the transition from the state $z'$ to the state $z$ during a proliferation-favorable glioma-tissue interaction. 

%At this point we may also think about including source terms for the cell density, i.e. proliferation and decay. This can be done for instance similarly to the approach in \cite{EHS}, where the interaction of cells with the surrounding tissue was considered to be at the onset of cell survival and mitosis. Here we modify that approach in order to account for the unfavorable effect of acidity on these processes. Thus, we supplement \eqref{KinEq1} with a term $\mathcal P(p,Q,h,\rho)$ to be specified later.\\
 
For the tissue dynamics we adopt the approach in \cite{Hillen} which was also employed in 
\cite{KelSur,LorSur,NieUrr}, with slight modifications. We look for the evolution of $q(t,x,\theta )$, hence want 
an equation of the form 
$$\partial _tq(t,x,\theta )=\tau (\theta ,p,q),$$
with $\tau $ satisfying the natural conditions:
$$\tau (\cdot, \cdot, 0)=0,\quad \int _{\mathbb S^{N-1}}\tau (\theta ,\cdot,\cdot)d\theta =0,\quad %\tau (\theta ,p,q)=\tau (-\theta ,p,q),\ 
\forall \ \theta \in \mathbb S^{N-1}.$$
%which come from $\frac{q(\theta )}{\omega }$ being a symmetric probability density function. \footnote{thus an even function w.r.t. $\theta$, since it is required to be a genuine probability kernel} 

We thus compare -as in \cite{Hillen}- the fiber orientation $\theta $ with the 
cell direction $\hat v$, and introduce the operator:
\begin{equation}\label{mean-projection}
\Pi [p](t,x,\theta):= \left \{\begin{array}{cl}
\frac{1}{\rho(t,x)}\int_Y\int _V |\theta \cdot \hat v| p(t,x,v,y)\ud v \ud y,&\text{ for undirected tissue}\\
\frac{1}{\rho(t,x)}\int_Y\int _V \theta \cdot \hat v\ p(t,x,v,y)\ud v \ud y,&\text{ for directed tissue,}
\end{array}\right .
\end{equation}
\noindent
which averages the projection of cell movement direction on the fiber orientation. It models the fact that cells preferentially degrade fibers which are nearly orthogonal to their movement direction % will provide the information about the tissue being degraded or not, depending on the orientation of $\hat v$ with respect to $\theta$ 
(for more details see \cite{Hillen}). It holds that $0\le \Pi [p]\le 1$ and $-1\le \Pi [p]\le 1$ for undirected and for directed tissue, respectively. Thereby, \textit{undirected} means that the fibers making up the tissue are symmetrical all along their axes, i.e. there is no 'up' and 'down' on such fibers, which translates into symmetry of the orientational distribution:
\begin{equation*}
q(t,x,-\theta )=q(t,x,\theta ),\quad \forall \ \theta \in \mathbb S^{N-1}.
\end{equation*}

If this is not the case, then the tissue is said to be \textit{directed}. These tissue properties reflect on the cell motility: While undirected fibers do not impose any supplementary bias on the (re)orientation of a cell, in a directed tissue the cells will have a preferred direction of advancement along the fibers. As DTI data do not provide any information about such directionality, we will consider in the following both types of tissue, which will lead to two different ways of performing the transition from the lower scales (micro and meso) to the macroscopic description of tumor dynamics in interaction with their fibrous and biochemical surroundings.

If we denote by $\textcrg (t,x,\theta )$ the mesoscopic density of tissue fibers with orientation $\theta $ we 
obtain for the evolution of this quantity the equation
\begin{equation}\label{tissue-evolution}
\partial_t\textcrg (t,x,\theta) =r_D(h)(\Pi [p](t,x,\theta)-1)\rho (t,x) \textcrg (t,x,\theta ),
\end{equation}
where $r_D$ is a nonnegative quantity characterizing the efficiency of fiber degradation (e.g., it 
can be considered to be proportional to the amount of matrix degrading enzymes\footnote {shortly MDEs} expressed by the cells, thus to the acidity $h$, since MDEs are actually known to be enhanced by an acidic
extracellular pH, see \cite{calorini,kato}). We then consider $r_D$ to be a monotonically increasing function of $h$ with $r_D(h)=r_0(h-h_0)_+$.  % $r_0=10$ (1/(mol\ sec)) versuchen.} \\

It would be important to have information about the directional distribution function $q(t,x,\theta)$, hence we express the latter 
with respect to the mesoscopic tissue density $\textcrg (t,x,\theta )$. Following \cite{Hillen}, this 
relationship takes the form
$$q(t,x,\theta )=\frac{\textcrg (t,x,\theta )}{\int _{\mathbb S^{N-1}}\textcrg (t,x,\theta )d\theta },
\qquad \textcrg \neq 0.$$
%\cb {Together with \eqref{tissue-evolution} this leads to an equation for $q(t,x,\theta )$:
%\begin{equation}
%\partial_tq(t,x,\theta )=r_D(h)\Big (\Pi [p](t,x,\theta)
%-\int_{\mathbb S^{N-1}} \Pi [p](t,x,\theta)q(t,x,\theta)d\theta \Big )\rho (t,x)q(t,x,\theta).\label{full-meso2}
%\end{equation}}

The macroscopic tissue density $Q(t,x)$ represents the volume fraction of tissue fibers, irrespective of their orientation, thus we also have the following relationship:
\begin{equation}
\textcrg (t,x,\theta )=q(t,x,\theta)Q(t,x), \quad t>0,\ x\in \R^N,\ \theta \in \mathbb S^{N-1}.
\end{equation}

Integrating  \eqref{tissue-evolution} w.r.t. $\theta \in \mathbb S^{N-1}$ we get
\begin{align}\label{eq:Q-1}
\partial_tQ=r_D(h)\rho Q\left(\int _{\mathbb S^{N-1}}\Pi[p](t,x,\theta)q(t,x,\theta )d\theta -1\right)=r_D(h)\rho \, Q \int _{\mathbb S^{N-1}} \big(\Pi [p] -1\big) q\, d\theta .
\end{align}

In the sequel the average of $\Pi[p]$ w.r.t. the distribution of fiber orientations will be denoted as in \cite{Hillen} by
\begin{equation}\label{eq:A}
A[p](t,x):=\int _{\mathbb S^{N-1}}\Pi[p](t,x,\theta)q(t,x,\theta )d\theta ,
\end{equation}
hence \eqref{eq:Q-1} becomes
\begin{align}\label{eq:Q}
\partial_tQ=r_D(h)\rho Q(A[p]-1).
\end{align}

Likewise, we can deduce an equation for the fiber directional distribution function:
\begin{align}\label{eq:fiber-q-gen}
\partial _tq(t,x,\theta )&=r_D(h)\rho (t,x) q(t,x,\theta )\left (\Pi[p](t,x,\theta)-A[p](t,x)\right ).%\int _{\mathbb S^{N-1}}\Pi[p](t,x,\theta)q(t,x,\theta )d\theta \right )\notag \\
\end{align}
%which \textit{in the case of directed tissue} becomes
%\begin{align}\label{eq:fiber-q-directed}
%\partial _tq(t,x,\theta )=r_D(h)\rho (t,x)q(t,x,\theta )\left (\Pi[p](t,x,\theta)-\frac{1}{s}\mathbb E_q(t,x)\cdot U(t,x)\right ).
%\end{align}

Eventually, the decay of both tissue and glioma (primarily due to acidity) generates necrotic tissue, large amounts of which are an indicative of poor survival prognosis \cite{Hammoud1996,Louis}. Therefore, the detection and assessment of necrosis is an important issue in therapy. Here we describe the dynamics of necrotic tissue density by
\begin{align}\label{eq1-necrotic}
\partial_tn&=r_D(h)\rho Q\int _{\mathbb S^{N-1}}(1-\Pi [p](t,x,\theta))q(t,x,\theta)d\theta +\mathcal F(h,\rho, Q)
%&=r_D(h)\rho Q(1-\frac{1}{s^2}\mathbb E_q\cdot U)+\mathcal F(h,\rho, Q),
\end{align}
where $\mathcal F(h,\rho,Q)$ describes glioma death due to hypoxia and is to be specified later. 

Thus, the full model for 
cell-tissue interactions and response to acidity is given by 
\eqref{KinEq1}, \eqref{eq:acid}, \eqref{eq:Q}, \eqref{eq:fiber-q-gen},  \eqref{eq1-necrotic}, supplemented with appropriate initial and boundary conditions.

As in previous works \cite{EHKS,EHS,EKS,KelSur} we assume $p$ to be compactly supported in the $(v,y)$-space. The initial conditions and the boundary conditions w.r.t. space will be addressed in the following sections.
The wellposedness of 
this problem (especially in less regular function spaces) is not trivial and also not the aim of this 
paper. When the space domain is the whole $\R^N$ the approach in \cite{KeSu11,KelSur,LorSur} could be used as a starting point in order to address this issue. 

\section{Towards population behavior: Upscaling}\label{sec:upscaling}

Our objective is to assess the macroscopic evolution of the tumor cell population interacting with the 
brain tissue. %To this aim we will investigate two types of scaling: a parabolic (in this section) and a hyperbolic one (in Section \ref{sec:hyp-scale}). 
%-----------\cb {This is not completely correct, has to be rewritten:}\\
As the latter 
exhibits anisotropy variability, with highly aligned regions alternating with areas of isotropic 
fiber distribution, the invasive behavior of glioma cells will be correspondingly -and locally- dominated by diffusion with or without drift. Therefore, it is desirable that 
our model includes a switch between these two kinds of 
motion, in the sense that the influence of drift together with diffusion can be potentiated against pure diffusion. These effects are built in via the 
turning rate $\lambda (x,y)$ given in Section \ref{sec:setup} above. 
Indeed, its 
dependence on the (local) fractional anisotropy $FA(x)$ and the amount of receptors bound to their ligands on the 
tissue fibers will be able to capture the alternation between the epochs of higher- or less-aligned tissue.\\
%Remark that the operator $L_0$ chosen in \eqref{L0} has the following properties:
\begin{enumerate}
	\item {\bf Mass conservation:} $\int _VL_0(p)(t,x,v,y) \ud v =0$.
	\item {\bf Self--adjointness:} $L_0$ is self--adjoint in $L^2(\frac{dv}{M(v)})$.
	\item\label{new4} {\bf Kernel of $L_0$:} $L_0(p)=0\ \Leftrightarrow \ p\in \langle M(v)\rangle $, thus $Ker(L_0)=\langle M(v)\rangle$, the space 
	generated by $M(v)$.
In particular, using the self--adjointness, we know that $M^\bot \in \langle M(v)\rangle^\bot $ iff  $\int_V M^\bot (v) dv =0$.
%	\item $L_0(-vM(v)) = vM(v)$.
\end{enumerate}

Before proceeding with the scaling we re-express (similarly to e.g. \cite{EHKS}) the subcellular dynamics in a more convenient way: Let us consider 
$y^*= f(Q):=\frac{k_+ Q}{k_-+k_+ Q }$, 
the steady-state of \eqref{y-dynamics0}, and denote by 
$$z:=y-y^*$$ 
the deviation of the current activity variable $y$ from the steady state. Then $z\in Z\subseteq [-y^*,1-y^*]$ and the microscopic dynamics \eqref{microlevel-dyn} turns into
\[
\frac{dx}{dt}=v;\quad \frac{dv}{dt}=\alpha \Bmb(v)\,  \big(\nabla _xq(t,x,\hat v)-\nabla_xh(t,x)\big); \quad \frac{dz}{dt}=-(k_+Q+k_-)z- f'(Q)(Q_t+v\esc \nabla_x Q).
\]

Then \eqref{KinEq1} becomes
\begin{align}\label{KinEq2}
\partial_t p &+  v \esc \nabla_x p - \partial_z \Big (((k_+Q+k_-)z+f'(Q)(Q_t+v\cdot \nabla Q))p\Big ) + \alpha  \div_v \Big (\mathbb B(v) (\nabla_x q(\hat v)-\nabla_xh) p\Big ) \nonumber \\
&= \lambda (x,z) L_0(p)+\mathcal P(p,Q,h,\rho),%+ a \lambda(x,y) L_1[\nabla_x q](p),
\end{align}
where the turning rate in terms of $z$ is given by 
$\lambda (x,z)=%\frac{ \kappa (z+y^*)}{FA(x)+z+y^*}=
\frac{\kappa (z+f(Q(t,x)))}{FA+z+f(Q(t,x))}$. 
To simplify the subsequent computations we 
linearize it as follows: 
$$\lambda (t,x,z)\simeq \lambda _0(t,x)+\lambda _1(t,x)z,\ \text{where} \  \lambda_0(t,x)=\frac{ \kappa f(Q)}{FA+f(Q)}\ \text{and}\  \lambda _1(t,x)=\frac{ \kappa FA}{(FA+f(Q))^2}.$$

\begin{rmk}\emph{
			Notice that the fractional anisotropy FA changes  dynamically, i.e. depend both on $t$ and $x$, since tissue evolution (here this means degradation) leads to local modifications of the water diffusion tensor and, correspondingly, of its eigenvalues. Indeed, the apparent diffusion tensor can even vanish locally, in which situation the cell diffusivity degenerates. In hitherto works \cite{Martina,CKSS,EHKS,EHS,EKS,Hunt,HS,jbabdi,Painter-Hillen} FA has been assumed to be time-independent, motivating that the resolution of DTI data does rarely go below voxels with sizes of  1 $mm^3$, which is rather rough compared with the size of glioma cell bodies (15-60 $\mu m$, \cite{kuche}). This simplifies the numerical handling and also avoids problems related to the possibly singular behavior of solutions to the macroscopic PDE, as systems with degenerating myopic diffusion and haptotaxis can lead to blow-up even in 1D  \cite{winkler,WiSu}. In Section \ref{sec:numerics}, however, we will consider several simulation scenarios, including, in turn,  evolving and time-stationary tissue, and compare their outcome.}
\end{rmk}

\subsection{Parabolic limit, undirected tissue}\label{sec:par-scaling}
The parabolic scaling corresponds formally to the change of variables $t\to \eps^2 t, x\to \eps x.$ We perform it and thereby rescale as in \cite{EHS} the source term $\mathcal P(p,Q,h,\rho)$ with $\eps^2$, in order to let it act on the correct time scale. Correspondingly, equation \eqref{KinEq2} becomes:
\begin{align}\label{EscEq}
&\eps ^2\partial_t p+\eps \nabla_x \cdot (vp)- \partial_z \Big (((k_+Q+k_-)z+f'(Q)(\eps ^2Q_t+\eps v\cdot \nabla Q))p\Big ) %\notag \\ 
%&
+\alpha \eps  \div_v \Big (\mathbb B(v) (\nabla_x q(\hat v)-\nabla_xh) p\Big )\notag \\
&\hspace*{1cm}= \lambda (t,x,z)L_0(p)+\eps^2\mu (\rho,h)Q(t,x)\int _Z\chi(x,z,z')p(t,x,v,z')dz'.%(\lambda_0 (x)+\lambda _1(x)z) L_0(p).
\end{align}

Next we consider the moments of $p$ with respect to $v$ and especially with $z$:
\begin{align*}
&m(t,x,v):=\int _Zp(t,x,v,z)dz, \quad m^z(t,x,v):=\int _Z zp(t,x,v,z)dz, \quad \rho ^z(t,x)=\int _Vm^z(t,x,v)dv\\
&\bar p(t,x,z):=\int _Vp(t,x,v,z)dv,\quad \rho (t,x):=\int _Vm(t,x,v)dv=\int _Z\bar p(t,x,z)dz
\end{align*}
and neglect the higher order moments with respect to $z$ by assuming very small deviations of the receptor binding dynamics from the steady-state, i.e. by assuming $z$ to be very small. As the subcellular dynamics is very fast in comparison to cell motion and proliferation, this is a reasonable assumption. 

Then from \eqref{EscEq} we obtain the moment equations:
\begin{subequations}
\begin{align}
& \eps ^2\partial_tm+\eps \nabla_x \cdot (vm)+\alpha \eps \div_v \Big (\mathbb B(v) (\nabla_x q(\hat v)-\nabla_xh) m\Big )=\lambda _0(t,x)(M(v)\rho -m)\notag \\
&\hspace*{1cm}+\lambda _1(t,x)(M(v)\rho ^z-m^z)
 +\eps^2\mu (\rho,h)Q(t,x)\int _Z\int _Z\chi(x,z,z')p(t,x,v,z')dz'dz\label{subeq:moments1}\\
& \eps ^2\partial_tm^z+\eps \nabla_x \cdot (vm^z)+\alpha \eps \div_v \Big (\mathbb B(v) (\nabla_x q(\hat v)-\nabla_xh) m^z\Big )+(k_+Q+k_-)m^z+\eps f'(Q)v\cdot \nabla Qm \notag \\
 &+\eps^2f'(Q)mQ_t=\lambda _0(t,x)(M(v)\rho ^z-m^z)+\eps^2\mu (\rho,h)Q(t,x)\int _Z\int _Zz\chi(x,z,z')p(t,x,v,z')dz'dz\label{subeq:moments2}
\end{align}
\end{subequations}
Performing the usual Hilbert expansion $p=\sum \limits _{k}\eps^k p_k$ and consequently the expansion of the moments:
\begin{align}\label{new3}
m=\sum \limits _{k}\eps ^km_k, \qquad m^z=\sum \limits _{k}\eps ^km_k^z,\qquad \rho=\sum \limits _{k}\eps ^k\rho _k,\qquad  \rho^z=\sum \limits _{k}\eps ^k\rho_k^z \end{align}
and identifying the powers of $\eps$ we obtain from \eqref{subeq:moments1} and \eqref{subeq:moments2} the following relationships:\\

\noindent
$\epsilon^0$ terms:
\begin{align*}
&0 = \lambda_0 (M(v)\rho _0-m_0) + \lambda_1 (M(v)\rho _0^z-m_0^z)\\
&(k_+Q+k_-) m_0^z = \lambda_0 (M(v)\rho _0^z-m_0^z)\qquad  \qquad \qquad \qquad\qquad\Rightarrow \quad \rho _0^z=m_0^z=0\quad \text{and}\quad M(v)\rho _0=m_0.
\end{align*}
\noindent
$\epsilon^1$ terms:
\begin{align*}
 \nabla_x\cdot (vm_0) &+\alpha \div_v\Big (\mathbb B(v) (\nabla_x q(\hat v)-\nabla_xh) m_0\Big )=\lambda _0(M(v)\rho_1-m_1)+\lambda _1(M(v)\rho_1^z-m_1^z)\\
 \nabla_x\cdot (vm_0^z) & +\alpha \div_v\Big (\mathbb B(v) (\nabla_x q(\hat v)-\nabla_xh) m_0^z\Big )\\ &+(k_+Q+k_-)m_1^z+f'(Q)v\cdot \nabla Qm_0=\lambda_0 (M(v)\rho _1^z-m_1^z).
\end{align*}

Integrating with respect to $v$ we obtain (for undirected tissue) 
\begin{subequations}\label{sus}
\begin{align}
\rho _1^z &=0,\quad m_1^z=-\frac{M(v)\rho_0}{k_+Q+k_-+\lambda _0}f'(Q)v\cdot \nabla Q,\\
m_1&=\frac{1}{\lambda _0}\Big [-\nabla _x\cdot (vM(v)\rho _0)- \alpha \div_v\Big (\mathbb B(v) (\nabla_x q(t,x,\hat v)-\nabla_xh)M(v)\rho _0\Big )-\lambda _1m_1^z\Big ]\\ &+M(v)\rho _1.\label{sus2}
\end{align}
\end{subequations}

\noindent
$\epsilon^2$ terms from \eqref{subeq:moments1}, after expanding $\mu $ about $\rho_0$ and integrating the equation with respect to $v$: 
\begin{align*}
 \partial_t\rho_0+\nabla _x\cdot \int _Vvm_1dv=\mu (\rho_0,h)Q\rho _0.
\end{align*}

From \eqref{sus} we can compute 
\begin{align*}
 \int _Vvm_1dv&=\frac{1}{\lambda _0}\Big [-\nabla _x\cdot (\int _Vv\otimes vM(v)dv\rho _0) \\ &-\alpha \int _Vv\div_v\Big (\mathbb B(v) (\nabla_x q(t,x,\hat v)-\nabla_xh(t,x))M(v)\rho _0\Big )dv\\
 &+\frac{\lambda _1f'(Q)}{k_+Q+k_-+\lambda _0}\int _Vv\otimes vM(v)dv\nabla Q\rho _0\Big ].
\end{align*}

We denote by
 \begin{equation}\label{eq:Eq}
 \tilde {\mathbb E}_q(t,x):=\int _Vv \frac{q(t,x,\hat v)}{\omega } \ud v=s\int_{\mathbb S^{N-1}}\theta q(t,x,\theta )d\theta =s\mathbb E_q(t,x)
 \end{equation}
% the mean fiber orientation 
and with
\begin{align}
	\mathbb D_T(t,x) := \ds \int _V(v-\tilde {\mathbb E}_q)\otimes (v-\tilde {\mathbb E}_q) M(t,x,v)\ud v=\ds \int _V(v-\tilde {\mathbb E}_q)\otimes (v-\tilde {\mathbb E}_q) \frac{q(t,x,\hat v)}{\omega } \ud v%=s^2 \int _{\mathbb S^{N-1}}\theta \otimes \theta q(t,x,\theta)d\theta 
	\label{tumor-diffusion-tensor}
\end{align} the so-called \textit{tumor diffusion tensor}\footnote{This designation is somewhat abusive, since here -unlike previous works \cite{EHKS,EHS,EKS,Painter-Hillen}- the turning rate is not involved in its expression, due to its dependence on the position. When writing out the myopic diffusion it becomes evident that the 'actual' diffusion tensor is $\mathbb D_T/\lambda _0$.}. In the situation with undirected tissue we have 
$\mathbb E_q(t,x)=0,$ %\int _Vv \frac{q(t,x,\hat v)}{\omega } \ud v=s\int_{\mathbb S^{N-1}}\theta q(t,x,\theta )d\theta =0,$$ 
thus in this case $\mathbb D_T(t,x)=s^2\mathbb V_q(t,x)$, where $\mathbb V_q$ denotes the variance-covariance matrix w.r.t. the fiber orientation distribution $q$. % introduced in Notation \ref{notation}.\\
With the notation
$$g(Q,\lambda _0)(t,x):=\frac{f'(Q)}{k_+Q+k_-+\lambda _0},$$ we have
\begin{align}\label{eq-cu-Sigma}
 \int _Vvm_1dv&=-\frac{1}{\lambda _0}\nabla _x\cdot (\mathbb D_T\rho _0)+\frac{\lambda _1}{\lambda _0}g(Q,\lambda _0)\mathbb D_T\nabla Q\rho _0+\frac{ \alpha }{\lambda _0}\Sigma (t,x)\rho _0,
\end{align}
with $\Sigma (t,x):=S_1(t,x;q)-S_2(t,x;h)$, where
\begin{subequations}\label{eqs:S}
\begin{align}
	&S_1(t,x;q):=\int _V\mathbb B(v) \nabla_x q(t,x,\hat v)M(t,x,v)dv=\frac{1}{\omega }\int _V\mathbb B(v) \nabla_x q(t,x,\hat v)q(t,x,\hat v)dv \label{haptotaxis-forces}\\
	&S_2(t,x;h):=\int _V\mathbb B(v) M(t,x,v)dv\ \nabla_x h(t,x)=s^2(\mathbb I_N-\mathbb V_q(t,x))\nabla_x h(t,x).
\end{align}
\end{subequations}

Then, the limiting macroscopic equation for the tumor cell density takes the form

\begin{eqnarray}\label{macro-eq}
 \partial_t\rho_0&=&\nabla \cdot \Big [\frac{1}{\lambda _0(x)}\Big (\nabla \cdot (\mathbb D_T\rho _0)
 -\lambda _1(x)g(Q,\lambda _0)\mathbb D_T\nabla Q\rho _0- \alpha S_1(t,x;q)\rho _0\nonumber \\ & &+\alpha S_2(t,x;h)\rho _0\Big )\Big ]+\mu (\rho_0,h)Q\rho _0.
\end{eqnarray}

The obtained  equation is of drift-diffusion type; the first term on the right hand side represents a non-Fickian, so-called myopic diffusion: the cells spread out according to information 
available in their immediate surroundings. The next two terms feature a sign opposite to the myopic diffusion and represent drift corrections of the diffusive part in the direction of the macroscopic and 
mesoscopic tissue gradients $\nabla Q$ and $\nabla q$, respectively. They could therefore be interpreted as haptotaxis-like terms. The first of them carries through the function $g$ the influence of the 
subcellular receptor binding dynamics to the surrounding tissue, while the remaining drift term accounts for the (mesoscopic) stress exerted by the cells on the tissue, the latter being described by 
the directional distribution of fibers $q$.  The drift term involving $S_2$ in \eqref{macro-eq} has the same sign with the diffusion, and contains a bias towards the opposite direction of acidity gradient $\nabla h$, thus describes a repellent pH-taxis. 
The information about the brain tissue structure, which is decisive for personalized predictions of the tumor space-time evolution, is contained in \eqref{macro-eq} 
via the tumor 'diffusion' tensor $\mathbb D_T$ and the vector $S_1(t,x;q)$, which in view of \eqref{microlevel-dyn} can be seen as a biomechanical interaction force between tumor cells and tissue. Thereby, the 
stress tensor $\mathbb B(v)$ contributes (together with the kernel $M(t,x,v)=\frac{q(t,x,\hat v)}{\omega }$) to the (mesoscopic) haptotactic sensitivity of the cells, which is expressed here in a velocity-averaged way. The last term in \eqref{macro-eq} encodes proliferation and decay of glioma embedded in tissue. The function $\mu (\rho_0,h)$ can be correspondingly chosen, e.g. in the form 
\begin{equation}\label{eq:choice-mu}	
	\mu (\rho_0,h)=\eta (1-\rho_0-n)(h_T-h)_+-\gamma h,
	\end{equation} which would correspond to a logistic type of growth and a hypoxia-triggered necrosis. %The latter is characterized by a Holling III type degradation with an acidity threshold $H$\footnote{corresponding e.g., to some critical pH for glioma; in humans the measured glioma pH has a mean value around $6.8$ and can be as low as $5.9$, see \cite{chiche,vaupel}} and an upper depletion level $b(h)$ proportional to the current proton concentration. 
	Thereby $\eta ,\gamma >0$ are constants and $h_T$ denotes an acidity threshold which is critical for the cancer cells: when the proton concentration exceeds it, they become hypoxic and cease proliferation. The choice of $\mu $ establishes the form of the last source term in \eqref{eq1-necrotic}. Thus, with \eqref{eq:choice-mu} we have in the equation for necrotic tissue
	\begin{equation}\label{eq:choice-F-rond}
	\mathcal F(h,\rho, Q)=\gamma h\rho Q.       %\eta \frac{b(h)\rho ^2 Q}{H^2+\rho^2}.
	\end{equation}

The role of the turning rate $\lambda $ is twofold: it connects the cell reorientations to the receptor binding 
kinetics and it also captures the effect of tissue anisotropy on the cells migrating through the tissue. While $\lambda _0$ influences all motility terms, $\lambda _1$ is specific for the haptotaxis component including subcellular level effects. The factor $\frac{1}{\lambda _0}$ is independent on $y$ and increasing with the space-varying fractional anisotropy $FA$, the effect of which is particularly accentuated in the term multiplied with $\lambda _1$. This means that $\lambda _1$ with the inherent $y$-dependence provides an anisotropy-triggered  switch between myopic diffusion with 'mesolevel' haptotaxis and migratory mode with enhanced haptotaxis, the latter supplementary providing bias in the direction of the tissue gradient. %expressed in a stationary and pure macroscopic way. 
Observe that an almost isotropic tissue lets $\lambda _0$ be near constantly $\kappa$ and nearly turns off $\lambda _1$ and therewith the influence of subcellular dynamics. In this case, without the effect of $S_1(t,x;q)$ there would only be myopic diffusion, as e.g., in \cite{Painter-Hillen}.

\subsubsection{Boundary conditions and the full macro-meso system}\label{BC-approx-projections}

So far we considered the space variable $x\in \R^N$, however the brain occupies a well delimited region inside the skull. Therefore we consider a bounded space domain $\Om \subset \R^N$ and assume it to have a smooth enough boundary. Through the rescaling $x\to \varepsilon x$ the domain on which \eqref{macro-eq} holds is $\tilde \Om =\varepsilon \Om$, having outer unit normal vector $\nu (x)$ at $x\in \partial \tilde \Om$. In order to determine the corresponding boundary conditions on $\partial \tilde \Om$ we assume that there is no normal mass flux across the boundary \cite{lemou}, which translates into the mesoscopic no-flux condition \cite{plaza}
\begin{equation}\label{BC-plaza}
\int _Vvp(t,x,v,z)\cdot \nu (x)\ dv=0,\qquad \text{for all }x\in \partial \tilde \Om,\ t>0.
\end{equation}

This condition actually means that the normal component of the macroscopic ensemble velocity $U(t,x)=\int _Vvm(t,x,v)dv$ of the tumor across the space boundary vanishes (the tumor cannot leave the brain). Following \cite{plaza} we write the boundary of the phase space as 
$$\partial \tilde \Om \times V\times Z=(\Gamma _+\cup \Gamma _-\cup \Gamma _0)\times Z,$$
where 
$$\Gamma _\pm :=\{(x,v)\in \partial \tilde \Om\times V\ :\ \pm v\cdot \nu (x)>0\},\quad \Gamma _0:=\{(x,v)\in \partial \tilde \Om\times V\ :\ v\cdot \nu (x)=0\}.$$

We assume that $\Gamma _0$ has zero measure w.r.t. the Lebesgue measure on $\partial \tilde \Om \times V$ and consider the trace spaces 
$$L^2_\pm:=L^2(\Gamma _\pm \times Z;|v\cdot \nu (x)|d\sigma (x)dvdz).$$

Moreover, $p$ is supposed to be regular enough so that we can define the traces $p|_{\Gamma _\pm \times Z}\in L^2_\pm$, and that for a fixed $t>0$ 
$$p|_{\partial \tilde \Om \times V\times Z}(t,x,v,z)=\lim _{\substack{\tilde x\in \tilde \Om \\
	\tilde x\to x}}p(t,\tilde x,z),\quad \text{for each }x\in \partial \tilde \Om.$$

Assuming that a regular Hilbert expansion is valid in $\tilde \Om $ we can therefore compute the trace by simply passing to the corresponding limit in the Hilbert expansions for $p(t,x,v,z)$ and accordingly also for the moments \eqref{new3}. %$m,m^z,\rho, \rho^z$. 
Thus, the no-flux condition \eqref{BC-plaza} becomes
$$\int _Vv(p_0(t,x,v,z)+\varepsilon p_1(t,x,v,z))dv\cdot \nu(x)+O(\varepsilon^2)=0,\quad x\in \partial \tilde \Om ,\ z\in Z,\ t>0.$$

This condition should not depend on $\varepsilon$, therefore we should have 
$$\int _Vvp_j(t,x,v)\cdot \nu (x)dv=0\quad \text{for all }j\ge 0,\ x\in \partial \tilde \Om,\ z\in Z,\ t>0.$$

Indeed, for $j=0$ we already have this condition satisfied for undirected tissue:
%\begin{align*}
%\int _Vvp_0(t,x,v,z)dv \cdot \nu (x)=0,
%\end{align*}
%which is satisfied if its integral over $Z$ nullifies, i.e.
\begin{align*}
\int _Z\int _Vvp_0(t,x,v,z)dvdz\cdot \nu (x)&=\int _Vvm_0(t,x,v)dv\cdot \nu (x)\\
&=\int _Vv\frac{q(t,x,\hat v)}{\om}\rho _0(t,x)dv\cdot \nu (x)\\
&=\rho _0\tilde {\mathbb E}_q(t,x)\cdot \nu (x)=0.
\end{align*}
%which holds for undirected tissue.\\

Next we look at the condition for $j=1$, more precisely at its integration w.r.t. $z$:
\begin{align*}
\int _Z\int _Vvp_1(t,x,v,z)dv dz\cdot \nu (x)&=\int _Vvm_1(t,x,v)dv\cdot \nu (x)\\
&\stackrel{\eqref{eq-cu-Sigma}}{=}\Big (-\frac{1}{\lambda _0}\nabla _x\cdot (\mathbb D_T\rho _0)+\frac{\lambda _1}{\lambda _0}g(Q,\lambda _0)\mathbb D_T\nabla Q\rho _0+\frac{\alpha }{\lambda _0}\Sigma (t,x)\rho _0\Big )\cdot \nu (x),
\end{align*}
which leads to a typical no-flux boundary condition 
\begin{equation}\label{BC-macro-cells}
\Big (-\frac{1}{\lambda _0}\nabla _x\cdot(\mathbb D_T\rho _0)+\frac{\lambda _1}{\lambda _0}g(Q,\lambda _0)\mathbb D_T\nabla Q\rho _0+\frac{\alpha }{\lambda _0}\Sigma (t,x)\rho _0\Big )\cdot \nu (x)=0\quad \text{on }\partial \tilde \Om,\ t>0
\end{equation}
for the macroscopic PDE \eqref{macro-eq}. This equation is coupled with the dynamics of acidity $h$ by way of $S_2(t,x;h)$ in $\Sigma (t,x)$. For this \textit{ab initio} macroscopic equation we can directly impose a no-flux condition:
\begin{equation}\label{BC-macro-acid}
D_H\nabla _xh=0\quad \text{on }\partial \tilde \Om,\ t>0.
\end{equation}

In previous works \cite{EHKS,EHS,EKS,Hillen,hillen-painter,HiSw} the quantities relating to the tissue, i.e. $q$ and (where applicable) $Q$ were assumed to be time-invariant, whereby $Q$ itself was estimated from the DTI data, as in \cite{EHKS,EHS,EKS,HS}. No tissue degradation was accounted for, and one could also include a proliferation term, e.g. like in \cite{EHS}. In this much simplified situation only coupling \eqref{macro-eq} with \eqref{eq:acid} and using the no-flux boundary conditions given above, the system takes the form of a Keller-Segel problem with some supplementary drift terms, all of which are linear in $\rho_0$. 

When the full evolution of the tissue becomes relevant, then \eqref{macro-eq} and \eqref{eq:acid} have to be supplementally coupled with the dynamics 
of $q$, as given by \eqref{eq:fiber-q-gen}, and with that of $Q$, given by \eqref{eq:Q-1}. The resulting (reaction-)diffusion-taxis system then characterizes cell, acidity, and tissue dynamics evolving on two scales (macroscopic and mesoscopic, respectively). Thereby, we deal with the macroscopic cell density in the first PDE, while \eqref{eq:fiber-q-gen} and \eqref{eq:Q-1} involve the mesoscopic quantity $p(t,x,v,y)$, which is inconvenient both for the analysis and the numerics.

Remark that, thanks to property \ref{new4} of $L_0$, the Hilbert expansion is equivalent to splitting $p$ as $p(t,x,v,y)=\bar p(t,x,y)M(t,x,v)+\eps M^\bot(t,x,v)$, with $\int _V M^\bot(t,x,v)dv=0$, i.e.  the Chapman-Enskog expansion used in  \cite{Hillen}. The leading order of the operator in \eqref{mean-projection}  leads to 
\begin{equation}\label{proj-undir-q}
\Pi _{a}[q](t,x,\theta )\simeq \left \{\begin{array}{cl}
\int _{\mathbb S^{N-1}}|\theta \cdot \theta '|q(t,x,\theta ')d\theta '&\quad\text{for undirected tissue}\\
\int _{\mathbb S^{N-1}}\theta \cdot \theta 'q(t,x,\theta ')d\theta '&\quad\text{for directed tissue,}
\end{array}\right .
\end{equation}
along with correspondingly rewriting \eqref{eq:A} as
\begin{equation}\label{rewrite-CE-eq:A}
A[q](t,x)=\int _{\mathbb S^{N-1}}\Pi _{a}[q](t,x,\theta )q(t,x,\theta )d\theta .
\end{equation}

Therewith we obtain from \eqref{eq:fiber-q-gen} the equation 
\begin{align}\label{rewrite-CE-eq:q}
\partial _tq(t,x,\theta )=r_D(h)\rho (t,x)q(t,x,\theta)\Big (\Pi _{a}[q](t,x,\theta )- A[q](t,x)\Big )
\end{align}
and from \eqref{eq:Q} 
\begin{align}\label{rewrite-CE-eq:Q}
\partial_tQ=r_D(h)\rho Q\left (A[q](t,x)-1\right ).
\end{align}

Thus, in order to determine the dynamics of tumor cells in interaction with the tissue they degrade and with the acidity they produce, we have to solve the macro-meso system
\begin{subequations}\label{macro-parabolic-full}
\begin{align}
& \partial_t\rho=\nabla _x\cdot \Big [\frac{1}{\lambda _0(t,x)}\Big (\nabla _x\cdot (\mathbb D_T\rho )
 -\lambda _1(t,x)g(Q,\lambda _0)\mathbb D_T\nabla Q\rho -\alpha S_1(t,x;q)\rho +\alpha S_2(t,x;h)\rho\Big )\Big ]\notag \\
& \hspace*{1cm}+\mu (\rho,h)Q\rho \label{cells-macro}\\
%&\partial _tq(t,x,\theta )=r_D(h)\Big (\Pi _{a}(t,x,\theta )-\int _{\mathbb S^{N-1}}\Pi _{a}(t,x,\theta )q(t,x,\theta )d\theta \Big )\rho (t,x)q(t,x,\theta) 
&\partial _tq(t,x,\theta )=r_D(h)\rho (t,x)q(t,x,\theta)\Big (\Pi _{a}[q](t,x,\theta )- A[q](t,x)\Big )
\label{tissue-meso}\\
%&\partial_tQ=r_D(h)\rho Q\int _{\mathbb S^{N-1}}(\Pi_{a} (t,x,\theta)-1)q(t,x,\theta )d\theta 
&\partial_tQ=r_D(h)\rho Q\left (A[q]-1\right )\label{tissue-macro}\\
%&h_t=D_H\Delta h+\frac{a\rho }{1+\rho}-b(1-\rho -n)(h-h_T)
&h_t=D_H\Delta h+\frac{a\rho }{1+\rho}-bQ(h-h_0)\label{acidity-macro}\\
%\partial_th(t,x)=D_H\Delta h+\frac{a\rho }{1+\rho}-bh\label{acidity-macro}\\
&\partial_tn=r_D(h)\rho Q\left (1-A [q]\right ) +\mathcal F(h,\rho, Q),
%&\cmg{\partial_tn(t,x)=r_D(h)\rho Q\int _{\mathbb S^{N-1}}(1-\Pi_{a} (t,x,\theta))q(t,x,\theta )d\theta +\mathcal F(h,\rho,Q),}
\label{nekrose-macro}
\end{align}
\end{subequations}
where $\mathbb D_T$ is as in \eqref{tumor-diffusion-tensor} with $\mathbb E_q=0$, with $S_1$ and $S_2$ as in \eqref{eqs:S}, $\Pi _{a}[q]$ as in \eqref{proj-undir-q}, $A[q]$ as in \eqref{rewrite-CE-eq:A}, $\mu $ as in \eqref{eq:choice-mu}, $\mathcal F$ as in \eqref{eq:choice-F-rond},  with boundary conditions \eqref{BC-macro-cells}, \eqref{BC-macro-acid}, and with given initial conditions $\rho(0,x)$, $q(0,x,\theta)$, $Q(0,x)$, $h(0,x)$, and $n(0,x)$.  These can be the tumor cell distribution (or an approximation of it) observed at diagnosis, 
the directional distribution of fibers obtained via DTI, some estimate of the macroscopic volume fraction of the tissue (e.g., most simply $FA$, as in \cite{CKSS,EHKS}), some (estimated) acidity distribution at diagnosis, and the necrotic tissue distribution, respectively. A tumor segmentation of the diagnosis image could be useful in assessing the latter.  

The mathematical handling of system \eqref{macro-parabolic-full} is nontrivial, both with respect to well posedness and numerics. 
The equations connect two modeling levels (macroscopic and mesoscopic) and the couplings via $q$ involved in the coefficients of all terms on the right hand side of \eqref{cells-macro} render the problem 
highly nonlinear. Moreover, the equation for $\rho$ features (along with the myopic diffusion) three types of taxis: 
\begin{itemize}
\item macroscopic haptotaxis towards $\nabla Q$, 
\item a new kind of mesoscopic haptotaxis (term with $S_1$), where the bias is actually given by $\nabla _xq^2$, and
\item pH-taxis, describing chemorepellence due to acidity (term with $S_2$).
\end{itemize}

\subsection{Moment closure for the fiber equation}
The system \eqref{macro-parabolic-full} is macroscopic with respect to the cell dynamics: 
by an asymptotic expansion of $p(t, x, v,y)$ around the local Maxwellian $q(t,x,\theta)$, we derived a model for the local cell density $\rho(t, x)$.
However, the tissue is modeled by the mesoscopic quantity $q(t,x,\theta)$. 
To reduce the level of detail in the tissue dynamics to match the rest of the model, we derive a low-order moment approximation to \eqref{tissue-meso}. 
For simplicity, we only consider the two-dimensional case. 

Recall the peanut distribution \cite{Hillen,Painter-Hillen}:
\begin{align*}
q(\theta) = \frac{1}{\ints{\theta\trans \DW \theta}} \theta\trans \DW \theta = \frac{1}{\pi \tr \DW} \theta\trans \DW \theta. 
\end{align*}
We interpret this as a $P_2$-approximation \cite{brunner2005two} \cite[III.5]{pomraning2005equations} under some additional constraints. 
A second-order basis of monomials is given by 
\begin{align*}
\basisFull = 
\begin{pmatrix}
\theta_x & \theta_y & \theta_x^2 & \theta_x\theta_y & \theta_y^2
\end{pmatrix}. 
\end{align*}
We drop the constant function from the basis, because it would introduce a linear dependence: $\theta_x^2 + \theta_y^2 = 1$. 
The $P_2$ ansatz is defined by the linear combination of basis functions
\begin{align*}
\mathfrak{q}(\theta) = \mplFull \cdot \basisFull, 
\end{align*} 
wherein $\mplFull$ is a vector of multipliers such that the moment constraints 
\begin{align*}
\momFull := \ints{\basisFull q} = \ints{\basisFull \mathfrak{q} } = \ints{\basisFull\basisFull\trans} \mathfrak{k}
\end{align*}
are fulfilled. 
Due to the symmetry constraints $\ints{\theta q} = 0$, we drop the first-order monomials $\theta_x, \theta_y$ from the basis. 
The remaining basis functions are then: 
\begin{align*}
\basis = \begin{pmatrix}
\theta_x^2 & \theta_x\theta_y & \theta_y^2
\end{pmatrix}. 
\end{align*}
With this choice it is easy to associate the multipliers $\mpl$ in the ansatz
\begin{align*}
\mathfrak{q} = \mpl \cdot \basis(\theta)
\end{align*}
with the components of the normalized water diffusion tensor: 
\begin{align*}
\mpl := \begin{pmatrix} \mplcomp[xx] & \mplcomp[xy] & \mplcomp[yy] \end{pmatrix} = \frac{1}{\pi \tr \DW }\begin{pmatrix} \DW[xx] &  2 \DW[xy] & \DW[yy] \end{pmatrix}. 
\end{align*}
Moreover, the moments $\mom := \ints{\basis q}$ are directly related to the components of the (normalized) tumor diffusion tensor $P = \ints{\theta\theta\trans q}$: 
\begin{align*}
\mom := \begin{pmatrix} \momcomp[xx] & \momcomp[xy] & \momcomp[yy] \end{pmatrix} = \begin{pmatrix} P_{xx} & P_{xy} & P_{yy} \end{pmatrix}. 
\end{align*}
We can translate multipliers to moments with the moment constraints $\mom = \ints{\basis \mathfrak{q}} = \ints{\basis \basis\trans} \mpl$. 
The transfer matrix is given by 
\begin{align*}
H &:= \ints{\basis\basis\trans} 
= \frac{\pi}{4}
\begin{pmatrix}
3 & 0 & 1 \\ 
0 & 1 & 0 \\
1 & 0 & 3
\end{pmatrix}. 
\end{align*} 
With these tools in hand, we derive a $P_2$-approximation of the tissue dynamics \eqref{tissue-meso} through the following steps: 
Insert the ansatz $\mathfrak{q}$ into \eqref{tissue-meso}, multiply by the basis $\basis$ and integrate over $\mathbb{S}^1$ to obtain
\begin{align}
\label{eq:tissue-moments-2d}
\partial_t \mom &= r_D(h) \rho \left( \ints{\basis \mathfrak{q} \Pi\left[\mathfrak{q}\right] } - \mom A\left[\mathfrak{q}\right] \right).
\end{align}
It remains to calculate the moments $\ints{\basis \mathfrak{q} \Pi\left[\mathfrak{q}\right] }$. 
Inserting the definitions of $\mathfrak{q}$ and $\Pi$, we obtain
\begin{align*}
\ints{\basis \mathfrak{q} \Pi\left[\mathfrak{q}\right] } = \int_{\mathbb S^{1}} \int_{\mathbb S^{1}} \basis(\theta) \mpl \cdot \basis(\theta) \mpl \cdot \basis(\theta') |\theta \cdot \theta'| \dd \theta' \dd \theta. 
\end{align*}
With the tensor 
\begin{align*}
\mathfrak{P}_{k, ij} = \int_{\mathbb S^{1}} \int_{\mathbb S^{1}} \basiscomp[k](\theta) \basiscomp[i](\theta) \basiscomp[j](\theta') |\theta \cdot \theta'| \dd \theta' \dd \theta
\end{align*}
we can write the $k$-th component $\ints{\basis \mathfrak{q} \Pi\left[\mathfrak{q}\right] }_k$ as 
\begin{align*}
\ints{\basis \mathfrak{q} \Pi\left[\mathfrak{q}\right] }_k = \mpl\trans \mathfrak{P}_k \mpl = \mom\trans \left(H\trans[-1] \mathfrak{P}_k H^{-1}\right) \mom. 
\end{align*}
Finally, we obtain $A\left[\mathfrak{q}\right]$ from the identity $ \theta_x^2 + \theta_y^2 = 1$: 
\begin{align*}
A\left[\mathfrak{q}\right] = \ints{1 \mathfrak{q} \Pi\left[\mathfrak{q}\right] } = \ints{(\theta_x^2 + \theta_y^2) \mathfrak{q} \Pi\left[\mathfrak{q}\right] }
= \ints{\basis \mathfrak{q} \Pi\left[\mathfrak{q}\right] }_{xx} + \ints{\basis \mathfrak{q} \Pi\left[\mathfrak{q}\right] }_{yy}. 
\end{align*}
The tensor $\mathfrak{P}_{k, ij}$ can be precomputed once and for all with a high-order quadrature.
To evaluate each component of the right-hand side of \eqref{eq:tissue-moments-2d} we need to compute only a quadratic form at run time.
Note that the normalization $\ints{q} = 1$ results in the loss of an additional degree of freedom. 
It holds $\tr P = 1$, therefore we can reconstruct $\momcomp[yy] = 1 - \momcomp[xx]$ and only need the evolve the two moments $\momcomp[xx], \momcomp[xy]$ in \eqref{eq:tissue-moments-2d}. 
\paragraph{Remark:}
\textit{In three dimensions, the previous considerations are completely analogous for the basis 
	$\basis = \begin{pmatrix} \theta_{x}^2 & \theta_{y}^2 & \theta_{z}^2 &  \theta_{x}\theta_{y} & \theta_{x}\theta_{z} & \theta_{y}\theta_{z} \end{pmatrix}$. }

\subsection{Hyperbolic scaling, directed tissue} 
Here we also aim to investigate the effect of reducing diffusivity that might not be experimentally consistent. In order to achieve this objective and for the sake of completeness, we propose in this subsection to perform a hyperbolic limit that will provide us with a macroscopic vision in which the terms of transport and potentials win the battle over diffusion. We will take advantage of most of the calculations in Subsection \ref{sec:par-scaling},  that we will omit in part so as not to be repetitive. Finally, in order to try to model a series of effects of a priori minor influence according to the experiments, not contemplated in the model variables, we carry out a "small" extension to the second order of the hyperbolic expansion and compare this double development on the scale with the parabolic approach.

In the following we address a version of \eqref{KinEq2} where we neglect the dependence of the turning rate on the subcellular dynamics, i.e. we consider it to be of the form $\lambda (x)=\frac{\kappa }{FA(x)+1}$ %. The consequence of this will be the alltogether exclusion of receptor binding dynamics from the system after integrating the kinetic transport equation w.r.t. the subcellular variable and correspondingly no term   
and are interested in the situation of directed tissue. 

%\subsection{Deducing the macroscopic drift PDE with diffusion-drift correction  for glioma cell density}

We thus consider the kinetic transport equation
\begin{align}\label{KinEq-hyperb}
\partial_t p &+  v \esc \nabla_x p - \partial_z \Big (((k_+Q+k_-)z+f'(Q)(Q_t+v\cdot \nabla Q))p\Big ) + \alpha  \div_v \Big (\mathbb B(v) (\nabla_x q(\hat v)-\nabla _xh) p\Big )\notag \\ 
&= \lambda (x) L_0(p)+\mathcal P(p,Q,h,\rho),%+ a \lambda(x,y) L_1[\nabla_x q](p),
\end{align}
with $L_0(p)=\bar p(t,x,z) \frac{q(x,\hat v)}{\omega } - p(t,x,v,z)$ having the same properties as in Subsection \ref{sec:par-scaling} and perform a hyperbolic scaling $t\to \eps t$, $x\to \eps x$, while the proliferation term is rescaled as previously with $\eps^2$. This leads to 
\begin{align}\label{KinEq-hyperb1}
\eps \partial_t p &+  \eps v \esc \nabla_x p - \partial_z \Big (((k_+Q+k_-)z+\eps f'(Q)(Q_t+v\cdot \nabla Q))p\Big ) + \alpha  \eps \div_v \Big (\mathbb B(v) (\nabla_x q(\hat v)-\nabla _xh) p\Big )\notag \\ 
&= \lambda (x) L_0(p)+\eps^2\mathcal P(p,Q,h,\rho).
\end{align}

We now consider a Chapman-Enskog expansion (equivalent here to the Hilbert one, as stated in Subsection \ref{sec:par-scaling}), i.e., a decomposition of $p$ into a $Ker(L_0)$-component and a $Ker(L_0)^\bot$-part, as follows:
%Consider on $L^2(\frac{dv}{\frac{q}{\om }})$ a Chapman-Enskog expansion: 
\begin{equation}\label{chap-ensk}
p(t,x,v,z)=\bar p(t,x,z)\frac{q(t,x,\hat v)}{\omega }+\eps p^\perp(t,x,v,z),
\end{equation}
where $p^\perp\in \left <\frac{q}{\omega}\right >^\perp$ verifies $\int _Vp^\perp(t,x,v,z)dv=0$. This decomposition leads to the corresponding expansion  for the moments of $p$, in particular: $m(t,x,v)=\rho (t,x) \frac{q(t,x,\hat v)}{\omega }+\eps m^\perp(t,x,v,z)$, and then, integrating
\eqref{KinEq-hyperb1} w.r.t. $z\in Z$ %and use the notation $m(t,x,v):=\int _Zp(t,x,v,z)dz$ and the above expansion giving $m(t,x,v)=\rho (t,x) \frac{q(t,x,\hat v)}{\omega }+\eps m^\perp(t,x,v)$ 
we obtain 
\begin{align}\label{KinEq-hyperb2}
%%\eps m_t+\eps v\cdot \nabla _xm+\alpha \eps \nabla_v\cdot (\mathbb B(v) (\nabla_x q(\hat v)-\nabla _xh)m)
\eps \partial_t\rho \frac{q}{\omega}+\eps \rho \partial_t(\frac{q}{\omega})+\eps ^2\partial_tm^\perp +\eps v\cdot \nabla_x(\rho \frac{q}{\omega})+\eps ^2v\cdot \nabla _xm^\perp+&\alpha \eps \nabla_v\cdot \Big (\mathbb B(v) (\nabla_x q(\hat v)-\nabla _xh)m\Big )\notag \\
%&=\lambda(x)L_0(m) \notag \\
&=\eps \lambda(x)L_0(m^\perp )+\eps^2\mu(\rho,h)Qm.
\end{align}

Integrate w.r.t. $v$ and divide by $\eps $ to get
\begin{equation}\label{KinEq-hyperb3}
\rho _t+\nabla _x\cdot (\rho \tilde {\mathbb E}_q+\eps \int _Vvm^\perp dv)=\eps \rho \mu (\rho,h)Q,
\end{equation}
where as before $\tilde {\mathbb E}_q(t,x)=\int _Vv\frac{q(t, x,\hat v)}{\omega }dv=s\mathbb E_q(t,x)$. 

Clearly \eqref{KinEq-hyperb3} is drift-dominated; however,  it is worth computing the ${\cal O}(\eps)$ correction with respect to the pure drift. From \eqref{KinEq-hyperb2} and \eqref{KinEq-hyperb3} follows
\begin{align}\label{KinEq-hyperb4}
\lambda (x)L_0(m^\perp)&=\eps \frac{q}{\omega}\rho \mu (\rho,h)Q-\frac{q}{\omega}\nabla_x\cdot \Big (\rho \tilde {\mathbb E}_q+\eps \int _Vvm^\perp dv\Big )+\rho \frac{\partial_tq}{\omega}+\eps \partial_tm^\perp +v\cdot \nabla _x(\rho \frac{q}{\omega}+\eps m^\perp)\notag \\
+&\alpha \nabla_v\cdot \Big (\mathbb B(v) (\nabla_x q(\hat v)-\nabla_xh)m\Big )-\eps\rho \mu (\rho,h)Q .
\end{align}

Now observe that the integral w.r.t. $v$ of the right hand side in \eqref{KinEq-hyperb4} vanishes, so that we can take the (pseudo)inverse of $L_0$ giving (at leading order)
\begin{align}
m^\perp\simeq &-\frac{1}{\lambda (x)}\Big [\frac{q}{\omega}(v-\tilde {\mathbb E}_q)\cdot \nabla \rho +\rho \Big (v\cdot \nabla _x\frac{q}{\omega}-\frac{q}{\omega}\nabla_x \cdot \tilde {\mathbb E}_q\Big )+\rho \frac{\partial_tq}{\omega} \\ &+\alpha \nabla _v\cdot\Big (\mathbb B(v) (\nabla_x q(\hat v)-\nabla_xh)m\Big ) \Big ],
\end{align}
hence 
\begin{align}\label{vm-orthog}
\int _Vvm^\perp dv&\simeq -\frac{1}{\lambda (x)}\Big (\int _Vv\otimes (v-\tilde{\mathbb E}_q)\frac{q}{\omega}dv\ \nabla \rho \\ &+\rho\Big (\nabla_x \cdot \int_Vv\otimes v\frac{q}{\omega}dv -\tilde{\mathbb E}_q\nabla_x \cdot \tilde{\mathbb E}_q\Big )+\rho \partial_t\tilde{\mathbb E}_q-\alpha \rho \Sigma (t,x)\Big )\notag \\
%&=-\frac{1}{\lambda (x)}\Big (\mathbb D_T \nabla_x \rho+\rho\nabla_x \cdot (\mathbb D_T-\tilde{\mathbb E}_q\otimes \tilde{\mathbb E}_q)-\rho \tilde{\mathbb E}_q\nabla_x \cdot \tilde{\mathbb E}_q\cb{+\rho \partial_t\tilde{\mathbb E}_q}-\alpha \rho \Sigma (t,x)\Big )\notag\\
&=-\frac{1}{\lambda (x)}\Big (\nabla _x\cdot (\mathbb D_T\rho)-\rho \tilde{\mathbb E}_q\nabla_x\cdot \tilde{\mathbb E}_q+\rho \partial_t\tilde{\mathbb E}_q-\alpha \rho \Sigma (t,x)\Big ),
\end{align}
where $\mathbb D_T(t,x)$ and $\Sigma (t,x):=S_1(t,x;q)-S_2(t,x;h)$ are as introduced in Section \ref{sec:par-scaling} and we observe that
\begin{align*}
\int _Vv\otimes (v-\tilde{\mathbb E}_q)\frac{q}{\omega}dv=\int _V(v-\tilde{\mathbb E}_q)\otimes (v-\tilde{\mathbb E}_q)\frac{q}{\omega}dv=\mathbb D_T.
\end{align*}

Together with \eqref{KinEq-hyperb3} this leads to the macroscopic PDE
\begin{equation}\label{macro-hypscale}
\rho_t+\nabla \cdot (\rho\tilde{\mathbb E}_q)=\eps \nabla \cdot \Bigg (\frac{1}{\lambda (x)}\Big (\nabla \cdot (\rho \mathbb D_T(t,x))+\rho(\partial_t\tilde{\mathbb E}_q-\tilde{\mathbb E}_q \nabla \cdot \tilde{\mathbb E}_q-\alpha \Sigma(t,x))\Big )\Bigg )+\eps \rho \mu (\rho,h)Q,
\end{equation}
which is drift-dominated. Notice that for $\mathbb E_q=0$, i.e. if the tissue is undirected, the correction term in \eqref{macro-hypscale} has the same form as the right hand side of \eqref{macro-eq} - except for the middle term therein, which got lost when integrating \eqref{KinEq-hyperb1} w.r.t. $z$. In fact, it was the assumption of the turning rate not depending on $z$ which effaced the whole influence of subcellular dynamics. The effect of this is not having the macroscopic haptotaxis term in the $\varepsilon$-correction on the right hand side. Nevertheless we still get the pH-taxis and (mesoscopic) haptotaxis correction terms contained in 
$$\varepsilon \nabla \cdot  \Big (\frac{\alpha}{\lambda (x)}\rho \Sigma (t,x)\Big )=\varepsilon \nabla \cdot  \Big (\frac{\alpha}{\lambda (x)}\rho (S_1(t,x;q)-S_2(t,x;h))\Big ).$$

Coupling \eqref{macro-hypscale} with the equations \eqref{tissue-meso} and \eqref{acidity-macro} describing tissue and acidity evolution, respectively, leads to challenges similar to those in Section \ref{sec:par-scaling}. 

We need to prescribe boundary conditions. To this aim we consider again a bounded, sufficiently smooth domain $\tilde \Om=\varepsilon \Om  \subset \R^N$ and start from a mesoscopic no-flux condition of the form \eqref{BC-plaza}. Using the Chapman-Enskog expansion \eqref{chap-ensk} of $p(t,x,v,z)$ we rewrite this condition as
\begin{align*}
\Big (\rho(t,x)\tilde{\mathbb E}_q(x)+\varepsilon \int _Z\int _Vvp^\perp(t,x,v,z)dvdz\Big )\cdot \nu (x)=0,\qquad \text{for all }x\in \partial \tilde \Om,\ t>0,
\end{align*}
which in virtue of \eqref{vm-orthog} leads to 
\begin{align}
\Big (-\rho\tilde{\mathbb E}_q+\frac{\varepsilon}{\lambda (x)}\Big (\nabla \cdot (\rho \mathbb D_T(t,x))+\rho(\partial_t\tilde{\mathbb E}_q-\tilde{\mathbb E}_q \nabla \cdot\tilde{\mathbb E}_q-\alpha \Sigma(t,x))\Big )\Big )\cdot \nu (x)=0,\quad x\in \partial \tilde \Om,\ t>0,
\end{align}
meaning that the normal macroscopic flux and its $\varepsilon $-correction vanish on the spatial boundary.\\

Thus, in the case with directed tissue and when no subcellular dynamics are taken into account, the macro-meso system to be solved becomes 

\begin{subequations}\label{macro-hyperbolic-full}
\begin{align}	
&\rho_t+s\nabla \cdot (\rho\mathbb E_q)=\eps \nabla \cdot \Bigg (\frac{s^2}{\lambda (x)}\Big (\nabla \cdot (\rho \mathbb V_q(t,x))+\rho(\frac{1}{s}\partial_t\mathbb E_q-\mathbb E_q \nabla\cdot  \mathbb E_q-\frac{\alpha}{s^2} \Sigma(t,x))\Big )\Bigg )+\eps \rho \mu (\rho,h)Q\label{macro-hyperbolic-full-cells}\\
&h_t=D_H\Delta h+\frac{a\rho }{1+\rho}-bQ(h-h_0)\label{macro-hyperbolic-full-acid}\\
%&\partial _tq(t,x,\theta )=r_D(h)\rho (t,x)q(t,x,\theta )\left (\Pi[p](t,x,\theta)-\frac{1}{s}\mathbb E_q(t,x)\cdot U(t,x)\right )
&\partial _tq(t,x,\theta )=r_D(h)\rho (t,x) q(t,x,\theta )\left (\Pi[p](t,x,\theta)-A[p](t,x)\right ),\label{macro-hyperbolic-full-tissue-q}\\
&\partial_tQ=r_D(h)\rho Q(A[p]-1)\label{macro-hyperbolic-full-tissue-Q}\\
&\partial_tn=r_D(h)\rho Q\left (1-A [p]\right ) +\mathcal F(h,\rho, Q),\label{macro-hyperbolic-full-necrotic}
\end{align}
\end{subequations}
with $\Pi[p]$ and $A[p]$ as given in \eqref{mean-projection} and \eqref{eq:A}, respectively. Observe that upon assuming the tisue to be undirected (i.e. that $\mathbb E_q=0$), \eqref{macro-hyperbolic-full-cells} takes the form of \eqref{cells-macro}, which was obtained by parabolic scaling, with the difference of the right hand side of the glioma dynamics being scaled here by $\eps$.

As for the parabolic limit, in \eqref{macro-hyperbolic-full} the dependence of the featured operators on the mesoscopic cell distribution function $p$ is inconvenient, so we use the same approach as in Subsection \ref{BC-approx-projections} to approximate $\Pi[p]$ and $A[p]$ by \eqref{proj-undir-q} and \eqref{rewrite-CE-eq:A}, respectively. Notice that in the present case (directed tissue) we can rewrite 
\begin{align*}
\Pi_a[q](t,x,\theta )&=\theta \cdot \mathbb E_q(t,x)\\
A[q](t,x)&=\mathbb E_q(t,x)\cdot \mathbb E_q(t,x)
\end{align*}

Consequently, \eqref{macro-hyperbolic-full-tissue-q} can be replaced by
\begin{align*}
\partial _tq(t,x,\theta )=r_D(h)\rho (t,x) q(t,x,\theta )(\theta \cdot \mathbb E_q(t,x)-\mathbb E_q^2(t,x)),
\end{align*}
hence we can compute
\begin{align*}
\partial_t\mathbb E_q=r_D(h)\rho (\mathbb V_q+\mathbb E_q\otimes \mathbb E_q-\mathbb E_q^2\mathbb I_N)\cdot \mathbb E_q
\end{align*}
and plug it into \eqref{macro-hyperbolic-full-cells}. Let us denote 
\begin{align*}
\mathbb T_q:=\frac{1}{s}r_D(h)\rho (\mathbb V_q+\mathbb E_q\otimes \mathbb E_q-\mathbb E_q^2\mathbb I_N)\cdot \mathbb E_q-\mathbb E_q\nabla \cdot \mathbb E_q.
\end{align*}

Then altogether the system becomes
\begin{subequations}\label{macro-hyperbolic-full-neu}
	\begin{align}	
&\rho_t+\nabla \cdot (\rho\mathbb E_q)=\eps \nabla \cdot \Bigg (\frac{s^2}{\lambda (x)}\Big (\nabla \cdot (\rho \mathbb V_q(t,x))+\rho(\mathbb T_q-\frac{\alpha}{s^2} \Sigma(t,x))\Big )\Bigg )+\eps \rho \mu (\rho,h)Q\label{macro-hyperbolic-full-cells-neu}\\
&h_t=D_H\Delta h+\frac{a\rho }{1+\rho}-bQ(h-h_0)\label{macro-hyperbolic-full-acid-neu}\\
&\partial _tq=r_D(h)\rho  q(\theta \cdot \mathbb E_q-\mathbb E_q^2)\label{macro-hyperbolic-full-tissue-q-neu}\\
&\partial_tQ=r_D(h)\rho Q(\mathbb E_q^2-1)\label{macro-hyperbolic-full-tissue-Q-neu}\\
&\partial_tn=r_D(h)\rho Q\left (1-\mathbb E_q^2\right ) +\mathcal F(h,\rho, Q),\label{macro-hyperbolic-full-necrotic-neu}
\end{align}
\end{subequations}
with $\theta \in \mathbb S^{N-1}$, $t>0$, and $x\in\tilde \Om $. The functions $\mu(\rho,h)$ and $\mathcal F(h,\rho, Q)$ are still those given in \eqref{eq:choice-mu} and \eqref{eq:choice-F-rond}, respectively. The equations are completed by initial conditions as in Subsection \ref{BC-approx-projections} and by the no-flux boundary conditions $\nabla h\cdot \nu=0$ on $\partial \tilde \Om$ and 
\begin{align}
\Big (-s\rho\mathbb E_q+\frac{s^2\varepsilon}{\lambda (x)}\Big (\nabla \cdot (\rho \mathbb V_q(t,x))+\rho(\mathbb T_q(t,x)-\frac{\alpha}{s^2} \Sigma(t,x))\Big )\Big )\cdot \nu (x)=0,\quad x\in \partial \tilde \Om,\ t>0.
\end{align}

\section{Numerics}\label{sec:numerics}
In this section the equations obtained by parabolic upscaling are solved numerically. Simulations are conducted for three scenarios of increasing complexity. 
In each scenario we describe a set of assumptions simplifying the equations. This allows for directly investigating the relative importance of the different effects included in our model. The results are presented in Subsection \ref{subsec:numerical results}. For convenience we state in brevity the exact macroscopic systems of PDEs for the considered scenarios:
\begin{enumerate}
	\item[\textbf{Scenario 1:}]
We assume that neither $q$ nor $Q$ depend on time, thus can be assessed from the DTI data. We follow previous works, starting from a pre-assigned directional distribution $q$ (peanut) and independently estimate $Q$ using the approach in \cite{EHS,Hunt,HS} relying on Brownian motion description of water molecule  diffusion.\footnote{Alternative, more precise ways to assess the (macroscopic and/or mesoscopic) tissue density from DTI data would be to perform density estimation by using statistical or variational methods, see e.g. \cite{descoteaux,FlJo,kim-rich,wong}. Here we do not address this issue, as our focus is on providing a continuous approach to modeling glioma growth and spread which is amenable to efficient computations and at the same time able to capture lower scale effects.} The system then reduces to \eqref{cells-macro}, \eqref{acidity-macro}, merely featuring myopic diffusion, pH-taxis, and transport terms with drift velocities computed from the data, together with the necrosis dynamics \eqref{nekrose-macro} with $r_D(h)=0$, i.e.
	
	\begin{subequations}
		\begin{align}
		\partial_t \rho =&\nabla _x\cdot \Big [\frac{1}{\lambda _0(x)}\Big (\nabla _x\cdot (\mathbb D_T\rho )
		-\lambda_1(x)g(Q,\lambda _0)\mathbb D_T\nabla Q\rho - \alpha S_1(t,x;q)\rho + \frac{\alpha}{h_0} S_2(t,x;h)\rho\Big )\Big ]\nonumber\\
		&+Q  \rho  \eta  (1-\rho-n)\frac{(h_T-h)_+}{h_0} -\gamma \frac{h}{h_0} \rho Q \\
		\partial_t h 	=&D_H
		\Delta h+\frac{a\rho }{1+\rho}-bQ(h-h_0)\\
		\partial_t n 	=& \mathcal F(h,\rho, Q).
		\end{align}
	\end{subequations}
	
\item[\textbf{Scenario 2:}]
 $Q$ evolves in time, meaning that the tissue is degrading irrespective of the fibre orientations (hence the anisotropy does not play any role in the degradation). Here $A[q]$ is zero. This corresponds to an indirect depletion, i.e. not upon contact with glioma cells, but rather through the acidity they produce. Consequently, the macroscopic system \eqref{macro-parabolic-full} simplifies to 
\begin{subequations}
	\begin{align}
	\partial_t \rho =&\nabla _x\cdot \Big [\frac{1}{\lambda _0(x)}\Big (\nabla _x\cdot (\mathbb D_T\rho )
	-\lambda_1(x)g(Q,\lambda _0)\mathbb D_T\nabla Q\rho - \alpha S_1(t,x;q)\rho + \alpha S_2(t,x;h)\rho\Big )\Big ]\nonumber\\
	&+Q  \rho  \eta  (1-\rho-n)\frac{(h_T-h)_+}{h_0} -\gamma \frac{h}{h_0} \rho Q \\
	\partial_t Q    =&   -r_D(h)\rho Q\\
	\partial_t h 	=&D_H\Delta h+\frac{a \rho}{1+\rho}-bQ (h-h_0)\\
	\partial_t n	=&r_D(h)\rho Q + \mathcal F(h,\rho, Q).
	\end{align}
\end{subequations}
\item[\textbf{Scenario 3:}]
Both $q$ and $Q$ evolve and the full system \eqref{macro-parabolic-full} is solved.
\end{enumerate}

\subsection{Numerical discretization}
The equations were discretized by the method of lines approach in all three cases.
For the spatial discretization we used a vertex centered finite volume method. The
mesh is naturally structured because of the underlying medical dataset.
The mesh width $h=2mm$ is imposed by the particular DTI dataset,
 which is publicly available \cite{camino}. 
The reconstructed diffusion tensors were averaged with a component-wise arithmetic mean at the control volume interfaces.
 The advective fluxes were approximated by a first order upwind discretization. The no-flux barrier between the brain
and the background were implemented by suppressing flux assembly at the relevant cell interfaces\footnote{here we mean by 'cell' the numerical discretization element and not the biological entity}.  
The integration of the reaction terms coupling the evolving fields was performed by a  first order midpoint rule at the cell centers.
 As an initial condition we used a small Gau\ss ian peak as an approximation of the tumor and took for the spatial distribution of the acidity field $h$ the uniform steady-state of \eqref{acidity-macro}. % the coupling term $h_{ic} = \frac{a}{b}(\frac{\rho}{1+\rho})+h_0$. 
The initial condition for necrotic matter was simply $n=0$. 
The time discretization was realized by an implicit Euler scheme. The resulting
nonlinear system was solved via Newton iteration which internally used a BiCGSTAB solver preconditioned via AMG with an SSOR smoother \cite{BiCGSTAB}.

\subsubsection{Software}
The implementation was performed within the DUNE software framework \cite{BBD,BBD2},
 specifically with the dune-pdelab discretization module\cite{dune24:16,dune-pdelab}. 
The BiCGSTAB solver and the AMG were provided by the iterative solver template library dune-istl \cite{dune-istl}.

\subsection{Choice of simulation parameters and coefficient functions}
In Table \ref{tab:parameters} we present the parameters used throughout all numerical simulations, their meaning, and their respective sources. We can also assess the parabolic scaling parameter 
\begin{equation*}
\eps =\frac{s}{\kappa X}\simeq 2\cdot 10^{-6},
\end{equation*}
where we use $X=0.06\ m$ as a reference length, corresponding to the side length of he plots in e.g., Figure \ref{fig:scenario3-fields}). 
This value indicates that the parabolic limit is an adequate approximation to the kinetic model for the chosen parameters.

 Let us also highlight here the role of the scaling parameter $ \alpha $. It is influenced by the position gradients of $q$ and $h$ (involved in $ \Sigma $) and hence should be a large quantity, say of the order $\frac{1}{\gamma} \eps^ {- \gamma} $ for some $ \gamma> 0 $. This is similar to what happens when we perform hyperbolic scaling in a fairly similar kinetic framework (see \cite {BCNS,PS}), where hydrodynamic limits with dominant drift are obtained, but where the small diffusion correction of a non-diffusive limit can be made explicit by displaying the dependency of $\eps $ on all scaling parameters. In the present context $\gamma $ should be a parameter related to smaller scales, such as the microtubule (MT) extension zones that are responsible for the biochemical and biomechanical exchange of cells with their environment \cite{Gradilla,Tom}.  The length of microtubules is between $5$ and $8$ cell units \cite{Portela}, whereas a glioma cell has a diameter of ca. $15 \mu m = 15\times 10^{-3 } mm$ \cite{bionumbers}. This means that the regions of MTs that would correspond to the tumor front have a length around $100 \mu m = 0.1mm$. Then, the parameter $\gamma$ will represent the ratio of such MT area  with respect to the whole tumor; here we take $\gamma = 0.001$. 
Therefore, and in view of the above choice of $\eps$, the parameter $ \alpha $ is taken here of the order of $10^3$.

The coefficient functions involved in the simulations are summarized below:
\begin{equation*}
\begin{split}
& \text{For all scenarios:}\quad \mathcal F(h,\rho, Q)=\gamma \frac{h}{h_0}\rho Q,\qquad \\[1ex]
& \lambda_0=\frac{\kappa f(Q)}{FA+f(Q)},\qquad \lambda_1=\frac{\kappa FA}{(FA+f(Q))^2},\qquad f(Q)=\frac{k_+Q}{k_+ Q + k_-}\\[1ex]
&f'(Q)=\frac{k_+ k_-}{(k_+  Q + k_-)^2},\qquad g(Q,\lambda _0)=\frac{f'(Q)}{k_+Q+k_- + \lambda _0},\qquad r_D(h)=r_0\left (\frac{h}{h_0}-1\right )_+.	\\[1ex]
& \text{For Scenario 1:}\quad Q(x)= 1-\Big(\frac{tr(\mathbb{D_W}(x))}{\lambda_{max}}\Big)^{\frac{3}{2}},\qquad  \mathbb D_T(x)=\frac{s^2}{(N+2)} \Big( \mathbb I_N + 2 \frac{\mathbb{D_W}(x)}{tr(\mathbb{D_W}(x))} \Big),
\end{split}
\end{equation*}
with $\lambda_{max}$ denoting the maximum eigenvalue of the water diffusion tensor $\mathbb D_W$.

\begin{table}[h]
	\small
	\begin{tabular}{c|l|l|l}
		
		\textbf{parameter} 	&\textbf{value}									& \textbf{meaning}							& \textbf{source}\\
		\hline
		$s$					& 50 ($\mu $m/h) ($=$ 1.389e-8 m/s)		& cell speed								& \cite{Milo}	\\
		$\alpha$			&$10^3$ ($1$)							& weight for advective terms				& estimated\\
		$\kappa$			&0.1	(1/s)								    & maximum turning frequency					& estimated			\\					
		$h_0$				&1.0e-7.2		(mol/l)							& healthy acidity value  					& \cite{vaupel}			\\ 
		$h_T$				&1.995e-07	(mol/l)						& threshold of possible proliferation   	& \cite{vaupel}			\\
		$D_H$				&$0.5e-03 \  (mm^2/s)$					& diffusion coefficient of protons			& \cite{lide}			\\
		$a$					&$2.2e-17$\ (mol\ $cm^3 (cells\cdot sec)^{-1}$) & acid production rate 		    	& \cite{martin}			\\
		$b$					&0.8e-04		(1/sec)	 				& proton buffering by healthy tissue	& \cite{Ma-Ga}			\\
		$\eta$				&0.26e-06 (1/sec) &  tumor proliferation rate										& \cite{mercap}				\\
		$k_{+}$				&0.034 (1/sec)								& attachment rate of tumor cells to tissue &\cite{lauffi}				\\
		$k_{-}$				&0.01 (1/sec)								& cell detachment rate 				& \cite{lauffi}				\\
		$\gamma$			&5e-08 	(1/sec)								& acid-induced death rate of tumor cells &	\cite{swanson}				\\
	%	$\omega$			&$s^{N-1}$								& normalization factor in $q$			& 			\\
		$r_0$				& 1.0e-6 (1/(mol\ sec))							& efficiency of fiber degradation											& estimated\\
		\hline
			\end{tabular}
	\caption{Full set of parameters for the numerical simulations}\label{tab:parameters}
\end{table}

\subsection{Numerical results} \label{subsec:numerical results}
We focus the presentation of the numerical results on those aspects
 where the new model and its extension, in particular by including tissue dynamics, produce new qualitative effects. 
Computations were conducted with a simulated time of $52$ weeks. 
We first present line plots from the coordinate origin through the tumor to the upper right of the domain in order to visualize and discuss the evolution of the involved fields.
 This helps to establish the general dynamics of the model in Subsection  \ref{subsubsec:dynamic}. 
 We present the final states of the involved fields in \ref{subsubsec:states} and investigate in Subsection \ref{subsubsec:dominant_effects} which advective effect is playing a dominant role with respect to the tumor extent and appearance. 
  We then directly compare the solutions of the three considered 
  scenarios in Subsection \ref{subsubsec:compare_scenarios}, and address in Subsection \ref{subsubsec:tumor_grading} the grading of a tumor upon relying on the numerical results, particularly with respect to the amount of necrotic matter. 
  We also suggest in Subsection \ref{subsubsec:dose_painting} how the simulations of the model could be used to  
  perform dose painting for radiation treatment.
  
\subsubsection{Scenario 3: overall dynamics}\label{subsubsec:dynamic}
\begin{figure}[hpt!]
	\def\relincludepath{data/postprocessing/}
	%\documentclass[crop=false]{standalone}
%
%\def\relincludepath{./}
%\input{figuremacros.tex}
%
%
%\begin{document}

\pgfplotsset{lineoutdefault/.style={
		scale only axis,
		enlargelimits=false,
		scale only axis,
		width=\figurewidth,
		height=0.2\figurewidth,
		title style = {font=\figurelabelfont{\figurewidth}},
		xlabel style = {font=\figurelabelfont{\figurewidth}}, 
		ylabel style = {font=\figurelabelfont{\figurewidth}}, 
		xticklabel style={font=\figureticklabelfont{\figurewidth}},
		yticklabel style={font=\figureticklabelfont{\figurewidth}},
		xticklabel=\pgfmathparse{100*\tick)}\pgfmathprintnumber{\pgfmathresult},
		xlabel={$x / cm$},
}}

\setdefaultfigurelayout{onecol}{
	\begin{tikzpicture}
		\begin{axis}[%
		lineoutdefault,
		ymin=0, ymax=1,
		title={$0$ weeks},
		legend style={at={(0.,1.05)},
			anchor=south west,
			font=\figureticklabelfont{\figurewidth}
		},
		legend columns=4,
		]
		\addplot[line width=1pt, red] table[x=x, y=z, col sep=comma]{\relincludepath Scenario3_lineout_AcidityCompareScenarios_3_u_00000.csv};
		\addlegendentry{$\rho$}
		\addplot[line width=1pt, blue] table[x=x, y=z, col sep=comma]{\relincludepath Scenario3_lineout_AcidityCompareScenarios_3_n_00000.csv};
		\addlegendentry{$n$}
		\addplot[line width=1pt, black] table[x=x, y=z, col sep=comma]{\relincludepath Scenario3_lineout_AcidityCompareScenarios_3_Q_00000.csv};
		\addlegendentry{$Q$}
		\end{axis}
		\begin{axis}[%
		lineoutdefault,
		axis y line*=right,
		legend style={at={(1.,1.05)},
			anchor=south east,
			font=\figureticklabelfont{\figurewidth}
		},
		ylabel={pH},
		]
		\addplot[line width=1pt, green] table[x=x, y expr=(-ln(\thisrowno{4})/ln(10)), col sep=comma]{\relincludepath Scenario3_lineout_AcidityCompareScenarios_3_h_00052.csv};
		\addlegendentry{$h$}
		\end{axis}
	\end{tikzpicture}
	
	\begin{tikzpicture}
		\begin{axis}[%
		lineoutdefault,
		ymin=0, ymax=1,
		title={$26$ weeks},
		]
		\addplot[line width=1pt, red] table[x=x, y=z, col sep=comma]{\relincludepath Scenario3_lineout_AcidityCompareScenarios_3_u_00052.csv};
		\addplot[line width=1pt, blue] table[x=x, y=z, col sep=comma]{\relincludepath Scenario3_lineout_AcidityCompareScenarios_3_n_00052.csv};
		\addplot[line width=1pt, black] table[x=x, y=z, col sep=comma]{\relincludepath Scenario3_lineout_AcidityCompareScenarios_3_Q_00052.csv};
		\end{axis}
		\begin{axis}[%
		lineoutdefault,
		axis y line*=right,
		ylabel={pH},
		]
		\addplot[line width=1pt, green] table[x=x, y expr=(-ln(\thisrowno{4})/ln(10)), col sep=comma]{\relincludepath Scenario3_lineout_AcidityCompareScenarios_3_h_00052.csv};
		\end{axis}
	\end{tikzpicture}
	
	\begin{tikzpicture}
		\begin{axis}[%
		lineoutdefault,
		ymin=0, ymax=1,
		title={$52$ weeks},
		]
		\addplot[line width=1pt, red] table[x=x, y=z, col sep=comma]{\relincludepath Scenario3_lineout_AcidityCompareScenarios_3_u_00104.csv};
		\addplot[line width=1pt, blue] table[x=x, y=z, col sep=comma]{\relincludepath Scenario3_lineout_AcidityCompareScenarios_3_n_00104.csv};
		\addplot[line width=1pt, black] table[x=x, y=z, col sep=comma]{\relincludepath Scenario3_lineout_AcidityCompareScenarios_3_Q_00104.csv};
		\end{axis}
		\begin{axis}[%
		lineoutdefault,
		axis y line*=right,
		ylabel={pH},
		]
		\addplot[line width=1pt, green] table[x=x, y expr=(-ln(\thisrowno{4})/ln(10)), col sep=comma]{\relincludepath Scenario3_lineout_AcidityCompareScenarios_3_h_00104.csv};
		\end{axis}
	\end{tikzpicture}
}
%\end{document}
	\caption{Profiles of tumor density $\rho$, acidity $h$ (as pH), necrosis $n$, 
		and tissue density $Q$ over 52 weeks. Scenario 3: the tissue is dynamically degraded depending on its local orientation and the acidity concentration; a necrotic core is formed and develops, becoming the main part of the neoplasm. Available in color online.}
	\label{fig:scenario3-lineout}
\end{figure}
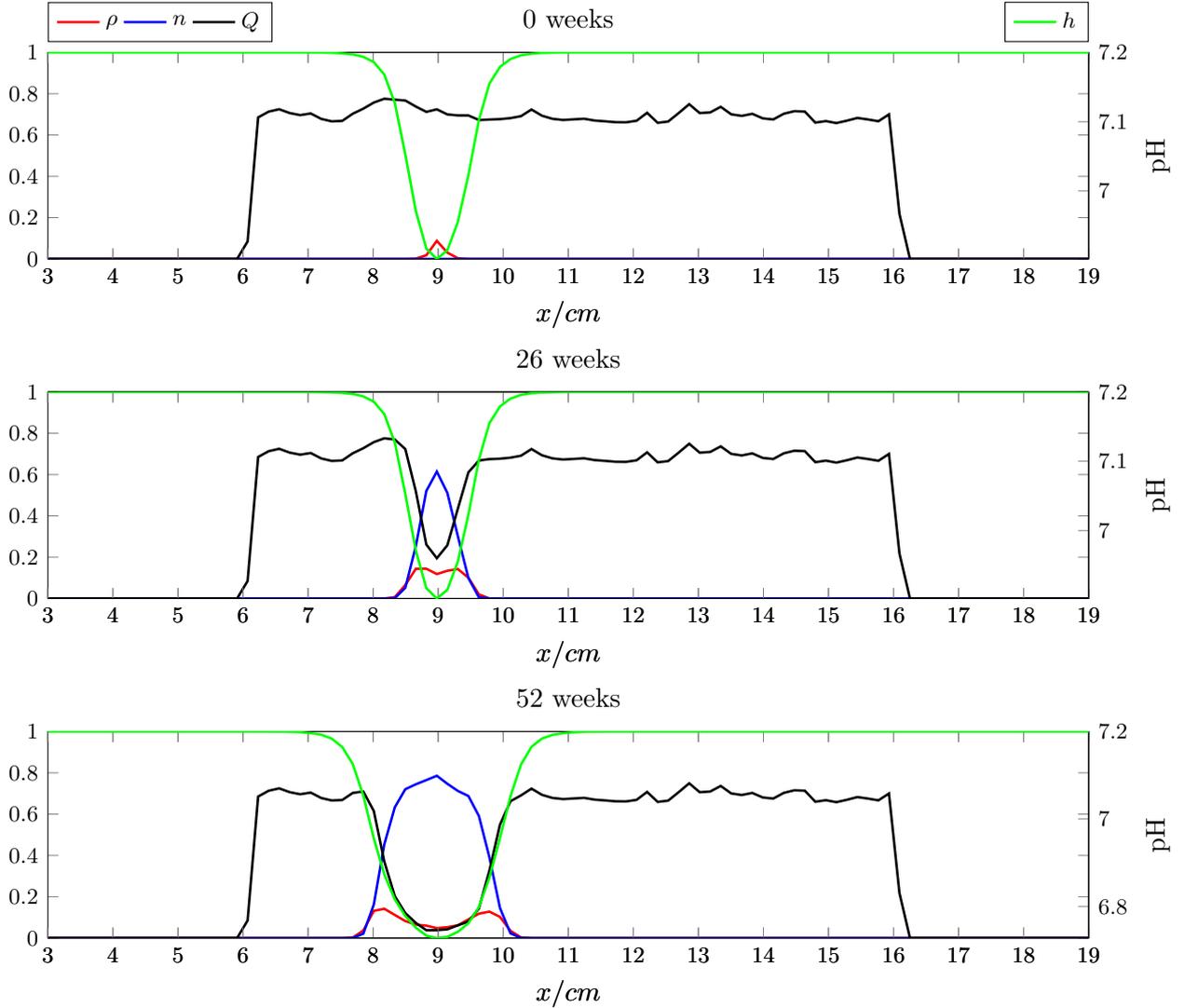

Figure \ref{fig:scenario3-lineout} shows the amplitude of the four fields $\rho,n,Q$ for densities of tumor cells, necrotic matter, and macroscopic tissue, respectively, and for the proton concentration field $h$ converted to its corresponding $pH$-value (right scale) at three distinct points in time. The uppermost plot indicates the initial condition we chose for the $\rho$- and $h$ fields, progressing then downwards with the simulation time.

As the tumor density $\rho $ increases due to the proliferation (thus also leading to an increase in acidity concentration) the cells start to accumulate at the interface with tissue due to pH-taxis and meso- and macroscopic haptotaxis. The sustained expression of protons (lower pH) leads to necrotic core formation and growth. The total neoplasm is considered to be made up by the densities of necrotic matter $n$ and living tumor cells $\rho$. At the final stage of the simulation, the former represents the largest part of the neoplasm ($n>0.6$), with a comparatively small density (approx. $\rho<0.2$) of living tumor cells. This is consistent with high grade tumors typically observed in clinical practice \cite{Hammoud1996,raza}. 

\subsubsection{Scenario 3: solution components at the end of simulation}\label{subsubsec:states}
\begin{figure}[hpt!]
	\def\relincludepath{data/postprocessing/}
	%\documentclass[crop=false]{standalone}
%
%\def\relincludepath{./}
%\input{figuremacros.tex}
%
%
%\begin{document}	
\setdefaultfigurelayout{twocol}{
	\begin{tikzpicture}
		\begin{groupplot}[group style={group size=2 by 2, horizontal sep = \figurehorizontalsep,  vertical sep = 2.5cm}, groupplotdefault, ]
			\lincolorplot{$\rho$}{sc3_u.png}{0}{1}{xmin=6, xmax=12, ymin=6, ymax=12}{xmin=6, xmax=12, ymin=6, ymax=12}
			\lincolorplot{$pH$}{sc3_h.png}{6.6}{7.2}{xmin=6, xmax=12, ymin=6, ymax=12}{xmin=6, xmax=12, ymin=6, ymax=12}
			\lincolorplot{$n$}{sc3_n.png}{0}{1}{xmin=6, xmax=12, ymin=6, ymax=12}{xmin=6, xmax=12, ymin=6, ymax=12}
			\lincolorplot{$Q$}{sc3_Q.png}{0}{1}{xmin=6, xmax=12, ymin=6, ymax=12}{xmin=6, xmax=12, ymin=6, ymax=12}
		\end{groupplot}			
	\end{tikzpicture}
}
%\end{document}
	\caption[Dynamic fields of solution components $\rho, pH=-\log_{10}(h), n, Q$ after 52 weeks in a computation for Scenario 3.]{Dynamic fields of solution components $\rho, pH=-\log_{10}(h), n, Q$ after 52 weeks in a computation for Scenario 3. Red arrows: main axes of $\mathbb{D}_W$. Green arrows: total advection vector.
	Top left: density $\rho$ of living tumor cells. 
	Top right: pH value calculated from the acidity field $pH = -\log_{10}(h)$.
	Bottom left: density $n$ of necrosis. 
	Bottom right: macroscopic density $Q$ of normal tissue. Available in color online.
	}
	\label{fig:scenario3-fields}
\end{figure}
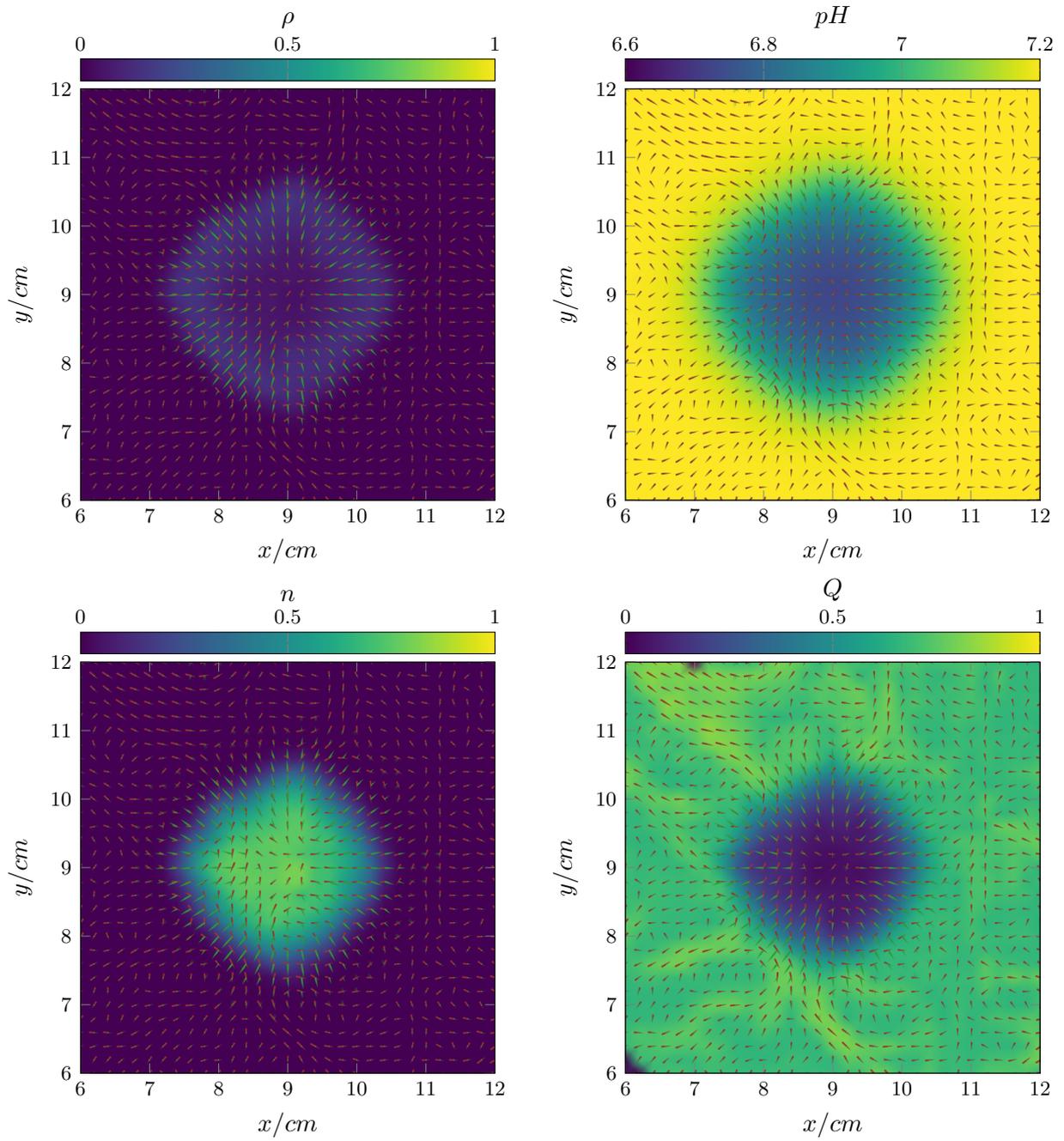

In Figure \ref{fig:scenario3-fields} we show the final states of the four fields of Scenario 3, overlain with the main axis of the water diffusion tensor $\mathbb{D}_w$ from the DTI data sets, as well as the direction of the combined advective fields. The dominant $S_2$ term describing repellent pH-taxis results in active transport of tumor mass away from the center. Similarly to the line plots presented earlier in Figure \ref{fig:scenario3-lineout}, we can identify a reduction of active tumor cells and tissue degradation within the acidic core area of the neoplasm, where necrosis is growing instead.

\subsubsection{Scenario 3: study of dominant advective effects}\label{subsubsec:dominant_effects}
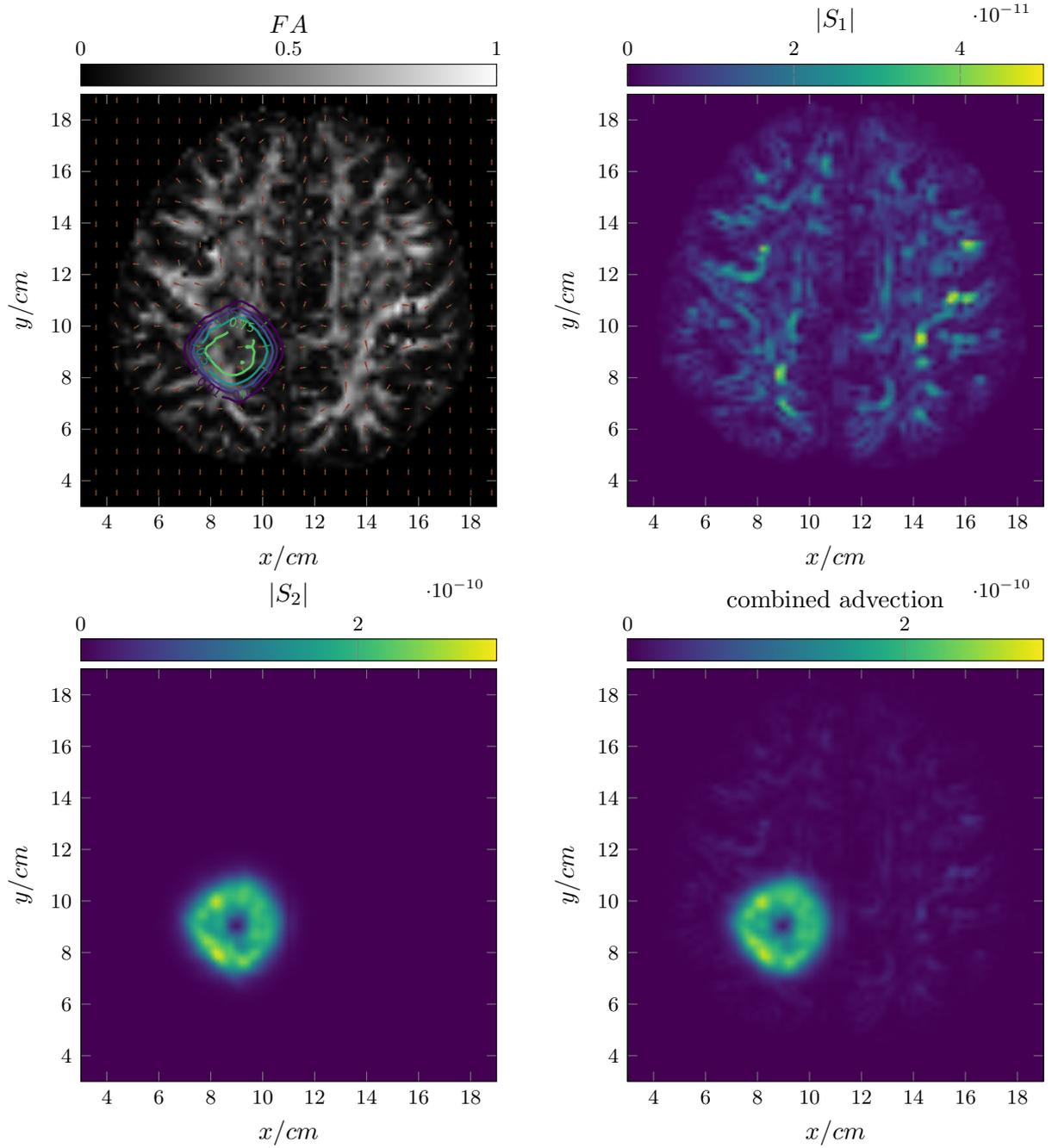
\begin{figure}[hpt!]
	\def\relincludepath{data/postprocessing/}
	%\documentclass[crop=false]{standalone}
%
%\def\relincludepath{./}
%\input{figuremacros.tex}
%
%
%\begin{document}	
\setdefaultfigurelayout{twocol}{
	\begin{tikzpicture}
		\begin{groupplot}[group style={group size=2 by 2, horizontal sep = \figurehorizontalsep,  vertical sep = 2.5cm}, groupplotdefault, ]
			\lincolorplot{$FA$}{sc3_u_full.png}{0}{1}{xmin=3, xmax=19, ymin=3, ymax=19,colorbar style={colormap name={cmgray}},}{xmin=3, xmax=19, ymin=3, ymax=19}
			\lincolorplot{$\vert S_1 \vert $}{sc3_S1_full.png}{0}{5e-11}{xmin=3, xmax=19, ymin=3, ymax=19}{xmin=3, xmax=19, ymin=3, ymax=19}
			\lincolorplot{$\vert S_2 \vert $}{sc3_S2_full.png}{0}{3e-10}{xmin=3, xmax=19, ymin=3, ymax=19}{xmin=3, xmax=19, ymin=3, ymax=19}
			\lincolorplot{combined advection}{sc3_adv_full.png}{0}{3e-10}{xmin=3, xmax=19, ymin=3, ymax=19}{xmin=3, xmax=19, ymin=3, ymax=19}
		\end{groupplot}			
	\end{tikzpicture}
}
%\end{document}
	\caption[Camino dataset and the resulting advective fields.]{Top left: fractional anisotropy of the camino dataset\cite{camino}. Contours: neoplasm density $\rho + n$ from Scenario 3 after 52 weeks. Red arrows: main axis of $\mathbb{D}_W$. Green arrows: total advection vector. 
Top right: magnitude of the $S_1$ term. Bottom left: magnitude of the $S_2$ term. 
	Bottom right: magnitude of all advective terms (myopic drift, macroscopic haptotaxis, mesoscopic haptotaxis $S_1$, repellent pH-taxis $S_2$). 
	The scales of the color maps differ between the plots. All advective terms contain the scaling factor $1/\lambda_0(x)$. Available in color online.}%	\textit{The chemotactic $S_2$ term dominates in the tumorous region.}
		\label{fig:scenario3-overview}
\end{figure}

% first theoverview plot:
To evaluate the relative importance of the four advective terms within the model,
we investigated their dynamically changing magnitudes.  We include the dynamic scaling term $1/\lambda_0(x)$ in these presentations. Figure \ref{fig:scenario3-overview} shows the fractional anisotropy of the underlying data set, the magnitudes of $S_1, S_2$, and the magnitude of all combined advective terms at 52 weeks simulation time. For the literature-based parameter set in Table \ref{tab:parameters} the chemotactic term $S_2$ is orders of magnitudes stronger than the other advective fields in this late stage of tumor progression. In the bottom right plot of Figure \ref{fig:scenario3-overview} it can be observed that the dominant advective effect at the proliferation front is given by $S_2$, while the other effects may only contribute significantly in the absence of acidity gradients, i.e. beyond the visible tumor margins. 

% On the diff-plot of the final states:
We also investigated in Figure \ref{fig:adv_tests} the relative importance of the different advective effects by successively excluding them in the simulation runs. That way, the total time-integrated effect of inclusion and exclusion of the terms could be investigated. The repellent pH-taxis described by the $S_2$-term was found to have the most dominant effect on the tumor density, but the haptotaxis terms also contribute to establishing low-density areas at the tumor edges. Given the infiltrative spread of glioma cells and post-treatment tumor recurrence which is mainly due to such relatively small cell aggregates located further away from the main tumor mass, such margins might be relevant for treatment planning, especially as far as radiotherapy is concerned (see also Subsection \ref{subsubsec:dose_painting}).
\begin{figure}[hbt!]
	\def\relincludepath{data/postprocessing/}
	% This plot compares the result of a scenario 3 simulation
% with other runs, where certain advective fields are switched off
% we present the absolute difference in those cases:	
%
%\documentclass[crop=false]{standalone}
%\def\relincludepath{./}
%\input{figuremacros.tex}
%\begin{document}	
	\setdefaultfigurelayout{twocol}{
		\begin{tikzpicture}
		\begin{groupplot}[group style={group size=2 by 2, horizontal sep = \figurehorizontalsep,  vertical sep = 2.5cm}, groupplotdefault, ]
		\lincolorplot{scenario 3: neoplasm: $\rho(x)+n(x)$, all adv. fields, 52 weeks}{sc3_neo.png}{0.0}{1.0}{xmin=6, xmax=12, ymin=6, ymax=12, colorbar style={colormap name={cmviridis}},}{xmin=6, xmax=12, ymin=6, ymax=12}
		\lincolorplot{scenario 3: no haptotaxis}{sc3_allAdv_vs_noHapto.png}{-1e-4}{1e-4}{xmin=6, xmax=12, ymin=6, ymax=12, colorbar style={colormap name={cmRdBu_r}},}{xmin=6, xmax=12, ymin=6, ymax=12}
		\lincolorplot{scenario 3 no $S_1$}{sc3_allAdv_vs_noS1.png}{-0.5}{0.5}{xmin=6, xmax=12, ymin=6, ymax=12, colorbar style={colormap name={cmRdBu_r}},}{xmin=6, xmax=12, ymin=6, ymax=12}
		\lincolorplot{scenario 3 no $S_2$}{sc3_allAdv_vs_noS2.png}{-0.8}{0.8}{xmin=6, xmax=12, ymin=6, ymax=12, colorbar style={colormap name={cmRdBu_r}},}{xmin=6, xmax=12, ymin=6, ymax=12}
		\end{groupplot}			
		\end{tikzpicture}
	}
%\end{document}
\caption[The effect of advective fields on the model predictions.]{The effect of advective fields on the model predictions. Red arrows: main axis of $\mathbb{D}_W$. Green arrows: total advection vector. All results are obtained for Scenario 3. 
 Top left: simulation result for the density $\rho+n$ of the neoplasm after 52 weeks. The rest of the plots show absolute differences in $\rho+ n$ between the full model and Scenario 3 without one of the adective terms. 
 Top right: no macroscopic haptotaxis.  Bottom left: no $S_1$ advection. Bottom right:  no $S_2$ advection.  The scales of the color maps differ between the plots.  Available in color online.
 } \label{fig:adv_tests}
\end{figure}

\subsubsection{Comparison between model scenarios}\label{subsubsec:compare_scenarios}
\begin{figure}[hpt!]
	\def\relincludepath{data/postprocessing/}
	%\documentclass[crop=false]{standalone}
%
%\def\relincludepath{./}
%\input{figuremacros.tex}
%
%
%\begin{document}	
\setdefaultfigurelayout{twocol}{
	\begin{tikzpicture}
		\begin{groupplot}[group style={group size=2 by 2, horizontal sep = \figurehorizontalsep,  vertical sep = 2.5cm}, groupplotdefault, ]
			\lincolorplot{Scenario 1: $\rho + n$}{sc1_neo.png}{0}{1}{xmin=6, xmax=12, ymin=6, ymax=12}{xmin=6, xmax=12, ymin=6, ymax=12}
			\lincolorplot{Scenario 3: $\rho + n$}{sc3_neo.png}{0}{1}{xmin=6, xmax=12, ymin=6, ymax=12}{xmin=6, xmax=12, ymin=6, ymax=12}
			\lincolorplot{Scenario 3 - Scenario 2}{sc3_diff32.png}{-0.02}{0.02}{xmin=6, xmax=12, ymin=6, ymax=12, colorbar style={colormap name={cmRdBu_r}},}{xmin=6, xmax=12, ymin=6, ymax=12}
			\lincolorplot{Scenario 3 - Scenario 1}{sc3_diff31.png}{-0.3}{0.3}{xmin=6, xmax=12, ymin=6, ymax=12, colorbar style={colormap name={cmRdBu_r}},}{xmin=6, xmax=12, ymin=6, ymax=12}
		\end{groupplot}			
	\end{tikzpicture}
}
%\end{document}
	\caption[Comparison of the scenarios.]{Comparison of the scenarios. Contours: density $\rho + n$ of the neoplasm computed in the framework of Scenario 3. Red arrows: main axis of $\mathbb{D}_W$. Green arrows: total advection vector.
	Top row: density $\rho + n$ at the final time (52 weeks) for Scenarios 1 and 3. 
	Bottom row: pointwise difference in $\rho + n$ between Scenarios 3 and 2 and difference between Scenarios 3 and 1. 
	Note: the scales of the color maps differ between the plots. Available in color online.} 
	\label{fig:scenario_compare}
\end{figure}
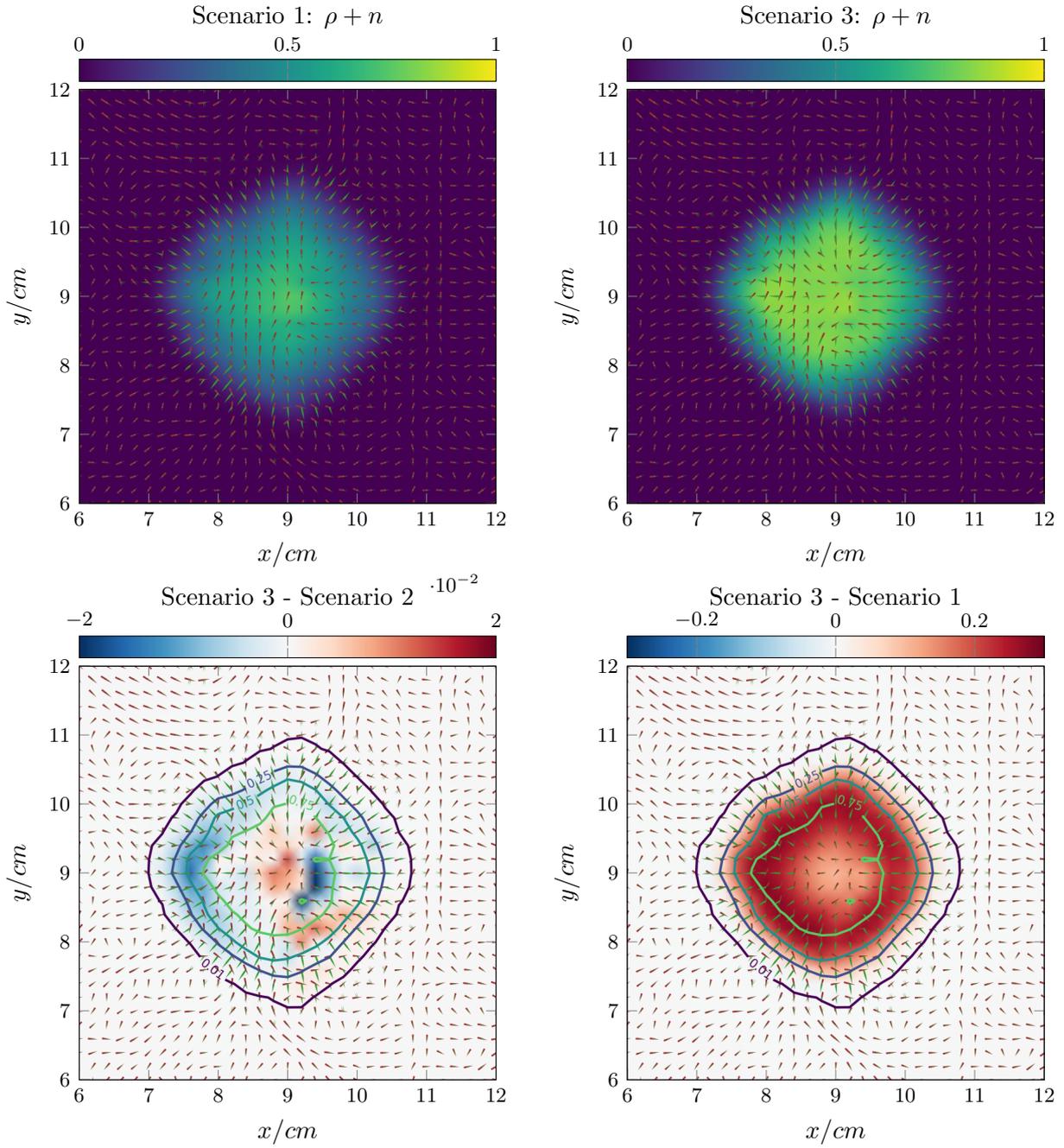

In the following we compare the simulation results for Scenarios 1-3 with identical initial condition, in order to find out whether there are any potentially relevant differences. Figure \ref{fig:scenario_compare} shows the density $\rho+n$ of the neoplasm after 52 weeks. As expected, Scenario 3 involving three-fold taxis and myopic diffusion predicts the largest tumor spread, with the most substantial difference being that w.r.t. Scenario 1. 
It is a modeling problem in itself to predict how living cells and necrotic matter  influence the gray scale images in different ways. A direct comparison of Scenarios 2 and 3 with medical images is therefore not possible without knowing the underlying mapping of the two species to the MRI gray scale imaging. However, the distinction of $\rho$ and $n$ allows deeper insights into the dynamics of the tumor expansion. The two plots in the second row of Figure \ref{fig:scenario_compare} illustrate the differences between the densities $\rho+n$ after 52 simulated weeks, computed upon using Scenarios 3-2, and 3-1. Thus, the latter comparison shows that taking into account the evolution of normal tissue is of high relevance, while considering the dynamics of mesoscopic tissue distribution $q$ (i.e the step from Scenario 2 to 3) is contributing less (please note the different scales of the color maps therein). Hence, the evolution of macroscopic tissue density $Q$ seems to contribute most to the prediction of tumor spread. 

\begin{figure}[hpt]
	\def\relincludepath{data/postprocessing/}
	%\documentclass[crop=false]{standalone}
%
%\def\relincludepath{./}
%\input{figuremacros.tex}
%
%
%\begin{document}	
	\setdefaultfigurelayout{twocol}{
		\begin{tikzpicture}
		\begin{groupplot}[group style={group size=2 by 2, horizontal sep = \figurehorizontalsep,  vertical sep = 2.5cm}, groupplotdefault, ]
		\lincolorplot{Initial $FA$}{sc3_FA_0.png}{0}{1}{xmin=6, xmax=12, ymin=6, ymax=12, colorbar style={colormap name={cmgray}},}{xmin=6, xmax=12, ymin=6, ymax=12}
		\lincolorplot{$FA$ after 52 weeks}{sc3_FA.png}{0}{1}{xmin=6, xmax=12, ymin=6, ymax=12, colorbar style={colormap name={cmgray}},}{xmin=6, xmax=12, ymin=6, ymax=12}
		\lincolorplot{$FA$ difference}{sc3_FA_diff.png}{-0.25}{0.25}{xmin=6, xmax=12, ymin=6, ymax=12, colorbar style={colormap name={cmRdBu_r}},}{xmin=6, xmax=12, ymin=6, ymax=12}
		\end{groupplot}			
		\end{tikzpicture}
	}
%\end{document}
	\caption[Tissue dynamics in Scenario 3.]{Tissue dynamics in Scenario 3. Contours: neoplasm density $\rho + n$ from Scenario 3. Red arrows: main axis of $\mathbb{D}_W$. Green arrows: total advection vector.
	Top left: gray scale plot of $FA$ at the start of the simulation. 
	Top right: $FA$ after 52 weeks, with visible reduction of anisotropy closer to the tumor centre. 
	Bottom left: difference in $FA$ between final and initial simulation times. Available in color online.} 
	\label{fig:tissue_dynamics}
\end{figure}
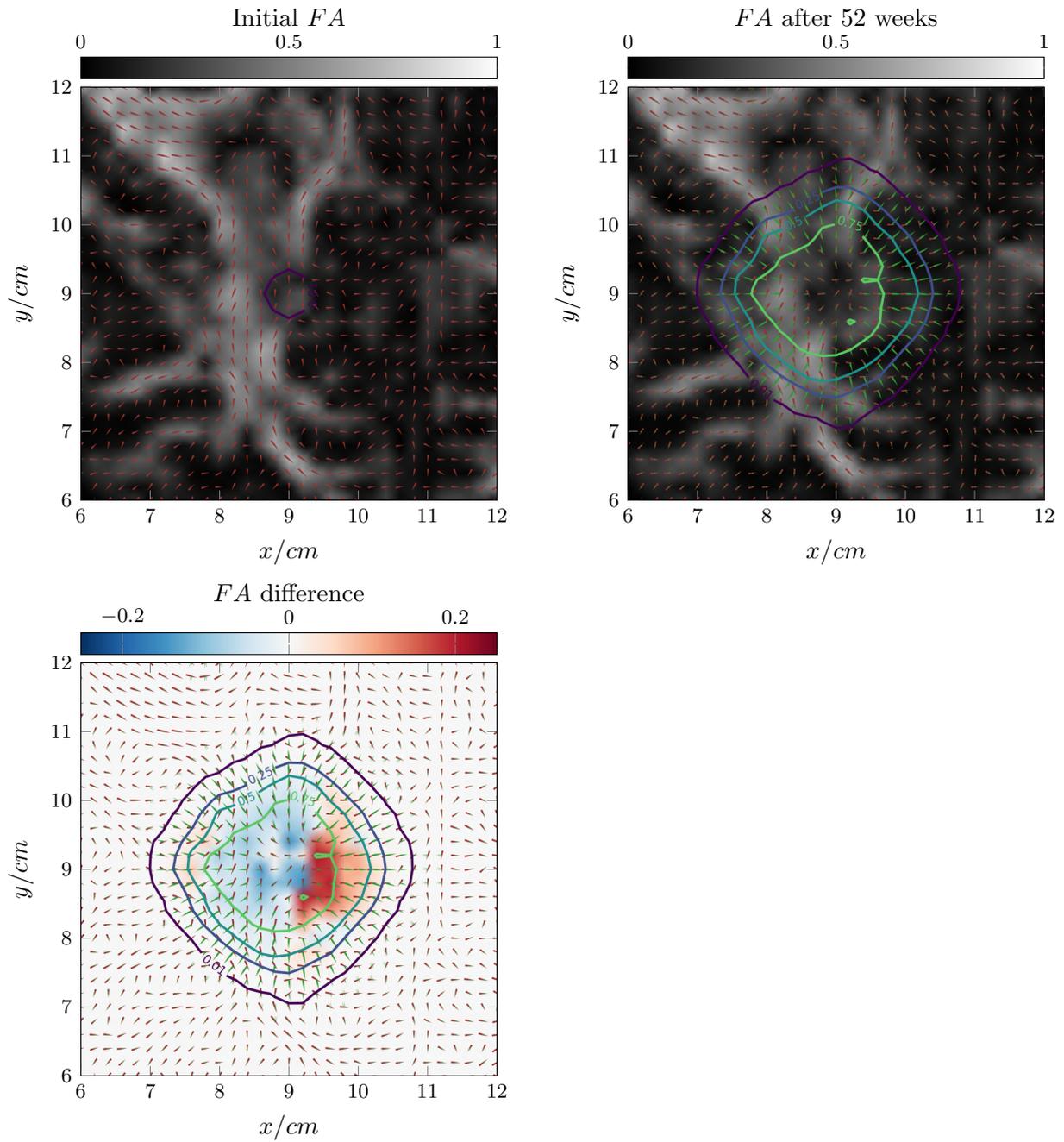

On the other hand, the dynamics of mesoscopic tissue $q$ contributes to substantially modifying the fractional anisotropy FA, which is a key observable in the DTI imaging of brain tumors. Our new model allows to directly model the depletion of brain structure and the therewith associated changes in FA, as indicated in Figure \ref{fig:tissue_dynamics}. The two plots of the first row show the initial fractional anisotropy and how it is diminished by the tumor growth after 52 weeks, again overlain by the direction of the total advective fields. The plot on the second row illustrates the difference in FA between the final and the initial simulation times. \\
The results indicate that tissue dynamics can be seen as a downstream effect, after tumor density and acidity have changed from baseline. The differences in $FA$ are therefore most prominent closer to the tumor center, and not so much at the invasion edge. Scenario 3 is more complex and requires more effort to treat, but beyond the differences put in evidence in Figures \ref{fig:scenario_compare} and \ref{fig:tissue_dynamics} it also provides a framework for a careful description of macroscopic tissue dynamics - along with its effects on the other solution components.

Still with the aim of investigating the influence of evolving tissue, we also compare a reduced version of Scenario 3 with a previous model obtained in \cite{Painter-Hillen} and accounting for the effect of tissue anisotropy on the migration of glioma cells, however with fixed tissue. Specifically, we compare the PDE  
\begin{align}\label{eq:PH-model}
\rho_t=\nabla \nabla : \left (\frac{1}{\lambda _0}\mathbb D_T\rho \right )+\eta \rho (1-\rho)
\end{align}
which is a slight modification of that in \cite{Painter-Hillen}, by adding the proliferation term and letting $\lambda_0$ be nonconstant, but depend on $x$ like in the rest of this paper, with the system
\begin{subequations}\label{eq:red_scenario3}
\begin{align}
&\rho_t=\nabla \cdot \left (\frac{1}{\lambda _0}\nabla \cdot (\mathbb D_T\rho)-\lambda _1g(Q,\lambda_0)\mathbb D_T\nabla Q\rho -\alpha S_1(q)\rho \right )+\eta \rho (1-\rho)\\
&q_t=r_0\rho q\left (\Pi_a[q]-A[q]\right)\\
&Q_t=r_0\rho Q\left (A[q]-1\right ),
\end{align}
\end{subequations}
both with the same initial conditions and no-flux boundary conditions and the same function $\lambda_0$. Figure \ref{fig:comp_PH_plots} shows the computation results. Note that none of the two model makes any distinction between active and necrotic tumor. The complete tumor mass is encoded in $\rho$. As before, the evolution of tissue leads to substantial changes in FA, although the dynamics of acidity and its influence are absent in these settings. The simulations predict a larger area of glioma spread when the cells are allowed to degrade the tissue; although there are rather small amounts of tumor cells in the enlarged regions, these might be relevant for tumor recurrence. Model \eqref{eq:PH-model} leads to higher cell densities at the core of the neoplasm, due to the migrating cells only performing myopic diffusion and not being supplementary driven by haptotaxis.

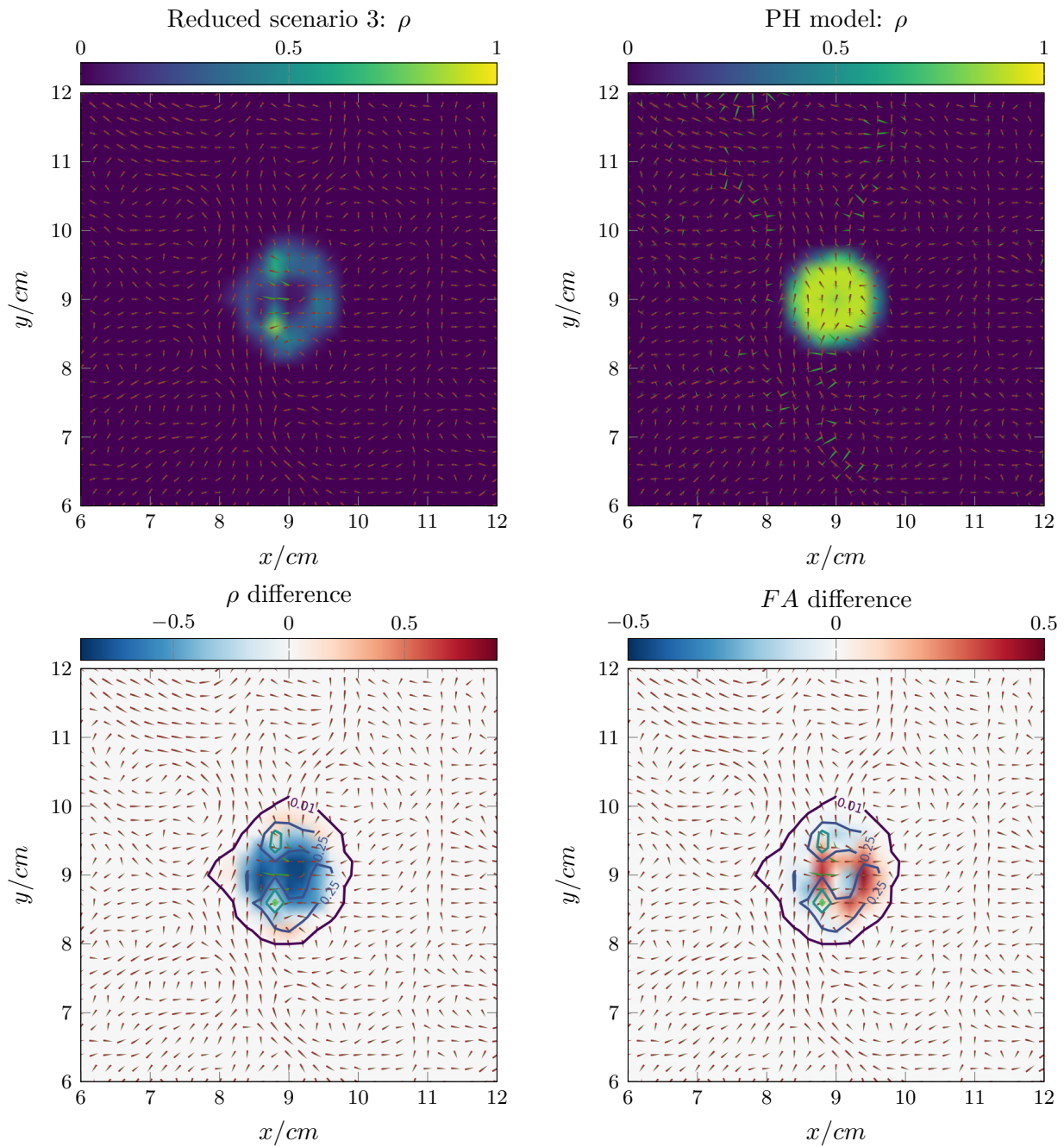
\begin{figure}[hpt!]
	\def\relincludepath{data/postprocessing/}
	%\documentclass[crop=false]{standalone}
%
%\def\relincludepath{./}
%\input{figuremacros.tex}
%
%
%\begin{document}	
	\setdefaultfigurelayout{twocol}{
		\begin{tikzpicture}
		\begin{groupplot}[group style={group size=2 by 2, horizontal sep = \figurehorizontalsep,  vertical sep = 2.5cm}, groupplotdefault, ]
		\lincolorplot{Reduced scenario 3: $\rho$}{sc3_only_tissue.png}{0}{1}{xmin=6, xmax=12, ymin=6, ymax=12, colorbar style={colormap name={cmviridis}},}{xmin=6, xmax=12, ymin=6, ymax=12}
		\lincolorplot{PH model: $\rho$}{sc3_ph.png}{0}{1}{xmin=6, xmax=12, ymin=6, ymax=12, colorbar style={colormap name={cmviridis}},}{xmin=6, xmax=12, ymin=6, ymax=12}
		\lincolorplot{$\rho$ difference}{sc3_diff_ph.png}{-0.9}{0.9}{xmin=6, xmax=12, ymin=6, ymax=12, colorbar style={colormap name={cmRdBu_r}},}{xmin=6, xmax=12, ymin=6, ymax=12}
		\lincolorplot{$FA$ difference}{sc3_only_tissue_FA_diff.png}{-0.5}{0.5}{xmin=6, xmax=12, ymin=6, ymax=12, colorbar style={colormap name={cmRdBu_r}},}{xmin=6, xmax=12, ymin=6, ymax=12}
		\end{groupplot}			
		\end{tikzpicture}
	}
%\end{document}
	\caption{Comparison between models \eqref{eq:red_scenario3} and \eqref{eq:PH-model}. Contours: tumor density $\rho$ from the computation of \eqref{eq:red_scenario3} (purple: $\rho=0.1$, blue: $\rho=0.25$, green: $\rho=0.5$). Red arrows: main axis of $\mathbb{D  }_W$. Green arrows: total advection vector. 
	Top left: tumor density $\rho$ at the final time (52 weeks) for model \eqref{eq:red_scenario3}. Top right: tumor density $\rho$ for model \eqref{eq:PH-model}. 
	Bottom left: difference in $\rho$ computed with model \eqref{eq:red_scenario3} and with model \eqref{eq:PH-model}.
	Bottom right: Difference in $FA$ between end and beginning of simulation, computed with model \eqref{eq:red_scenario3}. Available in color online. 
	}
\end{figure}\label{fig:comp_PH_plots}

\subsubsection{Tumor grading and importance of tissue evolution} \label{subsubsec:tumor_grading}
Tumor grade assessment is decisive for the treatment planning of glioma and for patient survival prognosis. Typically, glioma are classified according to cell activity and tumor aggressiveness, upon relying on a series of indicators, including histological patterns \cite{brat,wippold} and tumor composition. Among these, grading by the amount of necrosis relative to the whole tumor volume (both visible via biomedical imaging) has been addressed e.g., in \cite{Hammoud1996,kros,raza}. In fact, \cite{Hammoud1996} observed that 'location and volume of tumors were not statistically significant predictors of survival', which is in accordance to most conducted studies, as reviewed e.g. in \cite{Henker}.  We therefore address here only necrosis-based grading. In \cite{Hammoud1996} the percentage of necrotic matter visible on MRI is related to the total visible tumor volume. 
An attempt at tumor grading by way of connecting the results of numerical simulations for our model to this definition proves to be difficult without in-depth information on how the tissue types (e.g., obtained by segmentation) affect the MRI shading. The numerical results of our model do not contain any thresholding or shading effects, therefore we define the time-dependent grade $G(t)\in [0,1]$ of the simulated tumors via:
%Connecting the results of numerical simulations to this definition proves hard without in-depth information on how the tissue types effect the MRI shading. In the numerical results we have no thresholding or shading effects and define the grade $G \in [0,1]$ of the simulated tumors via:
\begin{equation}
G(t):= \frac{V_{n}(t)}{V_{n}(t) +V_{\rho}(t)}
\label{eq:tumor_grade_def}
\end{equation}
where $V_{n}(t)$ and $V_{\rho}(t)$ denote integrals over the whole space domain $\Omega$ of the necrosis and living cell densities $n(t,x)$  and $\rho(t,x)$, respectively. We investigate then the evolution of this quantity over time, upon being guided by the percentage classification in \cite{Hammoud1996}: $0<G<25\%$: grade 1; $25\%\le G< 50\%$: grade 2; $G\ge 50$: grade 3, with the highest grade corresponding to the most aggressive tumor and the poorest prognosis.
%The numerical simulations of the new model could be used to grade the tumor via the necrotic tissue for radiation therapy (Figure \ref{fig:necrotic_grading}).
\begin{figure}[hpt]
	\centering
	\includegraphics[width=\linewidth]{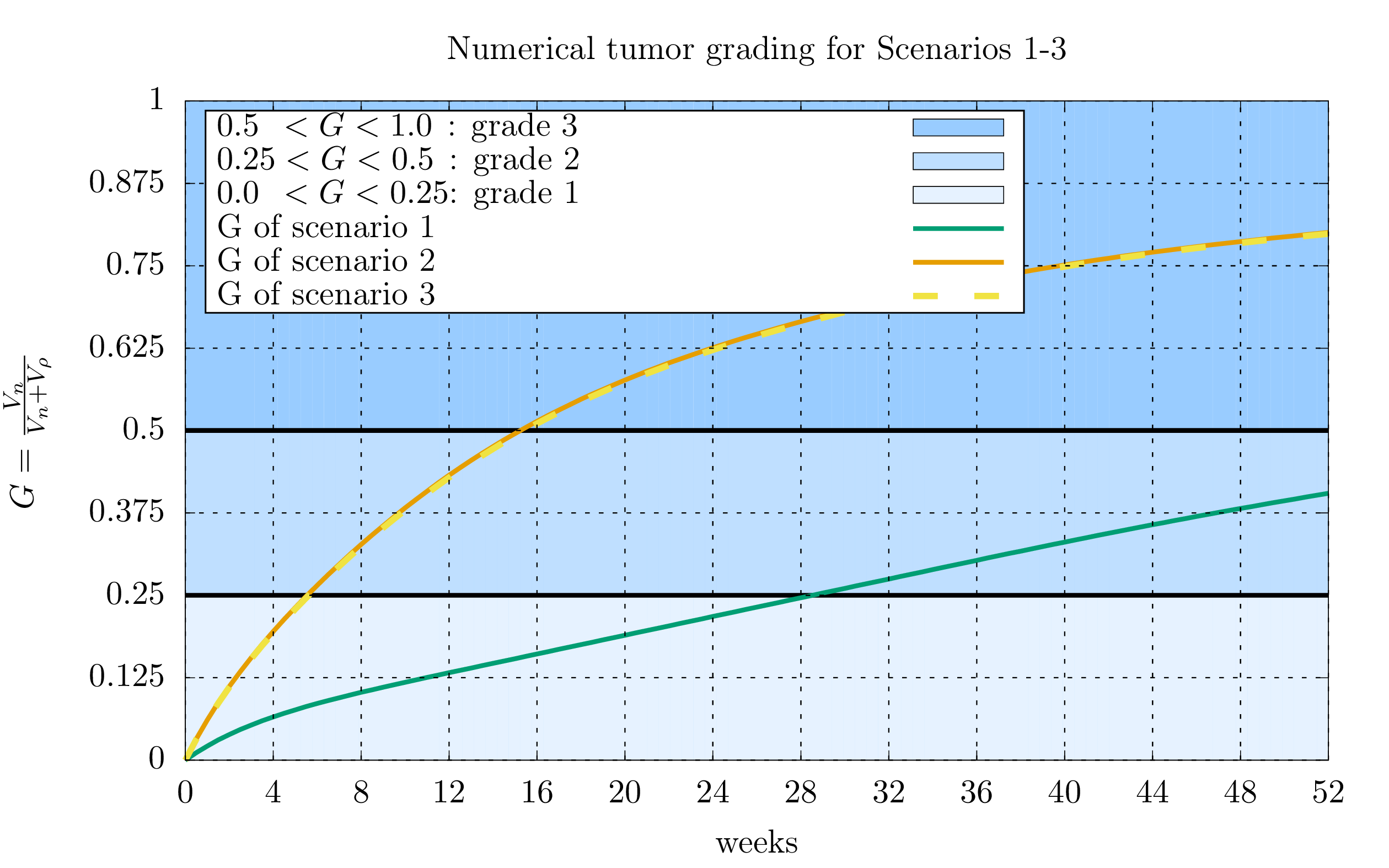} % We may use .pdf if file-size is not restricted
	\caption{Grade of tumor \eqref{eq:tumor_grade_def} for the three considered Scenarios. $G$ is essentially the same for Scenarios 2 and 3, whereas in Scenario 1 (with the same parameter choice) it exhibits  significantly slower progress. Full information is assumed, so that there is no visibility threshold. Scenario 1 predicts the progression from grade 1 towards grade 2 after about 28 weeks, whereas in Scenarios 2 and 3 it needs only about 5 weeks. The progression to grade 3 occurs  after about 16 weeks for Scenarios 2 and 3. For Scenario 1 it does not happen within the simulated time of 52 weeks (recall that the median survival time of patients with glioma is approx. 60 weeks \cite{wrensch}).} \label{fig:necrotic_grading}
\end{figure}

Figure \ref{fig:necrotic_grading} illustrates the evolution of the tumor grade for the three considered scenarios. We found only tiny differences between the grades for Scenarios 2 and 3, whereas the difference in tumor grade progression between Scenarios 1 and 2 is substantial. The definition \eqref{eq:tumor_grade_def} of grade $G$ effectively skips grade 0, as there is no threshold on the visibility of the necrotic tissue density $n$. The fact that Scenario 1 fails to predict tumor progression into grade 3 in due (biomedically realistic) time might be an indicator of Scenario 1 being insufficient for modeling glioma dynamics, thus endorsing the role played by tissue evolution. %, which would be expected in realistic cases, may be seen as an indication that the reduced scenario 1 is insufficient.

\subsubsection{Dose painting} \label{subsubsec:dose_painting}
%The inclusion of the necrotic density into the models allows us to simulate dose painting for radiation treatment. A possible approach could be to characterize different regions of the tumor necrotic tumor bulk via a levelset on the simulated necrotic density.
Beyond tumor grading based on assessment of necrotic matter and heterogeneity of the neoplasm, the model can provide valuable information about the extent and shape of the tumor and hence help establishing the contours for radiation treatment. The ability to predict space-time densities of clonogens as well as necrosis opens the way for more  accurate tumor segmentation and dose painting. Figure \ref{fig:dose_painting} shows a gray scale image of the computed neoplasm ($\rho +n$) for a simulated time of 52 weeks. Level sets for the density of living glioma cells are shown, thus providing information about the regions where higher radiation doses should be applied in order to eradicate the active cells, at the same time sparing areas where their density is lower (either as a consequence of necrosis - in the tumor core, or due to less cells having invaded the healthy tissue at the tumor margins).

\begin{center}
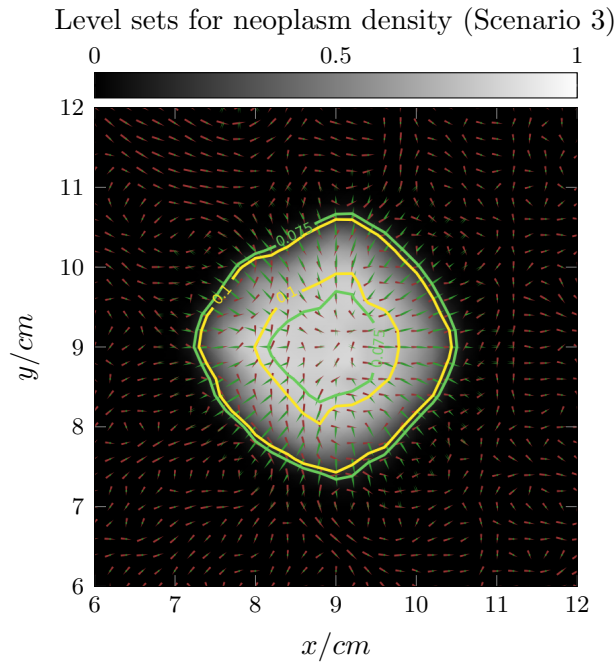
\begin{figure}[hpt]
	\centering
	\def\relincludepath{data/postprocessing/}
	%\documentclass[crop=false]{standalone}
%
%\def\relincludepath{./}
%\input{figuremacros.tex}
%
%\begin{document}	
\setdefaultfigurelayout{twocol}{
	\begin{tikzpicture}
		\begin{groupplot}[group style={group size=2 by 2, horizontal sep = \figurehorizontalsep,  vertical sep = 2.5cm}, groupplotdefault, ]
				\lincolorplot{Level sets for neoplasm density (Scenario 3)}{sc3_dose_painting_neo.png}{0}{1}{xmin=6, xmax=12, ymin=6, ymax=12, colorbar style={colormap name={cmgray}},}{xmin=6, xmax=12, ymin=6, ymax=12}
		\end{groupplot}			
	\end{tikzpicture}
}
%\end{document}
	\caption{Level sets for density $\rho$ of living tumor cells could be used for dose painting in radiotherapy. Gray scale image: density of neoplasm  $\rho + n$, computation from the reduced Scenario 3, simulation of 52 weeks.
	 Level sets: $\rho= 0.1$ (yellow) and $\rho=0.075$ (green).}
	\label{fig:dose_painting}
\end{figure}
\end{center}
%\cbl{The presented model includes more fields in the description of the tumor progression. The information on the acidity field $h(x)$ and the necrosis $n(x)$ could be used to develop more advanced dose painting procedures which take this new information into account.}

\section{Discussion} \label{sec:discussion}
%\cb {TODO: Comparison with previous models, advantages of this approach, what terms did we obtain and which terms not, explanation.}

We proposed in this paper a novel model for glioma invasion, which takes into account the interaction with the biochemical and biophysical tumor microenvironment, represented here by acidity acting as a repellent and, respectively, oriented tissue fibers guiding the cell migration. Both environmental components, most prominently the acidity, also influence the growth or depletion of tumor cells.

The modeling approach has a multiscale character: it starts from ODE descriptions of single cell (position, velocity, activity variable) dynamics, which are translated into a kinetic transport equation for the space-time distribution function of cells sharing the same regimes of velocity and activity variables, and eventually arrives at a macroscopic PDE for the evolution of the whole tumor as a population of such cells performing myopic diffusion and multiple taxis. The cell behavior is thereby dynamically coupled to that of the environment, the latter featuring a system of macroscopic and mesoscopic (integro-)differential equations with dependencies on space, time, and orientation of tissue fibers. We refer to \cite{KSSSL} for a review of existing models of cell migration with multiple taxis and therewith associated challenges. The diffusion and taxis coefficients of the PDE obtained here for the cell density and the source terms of the equations for meso- and macrocopic tissue and necrosis densities involve nonlocalities w.r.t. the orientations of tissue fibers and cell velocities; see \cite{CPSZ} for a review of models for cell migration involving this and other types of nonlocality. 

There are several alternative ways to include cell level environmental influences in a KTAP modeling framework, ultimately leading in the macroscopic limit to taxis terms:
\begin{itemize}
	\item[(i)] by using in the operator describing velocity reorientations turning rates which depend on the pathwise gradient of one or several signals, as done e.g. in \cite{LoPr20,OtHi};
	\item[(ii)] by describing the dynamics of activity variables for the cell state, which translates into corresponding transport terms in the kinetic equations and turning rates depending on the activity variable(s), see \cite{CKSS,EHKS,EHS,EKS,Hunt,HS} for glioma interacting with tissue and e.g., \cite{EO04,XueOth09} for swimming bacteria biased by some chemoattractant;
	\item[(iii)] by accounting for biochemical and/or biophysical effects translated into cell stress and forces acting on the cells and leading to transport terms w.r.t. the velocity variable in the kinetic PDE on the mesolevel, see \cite{CHP} for the case of cancer cells responding to a 'chemotaxis force' proportional to the gradient of a given chemoattractant concentration;
	\item[(iv)] by considering turning operators where the cell reorientation depends on the interaction with (some of) the environmental cues and using equilibrium distributions for the moment closure, as e.g. in \cite{CHP}.
\end{itemize}
While (i) and (iv) have a rather mesoscopic flavor, (ii) and (iii) refer closer to microscopic, single cell descriptions, therefore we employed in this work a combination of the latter: (ii) led to the macroscopic haptotaxis term, while (iii) with the description of cell stress and associated, speed-preserving forces acting on glioma cells yielded the terms characterizing repellent pH-taxis and mesoscopic haptotaxis. Notice that the coupling with evolving acidity and macro- and mesoscopic tissue led to genuine taxis, unlike previous models \cite{CKSS,EHKS,EHS,EKS,Hunt,HS} where the tactic cue (e.g., macroscopic tissue) was only space-dependent. To our knowledge dynamics of such complexity has not been treated in previous works, neither numerically nor from an analytical viewpoint. In \cite{Painter09} were performed interesting simulations on the kinetic level for a simpler model of cells migrating through the extracellular matrix with oriented fibers, which they degrade according to certain rules. Our closed macro-meso description \eqref{macro-parabolic-full} has the advantage of a lower dimensional phase space, thus allowing -among others- to accommodate even further possibly relevant biological effects with a rather detailed description via (ii) and/or (iii), without facing problems due to a prohibitive increase of the number of kinetic variables and therewith associated computational costs.

Concerning possible relationships between (i)-(iv), it has been proved in \cite{PeTaWa} that (ii) actually implied (iii) under certain assumptions on the subcellular dynamics for a model of bacterial run-and-tumble swimming. In \cite{KuSu20} such connection was investigated in a less rigorous way for a model describing glioma pseudopalisade patterning under the influence of repellent pH-taxis and anisotropic, but non-evolving tissue.

In order to pass from the KTAP framework to effective equations for the tumor spread we performed here a hyperbolic as well as a parabolic upscaling, according to the underlying tissue being directed or not. The latter is still not clarified from a biological viewpoint; the study (via numerical simulations) performed in \cite{KuSu20} for the mentioned model of glioma pseudopalisades revealed that such patterns which are invariably observed on histological samples of high grade tumors \cite{brat,wippold} formed in due time when using the system obtained in the parabolic limit, while the hyperbolic scaling led to a drift-dominated system predicting only a rapid shift of the initial aggregate of glioma cells towards the leading tissue orientation, without pattern development. This suggests that brain tissue might be undirected and induced us here to perform numerical simulations only for the equations obtained by parabolic scaling. 

%\cmg{We introduce in this paper two alternative perspectives on the different scales that connect the microscopic description with the macroscopic one of the GBM propagation models. The approximation given by the scale that leads to a linear diffusion does not seem the most appropriate from the point of view of an adaptation to the experimental reality, since it does not accurately capture the propagation front of the tumors. However, the purely hyperbolic scale radically eliminates a certain degree of uncertainty caused by not taking into account or not fully knowing about other processes involved in GB dynamics, such as certain biochemical (morphogenetic) or mechanical components (tissue heterogeneity, stiffness, pressure, etc.).\\
%\cmg{		Our proposal goes beyond the hyperbolic limit by calculating the following order in the asymptotic development with respect to the scale parameter. This order is physically relevant since the scale parameter contains information about the interaction between the micro and macro scales and can be incorporated from the experiments. It also provides a small correction to the hyperbolic limit that provides a better and more realistic approximation to the parabolic limit.\\
Our model makes use of DTI data and accounts for the evolution of brain tissue, transferring biomedical information through the directional distribution of tissue fibers. As mentioned before, this information belongs to both micro and macro worlds and they coexist in the macroscopic limit. In Subsection \ref{sec:par-scaling} we put in evidence the versatility of this model w.r.t. the influence of (spatially varying) tissue anisotropy in characterizing the tumor spread, thereby also considering the impact of hypoxia. In fact, our study of dominant advective effects suggests that pH-taxis is the dominating drift - at least as far as this model and its parameterization are concerned. In this work we used the approaches (ii) and (iii) to describe cell-tissue interactions and (iii) for the response of cells to acidity. This led to the anisotropy-triggered switch between myopic diffusion with 'meso-haptotaxis' and the migratory behavior with the supplementary 'macro-haptotaxis'. Instead, using (ii) for cell-acid interplay and correspondingly define the turning rate would conduct to a hypoxia-triggered switch. Several other options of combining (ii) and/or (iii) can be conceived as well, raising further questions about the relationship of these alternative approaches. How appropriate a particular model is for a specific problem eventually depends on a multitude of factors and can properly be assessed when validated with adequate patient data. The current high costs of collecting such data still keeps a reliable validation  out of reach, but the unprecedented development of biomedical technology during the last decade gives hope of it becoming available in the near future.

\section*{Acknowledgement} GC, CE, AK, CS, and MW acknowledge funding by the Federal Ministry of Education and Research BMBF in the project \textit{GlioMaTh} 05M2016. This work 
has been also partially supported by the MINECO-Feder (Spain) research grant number RTI2018-098850-B-I00, the Junta de Andaluc\'ia (Spain) Project PY18-RT-2422 \& A-FQM-311-UGR18 (JN, JS). We also thank our cooperation partners C. Berdel, Y. Dzierma, S. Knobe, W. Reith, and C. R\"ube from the Saarland University Hospital for their support and interesting meetings and discussions on glioma cancer and its treatment.

\end{document}